\documentclass[]{IEEEtran}
\usepackage[cmex10]{amsmath}
\usepackage{amsfonts,amssymb,amsthm}
\usepackage{graphicx,relsize,verbatim,enumerate,xcolor}
\usepackage{tikz}
\usepackage{pgfplots}
\usepackage{subfig}
\usepackage{flushend}

\usetikzlibrary{arrows,positioning,fit,backgrounds,shapes}
\usetikzlibrary{calc}

\newtheorem{definition}{Definition}
\newtheorem{theorem}{Theorem}[section]
\newtheorem{lemma}[theorem]{Lemma}
\newtheorem{corollary}[theorem]{Corollary}

\newcommand{\dist}[2][p]{\ensuremath{\MakeUppercase{#1}_{#2}(\MakeLowercase{#2})}}
\newcommand{\distns}[2][p]{\ensuremath{#1(#2)}}
\newcommand{\distnarg}[2][p]{\ensuremath{\MakeUppercase{#1}_{#2}}}
\newcommand{\distalt}[3][p]{\ensuremath{\MakeUppercase{#1}_{#2}(#3)}}

\newcommand{\rightratearrow}[5]{
\draw ([yshift=#3] #1) to node [midway,above] {#5} ([yshift=#3,xshift=-#4] #2) -- ++(0,#3) -- (#2);
\draw ([yshift=-#3] #1) -- ([yshift=-#3,xshift=-#4] #2) -- ++(0,-#3) -- (#2);
                            }

\newcommand{\downratearrow}[6]{
\draw ([xshift=#3] #1) to node[midway,minimum width=0pt,inner sep=0pt,right=#6] {#5} ([xshift=#3,yshift=#4] #2) -- ++(#3,0) -- (#2);
\draw ([xshift=-#3] #1) -- ([xshift=-#3,yshift=#4] #2) -- ++(-#3,0) -- (#2);
}

\newcommand{\upratearrow}[6]{
\draw ([xshift=#3] #1) to node[midway,minimum width=0pt,inner sep=0pt,right=#6] {#5} ([xshift=#3,yshift=-#4] #2) -- ++(#3,0) -- (#2);
\draw ([xshift=-#3] #1) -- ([xshift=-#3,yshift=-#4] #2) -- ++(-#3,0) -- (#2);
}

\newcommand{\downrightratearrow}[4]{
\draw ([yshift=#3] #1) to [out=0,in=90] ([xshift=#3,yshift=#4] #2) -- ++(#3,0) {[line join = bevel] -- (#2)} -- ++(-#3-#3,#4)--([xshift=-#3,yshift=#4] #2) to [out=90,in=0] ([yshift=-#3] #1);
                            }

\newcommand{\downleftratearrow}[4]{
\draw ([yshift=-#3] #1) to [out=180,in=90] ([xshift=#3,yshift=#4] #2) -- ++(#3,0) {[line join = bevel] -- (#2)} -- ++(-#3-#3,#4)--([xshift=-#3,yshift=#4] #2) to [out=90,in=180] ([yshift=#3] #1);
                            }

\tikzstyle{arw}=[->,>=latex]
\tikzstyle{node}=[rectangle,draw,outer sep=0pt,minimum width=1.7cm, minimum height=8mm]

\begin{document}

\title{Distributed Channel Synthesis}

\author{
  Paul~Cuff
  \thanks{This work is supported by the National Science Foundation (grant CCF-1116013) and the Air Force Office of Scientific Research (grant FA9550-12-1-0196).}
  \thanks{P. Cuff (cuff@princeton.edu) is with Department of Electrical Engineering at Princeton University.}
  \thanks{This paper was presented in part at ISIT 2008 \cite{cuff08-channel-synthesis-ISIT}.}
}

\maketitle

\begin{abstract}
  Two familiar notions of correlation are rediscovered as the extreme operating points for distributed synthesis of a discrete memoryless channel, in which a stochastic channel output is generated based on a compressed description of the channel input.  Wyner's common information is the minimum description rate needed.  However, when common randomness independent of the input is available, the necessary description rate reduces to Shannon's mutual information.  This work characterizes the optimal trade-off between the amount of common randomness used and the required rate of description.  We also include a number of related derivations, including the effect of limited local randomness, rate requirements for secrecy, applications to game theory, and new insights into common information duality.

  Our proof makes use of a soft covering lemma, known in the literature for its role in quantifying the resolvability of a channel.  The direct proof (achievability) constructs a feasible joint distribution over all parts of the system using a soft covering, from which the behavior of the encoder and decoder is inferred, with no explicit reference to joint typicality or binning.  Of auxiliary interest, this work also generalizes and strengthens this soft covering tool.
\end{abstract}

\begin{IEEEkeywords}
  Channel simulation, channel synthesis, soft covering, common information, random number generator, resolvability, reverse Shannon theorem, total variation distance.
\end{IEEEkeywords}

\section{Introduction}
\label{section introduction}

\IEEEPARstart{W}{hat} is the intrinsic connection between correlated random variables?  How much interaction is necessary to create correlation?  These are some of the inquiries that are illuminated by the distributed channel synthesis problem, introduced as follows:  An observer ({\em encoder}) of a random i.i.d. source sequence $X_1,X_2,...$ describes the sequence to a distant random number generator ({\em decoder}) that produces $Y_1,Y_2,...$.  What is the minimum rate of description needed to achieve a joint distribution that is statistically indistinguishable (as measured by total variation) from the distribution induced by a memoryless channel?

The nature of distributed channel synthesis is quite different than most problems in communication and source coding.  The objective of mimicking a random process is significantly more stringent than, say, producing an output sequence that is empirically correlated (jointly typical) with the input sequence.  Here we require the resulting input-output pairs $(X_t,Y_t)$ to be nearly i.i.d. according to the joint distribution that a prescribed memoryless channel would imply.  In previous work \cite{cuff-permuter-cover10-coordination-capacity} we define two notions of coordination that distinguish this important point.  This work is the ``strong coordination'' of \cite{cuff-permuter-cover10-coordination-capacity}.

Remarkably, random bits available in common to both the encoder and decoder play a non-trivial role in channel synthesis.  Because of the unusual nature of the problem, common randomness can replace some (yet not all) of the communication, providing a stochastic connection between the encoder and the decoder.  This stems from an important property of channel synthesis---unpredictability.  A properly synthesized channel will produce random outputs, free from perceivable patterns to all who do not see the communication and common randomness.

A particularly enticing use of distributed channel synthesis is in interactive adversarial settings.  In the context of game theory, correlated strategies can be advantageous to cooperating participants.  Correlation constraints on actions have been considered in the literature (e.g. \cite{anantharam-borkar07-counterexample} and \cite{gilpin-sandholm08-game-incomplete-information-AAMAS}).  We discuss the role of our channel synthesis results and the connection to secrecy systems in \S\ref{subsection game theory}.  In many repeated game settings, distributed channel synthesis provides the optimal means of communication.

The distributed channel synthesis problem provides a fresh look at correlation.  Many fruitful efforts have been made to quantify correlation between two random variables.  Each quantity is justified by the operational questions that it answers.  Covariance dictates the mean squared-error in linear estimation.  Shannon's mutual information is the descriptive savings for lossless compression due to side information and the additional growth rate of wealth in investing.  G\'{a}cs and K\"{o}rner's common information \cite{gacs-korner73-common-information} is the number of common random bits that can be extracted from correlated random variables.  It is less than mutual information.  Wyner's common information \cite{wyner75-common-information} is the number of common random bits needed to generate correlated random variables and is greater than mutual information.

In the distributed channel synthesis problem, two quantities emerge as extreme points --- Shannon's mutual information and Wyner's common information.  Without common randomness, the required communication rate is Wyner's common information $C(X;Y)$, consistent with Wyner's result in \cite{wyner75-common-information} concerning the minimum connection needed to generate correlated random variables.  However, when enough common randomness is available, the communication requirement is reduced to the mutual information $I(X;Y)$ (consistent with \cite{bennett-shor-smolin-thapliyal02-reverse-shannon-theorem}).  These extremes are evident in the main result, Theorem~\ref{theorem main result}, and common information is discussed further in \S\ref{subsection common information duality}.

Channel synthesis has emerged recently as a concept of interest in quantum and classical information theory.  Soljanin \cite{soljanin02} studied this in the context of quantum compression with unlimited common randomness.  Bennett et. al. introduced a ``reverse Shannon theorem'' \cite{bennett-shor-smolin-thapliyal02-reverse-shannon-theorem} (see also \cite{bennett-shor-smolin-thapliyal99-pre-reverse-shannon}, \cite{bennett-devetak-harrow-shor-winter09-channel-synthesis-Preprint}, and \cite{berta-christandl-renner11-quantum-reverse-shannon-theorem}) which states that all channels of the same capacity are equally valuable.  If one ignores encoding complexity and has unlimited common randomness available, then any memoryless channel can be used to synthesize any other channel of lower capacity.  Referring to Shannon's theorem as the reduction of a noisy channel to a noise-free one, their reverse Shannon theorem uses a noise-free channel to synthesize a noisy one.  This is precisely the problem considered in the present paper as well; however, we consider common randomness also to be a limited resource, yielding a trade-off between the use of communication and common randomness.

Limited common randomness for distributed channel synthesis was considered by Winter in \cite{winter02-reverse-shannon-finite-randomness-Arxiv} for a certain extremal operating point rather than the entire optimal trade-off.  Winter's communication scheme does not immediately generalize, though some of the proof methods were similar to ours.  He then connected these quantities (communication and common randomness for channel synthesis) to so-called ``extrinsic'' and ``intrinsic'' data in quantum measurements \cite{winter04-quantum-measurements}.  Further work on quantum measurements and exact channel synthesis can be found in \cite{wilde-hayden-buscemi-hsieh12-simulating-quantum-measurements}, \cite{harsha-jain-mcallester-radhakrishnan07-exact-synthesis-rejection-sampling-ICC}, and \cite{cubitt-leung-matthews-winter11-exact-synthesis}.

In addition to the main result, an emphasis of this paper is the proof technique.  Our construction of optimal codecs in \S\ref{section achievability} is unusual in that we don't begin by stating the behavior of the encoder in an explicit, causal manner but instead construct a joint distribution and infer the encoder behavior from it.  The tool that we use is the soft covering lemma of \S\ref{section:cloud}.  This is essentially the same tool used for the achievability proofs of Wyner's common information \cite{wyner75-common-information}, the resolvability of a channel \cite{han-verdu93-output-statistics}, and other results in the literature.  In \S\ref{section:soft covering continued}, we develop and generalize this tool, showing for example how it can be extended to superposition codebooks, similar to \cite{gohari-anantharam11-multiterminal-strong-coordination-ITW}.  Recently an alternative proof tool has been proposed in \cite{yassaee-aref-gohari12-random-binning-ISIT} which uses a random binning construction to take the role of soft covering.

We provide the main result and examples in \S\ref{section main result} followed by a number of extensions to the basic distributed channel synthesis problem in \S\ref{section extensions}.  Other extensions to this problem can be found in the recent literature (e.g. \cite{gohari-anantharam11-multiterminal-strong-coordination-ITW}, \cite{yassaee-gohari-aref12-channel-simulation-interactive-ISIT}, \cite{haddadpour-yassaee-gohari-aref12-coordination-via-relay-ISIT}, \cite{satpathy-cuff13-cascade-ISIT}), building on our introduction of the problem in \cite{cuff08-channel-synthesis-ISIT}, \cite{cuff09-dissertation}, and \cite{cuff-permuter-cover10-coordination-capacity}.

Let us note that ultimately the main result of the present paper has been solved concurrently and independently by Bennett, Devetak, Harrow, Shor, and Winter and can be found in preprint form in \cite{bennett-devetak-harrow-shor-winter09-channel-synthesis-Preprint}.  A presentation given by Bennett \cite{bennett-devetak-harrow-shor-winter07-reverse-shannon-Presentation} of unpublished work, unknown to us at the time, contains the complete trade-off between communication and common randomness and occurred at roughly the same time as our publication in \cite{cuff08-channel-synthesis-ISIT}.

\section{Main Result}
\label{section main result}

\subsection{Distributed Channel Synthesis}

\begin{figure}[ht]
  \begin{center}
    \begin{tikzpicture}[node distance=2cm]
      \node (src) [coordinate] {};

      \begin{scope}[gray]
      \node (bbox1) [coordinate,right=1cm of src] {};
      \node (enc) [node,right =5mm of bbox1] {Encoder};
      \node (dummy) [coordinate, right=1cm of enc] {};
      \node (rnd) [rectangle,minimum width=7mm,above=1cm of dummy] {$R_0$};
      \node (dec) [node,right=2cm of enc] {Decoder};
      \node (bbox2) [coordinate,right=5mm of dec] {};
      \end{scope}

      \node [rectangle,above =1.7cm of dummy] {Memoryless Channel $Q_{Y|X}$}; 

      \node (out) [coordinate,right=1cm of bbox2] {};

      \draw[arw] (src) to node [midway,above] {$X$} (bbox1);

      \begin{scope}[gray]
      \draw[arw] (bbox1) to (enc);
      \downrightratearrow{rnd.east}{dec.north}{1.5pt}{3pt}
      \downleftratearrow{rnd.west}{enc.north}{1.5pt}{3pt}
      \rightratearrow{enc.east}{dec.west}{1.5pt}{3pt}{$R$}
      \draw[arw] (dec) to (bbox2);
      \end{scope}

      \draw[arw] (bbox2) to node [midway,above] {$Y$} (out);
      \draw[dashed] ([yshift=-1cm] bbox1) rectangle ([yshift=2.5cm] bbox2);
    \end{tikzpicture}
    \caption{{\em Main Setup:}  We synthesize a memoryless channel across a distance by making use of communication and common randomness.  The necessary and sufficient rates are given in Theorem~\ref{theorem main result}.}
    \label{figure main setup}
  \end{center}
\end{figure}
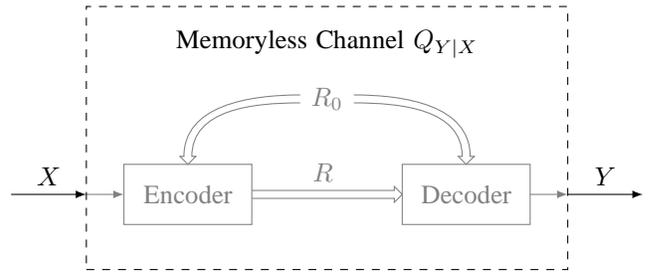

Let $\{X_i\}$ be a discrete i.i.d. random process, distributed according to \distnarg[q]{X}.  The dashed box in Fig.~\ref{figure main setup} represents a system that is designed to operate as if it were a memoryless channel defined by the conditional probability mass function \distnarg[q]{Y|X}.  However, the internal components of the box are constrained.  Suppose the input and output of the channel are not co-located.  An encoder which observes the channel input and a decoder which produces the channel output use communication and common randomness to synthesize the channel.  The synthesis is successful if it cannot be distinguished through a statistical test from the memoryless channel that it is designed to mimic.  This requirement will be clarified in \S\ref{subsection total variation}.

The distributed  channel synthesis problem asks what resources are needed to successfully accomplish this channel synthesis.  The resources come in the form of a message $J$ that is transmitted from the encoder to the decoder and common randomness $K$ that is independent of the channel input.  This work characterizes the required bit-rates of this communication and common randomness.

We allow the system to operate on blocks of $n$ inputs at a time, producing $n$ channel outputs.  This is the standard block encoding used in communication and compression.  However, even within each block, the system mimics a memoryless channel, where the $n$ outputs are conditionally independent given the $n$ inputs.

\begin{definition}
The {\em desired input-output distribution} for a block-length $n$ is the product distribution on the pair of sequences $X^n$ and $Y^n$ specified by the probability mass function $\prod_{t=1}^n \distalt[q]{X}{x_t} \distalt[q]{Y|X}{y_i|x_i}$.  We abbreviate this simply as
\begin{equation}
\prod \distnarg[q]{X} \distnarg[q]{Y|X}.
\end{equation}
\end{definition}

\subsection{Encoder and Decoder}

For a block-length $n$, the encoder produces a description of the source sequence $X^n$ at rate $R$, represented by $J \in [2^{nR}] \triangleq \{1,...,2^{nR}\}$.  A random variable $K$, uniformly distributed on $[2^{nR_0}]$ and independent of $X^n$, represents the common random bits at rate $R_0$ known at both the encoder and decoder.  The decoder generates a channel output $Y^n$ based only on $J$ and $K$.

The encoder and decoder are free to use randomization, and indeed they benefit from doing so.    Accordingly, the encoder and decoder are described by conditional probability mass functions.
\vskip .5em

\begin{center}
\begin{tabular}{ll}
Encoder: & \distnarg[f]{J|X^n,K} (a probability distribution), \\
Decoder: & \distnarg[g]{Y^n|J,K} (a probability distribution).
\end{tabular}
\end{center}

\begin{definition}
An $(R, R_0, n)$ {\em channel synthesis code} for input alphabet ${\cal X}$ and output alphabet ${\cal Y}$ consists of an encoder \distnarg[f]{J|X^n,K} and a decoder \distnarg[g]{Y^n|J,K} defined on the supports $X^n \in {\cal X}^n$, $Y^n \in {\cal Y}^n$, $J \in [2^{nR}]$, and $K \in [2^{nR_0}]$.
\end{definition}

\subsection{Induced Distribution}

Aside from the common randomness $K$, the behavior of the encoder and the decoder are independent.  Therefore, the combined behavior of the encoder and decoder results in a conditional distribution of the message $J$ and output $Y^n$ given by
\begin{eqnarray}
\distnarg{Y^n,J|X^n,K} & = & \distnarg[f]{J|X^n,K} \; \distnarg[g]{Y^n|J,K}.
\end{eqnarray}

\begin{definition}
The {\em induced joint distribution} of an $(R, R_0, n)$ channel synthesis code is the joint distribution on the quadruple $(X^n,Y^n,J,K)$ resulting from applying the encoder and decoder to the channel input and common randomness.  In other words, it is the probability mass function
\begin{eqnarray}
\distnarg{X^n,Y^n,J,K} & = & \distnarg{Y^n,J|X^n,K} \; \distnarg{X^n,K}, \label{definition induced distribution}
\end{eqnarray}
where, by definition of the problem,
\begin{eqnarray}
\dist{X^n,K} & = & \frac{1}{2^{nR_0}} \prod_{t=1}^n \distalt[q]{X}{x_t}.
\end{eqnarray}
\end{definition}

\begin{definition}
The {\em induced input-output distribution} is the marginal distribution of the induced joint distribution, assigning joint probabilities to only the input $X^n$ and the output $Y^n$ as
\begin{eqnarray}
\dist{X^n,Y^n} & = & \sum_{j,k} \dist{X^n,Y^n,J,K}.
\end{eqnarray}
\end{definition}

\subsection{Tolerance}
\label{subsection tolerance}

We say that the memoryless channel specified by \distnarg[q]{Y|X} can be synthesized with rates $(R,R_0)$ for input distribution \distnarg[q]{X} if there exists an $(R,R_0,n)$ channel synthesis code that induces the desired input-output distribution $\prod \distnarg[q]{X} \distnarg[q]{Y|X}$.  However, we actually tolerate some error.  If we require exact synthesis then we forfeit some of the substantial benefit that compression provides.  For example, consider distributed synthesis of the identity channel, which is equivalent to lossless compression.  ``Near lossless'' compression of $\{X_i\}$ can be achieved with a rate of $R = H(X)$, but exact lossless compression (and exact synthesis of the identity channel) requires $R = \log |{\cal X}|$.

Rather than tolerate error in the channel synthesis, we might instead ask for exact synthesis using variable length communication, just as variable length codes such as Huffman codes allow for exact lossless compression while achieving efficient average description lengths.  For distributed channel synthesis, a simple adaptation to block encoding would be to use an efficient channel synthesis code most of the time, which nearly synthesizes the channel, and with a small probability use an inefficient channel synthesis code (uncompressed communication) to implement the needed correction.  This is similar to the approach taken in \cite{bennett-shor-smolin-thapliyal02-reverse-shannon-theorem}, where exact synthesis is achieved using variable rate communication and an unlimited supply of common randomness.  Also, exact synthesis is achieved in \cite{harsha-jain-mcallester-radhakrishnan07-exact-synthesis-rejection-sampling-ICC} using rejection sampling.  But the steps for achieving efficient exact synthesis in our case of limited common randomness are not immediately obvious.

Instead of exact synthesis, we tolerate an arbitrarily small error measured by {\em total variation}.  In \S\ref{subsection total variation} we define and discuss total variation as a meaningful metric of tolerance.  For now, we move directly to the main definition for this work.

\begin{definition}
\label{definition achievability}
A pair of rates $(R, R_0)$ is {\em achievable} for synthesizing a memoryless channel specified by \distnarg[q]{Y|X} with input distribution \distnarg[q]{X} if there exists a sequence of $(R,R_0,n)$ channel synthesis codes, for $n=1,2,...$, where
\begin{eqnarray}
\lim_{n \to \infty} \left\| \distnarg{X^n,Y^n} - \prod \distnarg[q]{X} \distnarg[q]{Y|X} \right\|_{TV} & = & 0. \label{eq:tv to zero}
\end{eqnarray}
Let ${\cal C}$ be the closure of the set of achievable rate pairs $(R, R_0)$:\footnote{We deal with the closure because our proof does not handle the boundary points.}
\begin{eqnarray}
\mkern-12mu {\cal C} & \triangleq & \mbox{Closure} \left\{ \mbox{Achievable } (R, R_0) \mbox{ for } \distnarg[q]{X}, \distnarg[q]{Y|X} \right\}.
\end{eqnarray}
\end{definition}

\subsection{Main Result}

The main result of this paper characterizes the rate of communication and rate of common randomness needed to synthesize a discrete memoryless channel \distnarg[q]{Y|X} with an i.i.d. input distribution \distnarg[q]{X}.  This characterization is given in the definition of the following set ${\cal S}$:
\begin{eqnarray}
\label{definition S}
  {\cal S} & \triangleq & \left\{
  \begin{array}{rcl}
    (R,R_0) \in {\cal R}^2 & : & \exists \; \distnarg{X,Y,U} \in {\cal D} \mbox{ s.t.} \\
    R & \geq & I(X;U), \\
    R_0 + R & \geq & I(X,Y;U).
  \end{array}
  \right\},
\end{eqnarray}
where
\begin{eqnarray}
\label{definition D}
  {\cal D} & \triangleq & \left\{
  \begin{array}{rcl}
    \distnarg{X,Y,U} & : & (X,Y) \sim \distnarg[q]{X}\distnarg[q]{Y|X}, \\
    & & X - U - Y \mbox{ Markov}, \\
    & & |{\cal U}| \leq |{\cal X}| |{\cal Y}| + 1.
  \end{array}
  \right\}.
\end{eqnarray}

\begin{theorem}
\label{theorem main result}
For a discrete memoryless channel,
\begin{eqnarray}
{\cal C} & = & {\cal S}.
\end{eqnarray}
Furthermore, the total variation of \eqref{eq:tv to zero} decays exponentially fast with $n$ in the interior of ${\cal C}$.
\end{theorem}

\begin{figure}[ht]
  \begin{center}
    \begin{tikzpicture}[scale=4,
      dot/.style={draw,fill=black,circle,minimum size=1mm,inner sep=0pt},arw/.style={->,>=stealth}]
      \draw[arw] (0,0) to (0,1.1) node[above] {$R_0$};
      \draw[arw] (0,0) to (1.1,0) node[right]{$R$};
      \foreach \p in {75}
      {
       \draw[domain=(1-\p/100+0.00001):(2*(1-\p/100)),smooth,variable=\r] plot ({\r},{-\p/100*ln(\p/100)/ln(2)-(1-\p/100)*ln(1-\p/100)/ln(2) + \r * ((1-\p/100)/\r)*ln((1-\p/100)/\r)/ln(2) + \r * ((\r-1+\p/100)/\r) * ln((\r-1+\p/100)/\r)/ln(2)});
       \draw plot coordinates {(1-\p/100,{-\p/100*ln(\p/100)/ln(2)-(1-\p/100)*ln(1-\p/100)/ln(2)}) (1-\p/100,1.1)};
       \draw plot coordinates {({2*(1-\p/100)},{-\p/100*ln(\p/100)/ln(2)-(1-\p/100)*ln(1-\p/100)/ln(2) - 2*(1-\p/100)}) ({-\p/100*ln(\p/100)/ln(2)-(1-\p/100)*ln(1-\p/100)/ln(2)},0)};
      }
      \draw[dotted] (0.25,0) to (${-.75*ln(.75)/ln(2)-(.25)*ln(.25)/ln(2)}*(0,1)+0.25*(1,0)$);
      \draw (0.25,0) -- ++(0,-1/2pt) node[anchor=north,font=\small] {$I(X;Y)$};
      \draw (${-.75*ln(.75)/ln(2)-(.25)*ln(.25)/ln(2)}*(1,0)$) -- ++(0,-1/2pt) node[anchor=north,font=\small] {$C(X;Y)$};
    \end{tikzpicture}
  \caption{{\em Main Result:}  Theorem~\ref{theorem main result} gives a trade-off between the rate of communication and the rate of common randomness required to synthesize a discrete memoryless channel.  At the extremes, with no common randomness the communication rate requirement is Wyner's common information $C(X;Y)$, and the requirement reduces to the mutual information $I(X;Y)$ when unlimited common randomness is available.}
  \label{figure rate region extremes}
  \end{center}
\end{figure}
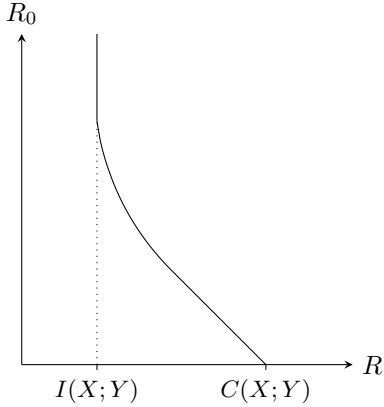

Two extreme points of the rate region ${\cal S}$ for distributed channel synthesis are manifested directly in the inequalities of \eqref{definition S} and illustrated in Fig.~\ref{figure rate region extremes}.  If $R_0 = 0$, the second inequality dominates, and the minimum communication rate $R$ is Wyner's common information \cite{wyner75-common-information}, defined as
\begin{eqnarray}
\label{definition common information}
  C(X;Y) & \triangleq & \min_{U \; : \; X - U - Y} I(X,Y;U).
\end{eqnarray}
At the other extreme, we see that with unlimited common randomness, the communication requirement reduces to $R \geq I(X;Y)$.  To see this, notice that the data processing inequality, which yields $I(X;U) \geq I(X;Y)$, can be met with equality by selecting $U = Y$.  Furthermore, this implies that $I(X,Y;Y) - R = H(Y) - I(X;Y) = H(Y|X)$ is a sufficient rate of common randomness to minimize the communication rate requirement.  It turns out, as is shown in \cite{cuff-permuter-cover10-coordination-capacity}, that sometimes a different choice of $U$ also minimizes the communication requirement while requiring even less common randomness.  The smallest amount of common randomness (after which additional common randomness does not benefit) is referred to as {\em necessary conditional entropy} in \cite{cuff-permuter-cover10-coordination-capacity}:
\begin{eqnarray}
\label{definition necessary entropy}
  H(Y \dag X) & \triangleq & \min_{f \; : \; X - f(Y) - Y} H(f(Y)|X).
\end{eqnarray}

Proof of Theorem~\ref{theorem main result} is the subject of sections~\ref{section:cloud}, \ref{section achievability}, and \ref{section converse}.  The achievability proof holds for general memoryless sources and channels, not only those that are discrete.  However, the converse presented in this work is specific to finite alphabets.  A general converse might arise from a careful analysis of ${\cal S}$.  For example, it would be sufficient to show that $\cap_{[Q]} {\cal S}_{[Q]} \subset {\cal S}$, where $[Q]$ represents the distribution of a finite quantization of $X$ and $Y$ under the desired distribution \distnarg[q]{X,Y}, and ${\cal S}_{[Q]}$ is defined as ${\cal S}$ in \eqref{definition S} and \eqref{definition D} but with respect to $[Q]$.  By Theorem~\ref{theorem main result} and first principles, ${\cal C} \subset {\cal S}_{[Q]}$ for any $[Q]$.

\subsection{Example:  Erasure Channel}
\label{section example erasure channel}

Let \distnarg[q]{X} be the binary symmetric distribution (i.e. Bernoulli-half), and consider the symmetric erasure channel \distnarg[q]{Y|X} with erasure probability $p$.  We now find the optimal distributed channel synthesis rates $(R,R_0) \in {\cal S}$ by considering Markov distributions in ${\cal D}$.  Fortunately, the sparsity in the joint distribution \distnarg[q]{X,Y} simplifies this optimization.

\begin{figure}[ht]
  \begin{center}
    \begin{tikzpicture}[node distance=2cm]
      \node (x0) [rectangle] at (0,0) {$0$};
      \node (labelX) [rectangle,above=3mm of x0] {$X$};
      \node (u0) [rectangle, right=of x0] {$0$};
      \node (labelY) [rectangle,above=3mm of u0] {$U$};
      \node (ue) [rectangle, below=7mm of u0] {$\mathsf{e}$};
      \node (u1) [rectangle, below=7mm of ue] {$1$};
      \node (y0) [rectangle, right=of u0] {$0$};
      \node (labelY) [rectangle,above=3mm of y0] {$Y$};
      \node (ye) [rectangle, right=of ue] {$\mathsf{e}$};
      \node (y1) [rectangle, right=of u1] {$1$};
      \node (x1) [rectangle, left=of u1] {$1$};

      \draw [arw] (x0) to (u0);
      \draw [arw] (x0) to node [midway,above] {$p_1$} (ue);
      \draw [arw] (x1) to node [midway,above] {$p_1$} (ue);
      \draw [arw] (x1) to (u1);
      \draw [arw] (u0) to (y0);
      \draw [arw] (u0) to node [midway,above] {$p_2$} (ye);
      \draw [arw] (ue) to (ye);
      \draw [arw] (u1) to node [midway,above] {$p_2$} (ye);
      \draw [arw] (u1) to (y1);
    \end{tikzpicture}
  \caption{{\em Concatenated Erasure Channels:}  Any optimal point in the rate region ${\cal S}$ in \eqref{definition S} for the symmetric erasure channel is achieved with a choice of $U$ that constitutes a concatenation of two symmetric erasure channels.}
  \label{figure erasure cascade}
  \end{center}
\end{figure}
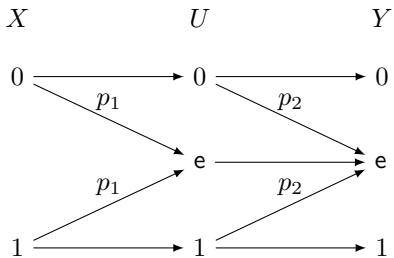

In the appendix we show that the optimizing choices of \distnarg{X,Y,U} will be the concatenation of two erasure channels depicted in Fig.~\ref{figure erasure cascade}.  We are left with two parameters---the erasure probability $p_1$ of the first channel \distnarg{U|X} and the erasure probability $p_2$ of the second channel \distnarg{Y|U}---and one constraint:  $(1-p_1) (1-p_2) = (1-p)$.  By labeling $r = (1-p_1)$ and evaluating the mutual information terms, we obtain the following achievable rate region:
\begin{eqnarray}
\label{eq:S for erasure channel}
  \mkern-24mu {\cal S} & = & \left\{
  \begin{array}{rcl}
    (R,R_0) & \in & {\cal R}^2 : \\
    \exists \; r & \in & [1-p,r^*] \mbox{ such that} \\
    R & \geq & r \mbox{ bits}, \\
    R_0 + R & \geq & h(p) \\ & & + r \left( 1 - h \left( \frac{1-p}{r} \right) \right) \mbox{ bits}.
  \end{array}
  \right\},
\end{eqnarray}
where $r^* = \min\{2(1-p),1\}$ and $h(\cdot)$ is the binary entropy function.  Choices of $r > r^*$ are suboptimal.

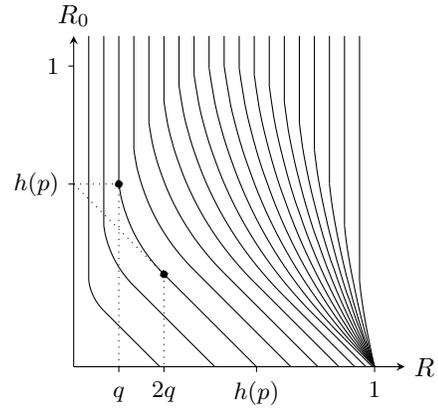
\begin{figure}[ht]
  \begin{center}
    \begin{tikzpicture}[scale=4,
      dot/.style={draw,fill=black,circle,minimum size=1mm,inner sep=0pt},arw/.style={->,>=stealth}]
      \draw[arw] (0,0) to (0,1.1) node[above] {$R_0$};
      \draw[arw] (0,0) to (1.1,0) node[right]{$R$};
      \foreach \p in {55,60,...,95}
      {
       \draw[domain=(1-\p/100+0.00001):(2*(1-\p/100)),smooth,variable=\r] plot ({\r},{-\p/100*ln(\p/100)/ln(2)-(1-\p/100)*ln(1-\p/100)/ln(2) + \r * ((1-\p/100)/\r)*ln((1-\p/100)/\r)/ln(2) + \r * ((\r-1+\p/100)/\r) * ln((\r-1+\p/100)/\r)/ln(2)});
       \draw plot coordinates {(1-\p/100,{-\p/100*ln(\p/100)/ln(2)-(1-\p/100)*ln(1-\p/100)/ln(2)}) (1-\p/100,1.1)};
       \draw plot coordinates {({2*(1-\p/100)},{-\p/100*ln(\p/100)/ln(2)-(1-\p/100)*ln(1-\p/100)/ln(2) - 2*(1-\p/100)}) ({-\p/100*ln(\p/100)/ln(2)-(1-\p/100)*ln(1-\p/100)/ln(2)},0)};
      }
      \foreach \p in {5,10,...,50}
      {
       \draw[domain=(1-\p/100+0.00001):1,smooth,variable=\r] plot ({\r},{-\p/100*ln(\p/100)/ln(2)-(1-\p/100)*ln(1-\p/100)/ln(2) + \r * ((1-\p/100)/\r)*ln((1-\p/100)/\r)/ln(2) + \r * ((\r-1+\p/100)/\r) * ln((\r-1+\p/100)/\r)/ln(2)});
       \draw plot coordinates {(1-\p/100,{-\p/100*ln(\p/100)/ln(2)-(1-\p/100)*ln(1-\p/100)/ln(2)}) (1-\p/100,1.1)};
      }
      \draw (${-.85*ln(.85)/ln(2)-(.15)*ln(.15)/ln(2)}*(0,1)$) -- ++(-1/2pt,0) node[anchor=east,font=\small] {$h(p)$};
      \draw[dotted] (${-.85*ln(.85)/ln(2)-(.15)*ln(.15)/ln(2)}*(0,1)$) to (${-.85*ln(.85)/ln(2)-(.15)*ln(.15)/ln(2)}*(0,1)+0.15*(1,0)$) node[dot] {};
      \draw[dotted] (0.15,0) to (${-.85*ln(.85)/ln(2)-(.15)*ln(.15)/ln(2)}*(0,1)+0.15*(1,0)$);
      \draw (0.15,0) -- ++(0,-1/2pt) node[anchor=north,font=\small] {};
      \node (qlabel) [rectangle,anchor=north,font=\small] at (.15,-1pt) {$q$};
      \draw[dotted] (${-.85*ln(.85)/ln(2)-(.15)*ln(.15)/ln(2)}*(0,1)$) to (${-.85*ln(.85)/ln(2)-(.15)*ln(.15)/ln(2)-0.3}*(0,1)+0.3*(1,0)$) node[dot] {};
      \draw[dotted] (${-.85*ln(.85)/ln(2)-(.15)*ln(.15)/ln(2)-0.3}*(0,1)+0.3*(1,0)$) to (0.3,0);
      \draw (0.3,0) -- ++(0,-1/2pt) node[anchor=north,font=\small] {$2q$};
      \draw (${-.85*ln(.85)/ln(2)-(.15)*ln(.15)/ln(2)}*(1,0)$) -- ++(0,-1/2pt) node[anchor=north,font=\small] {$h(p)$};
      \draw (1,0) -- ++(0,-1/2pt) node[anchor=north,font=\small] {$1$};
      \draw (0,1) -- ++(-1/2pt,0) node[anchor=east,font=\small] {$1$};
    \end{tikzpicture}
    \caption{{\em Erasure Channel Rate Regions:}  The boundaries of the achievable rate regions for synthesis of the $p$-erasure channel with symmetric inputs are shown for $p = .05, .1, .15,..., .9, .95$ from right to left.  Transition points on the curve for $p=.85$ are labeled, where $q = 1-p$.}
    \label{figure erasure channel region}
  \end{center}
\end{figure}

{\em Common Information:}  The common information $C_{Q}(X;Y)$ is found by evaluating the second inequality of \eqref{eq:S for erasure channel} (representing $I(X,Y;U)$) at $r = r^*$.  For erasure probabilities $p\leq1/2$ we see that $r^*=1$, which is equivalent to choosing $U = X$.  The common information in this case is $C_{Q}(X;Y) = 1$ bit.  For erasure probabilities $p>1/2$ we get $r^* = 2(1-p)$, which is equivalent to choosing the channel \distnarg{Y|U} to have 50\% erasures.  The common information in this case is $C_{Q}(X;Y) = h(p)$.  Notice that it is easily verified in Fig.~\ref{figure erasure cascade} that $H(X,Y|U=u) \leq 1$ bit for each $u$, which is achieved with equality by $r^*$.  To summarize:
\begin{eqnarray}
  C_Q(X;Y) & = & \left\{
  \begin{array}{ll}
    1 \mbox{ bit}, & p<1/2, \\
    h(p), & p \geq 1/2.
  \end{array}
  \right.
\end{eqnarray}

{\em Minimum Communication:}  The minimum communication rate required in the presence of enough common randomness is $R \geq I_Q (X;Y) = 1 - p$ bits, and the rate of common randomness needed to achieve this is $R_0 \geq H_Q(Y \dag X) = H_Q (Y|X) = h(p)$.  This operating point corresponds to a simple synthesis strategy.  The common randomness can be used to generate a list of erasure locations, and the encoder can then transmit the non-erased bits.

\subsection{Example:  Reverse Erasure Channel}

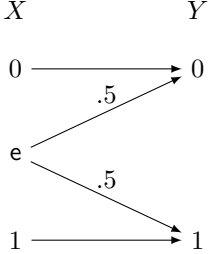
\begin{figure}[ht]
  \begin{center}
    \begin{tikzpicture}[node distance=2cm]
      \node (x0) [rectangle] at (0,0) {$0$};
      \node (labelX) [rectangle,above=3mm of x0] {$X$};
      \node (y0) [rectangle, right=of x0] {$0$};
      \node (labelY) [rectangle,above=3mm of y0] {$Y$};
      \node (xe) [rectangle, below=7mm of x0] {$\mathsf{e}$};
      \node (y1) [rectangle, right= of x1] {$1$};
      \node (x1) [rectangle, below=7mm of xe] {$1$};

      \draw [arw] (x0) to (y0);
      \draw [arw] (xe) to node [midway,above] {$.5$} (y0);
      \draw [arw] (xe) to node [midway,above] {$.5$} (y1);
      \draw [arw] (x1) to (y1);
    \end{tikzpicture}
  \caption{{\em Reverse Erasure Channel.}}
  \label{figure reverse erasure channel}
  \end{center}
\end{figure}

Now consider the reverse of the erasure channel example, by switching the input and output, as depicted in Fig.~\ref{figure reverse erasure channel}.  The channel input distribution \distnarg[q]{X} is symmetric on the set $\{0,e,1\}$ with probability $p$ of erasure.  The channel sorts the erasures randomly into $0$'s and $1$'s.

The same derivation and parameterizations as above (erasure channel) hold for this example as well.  The only modifications to ${\cal S}$ are the range of the optimal values of $r$ and the necessary update to the first inequality:
\begin{eqnarray}
\label{eq:S for reverse erasure channel}
  {\cal S} & = & \left\{
  \begin{array}{rcl}
    (R,R_0) & \in & {\cal R}^2 : \\
    \exists \; r & \in & [r^*,1] \mbox{ such that} \\
    R & \geq & h(p) - r h \left( \frac{1-p}{r} \right) \\ & & + (1-p) \mbox{ bits}, \\
    R_0 + R & \geq & h(p) - r h \left( \frac{1-p}{r} \right) \\ & & + r \mbox{ bits}.
  \end{array}
  \right\},
\end{eqnarray}
where $r^* = \min\{2(1-p),1\}$ and $h(\cdot)$ is the binary entropy function.  Choices of $r < r^*$ are suboptimal.

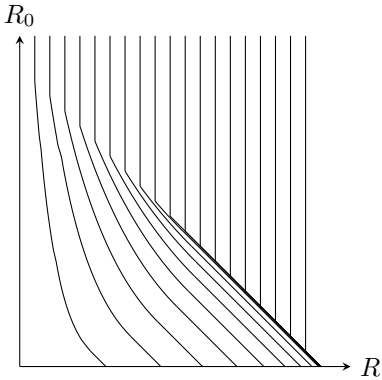
\begin{figure}[ht]
  \begin{center}
    \begin{tikzpicture}[scale=4,
      dot/.style={draw,fill=black,circle,minimum size=0.5mm,inner sep=0pt},arw/.style={->,>=stealth}]
      \draw[arw] (0,0) to (0,1.1) node[above] {$R_0$};
      \draw[arw] (0,0) to (1.1,0) node[right]{$R$};

      \foreach \p in {5,10,...,50}
      {
      \draw plot coordinates {(1,0) (1-\p/100,\p/100) (1-\p/100,1.1)};
      }
      \foreach \p in {55,60,...,95}
      {
      \draw[domain=2*(1-\p/100):1,smooth,variable=\r] plot ({-\p/100*ln(\p/100)/ln(2)-(1-\p/100)*ln(1-\p/100)/ln(2) + \r * ((1-\p/100)/\r)*ln((1-\p/100)/\r)/ln(2) + \r * ((\r-1+\p/100)/\r) * ln((\r-1+\p/100)/\r)/ln(2)+(1-\p/100)},{\r-(1-\p/100)});
      \draw plot coordinates {(1-\p/100,\p/100) (1-\p/100,1.1)};
      \draw plot coordinates {({-\p/100*ln(\p/100)/ln(2)-(1-\p/100)*ln(1-\p/100)/ln(2)-(1-\p/100)},1-\p/100) ({-\p/100*ln(\p/100)/ln(2)-(1-\p/100)*ln(1-\p/100)/ln(2)},0)};
      }
    \end{tikzpicture}
    \caption{{\em Reverse Erasure Channel Rate Region:}  The boundaries of the achievable rate regions for synthesis of the reverse $p$-erasure channel with symmetric inputs are shown for $p = .05, .1, .15,..., .9, .95$ from right to left.}
    \label{figure reverse erasure channel region}
  \end{center}
\end{figure}

\subsection{Example:  Scatter Channel}
\label{section example scatter channel}

Consider a channel \distnarg[q]{Y|X} which acts on an input $X \in \{1,2,...,m\}$ and produces an output uniformly at random from the same set excluding $X$.  That is,
\begin{eqnarray}
\label{definition scatter channel}
  \dist[q]{Y|X} & = & \frac{1}{m-1} \; \mathbf{1} (x \neq y),
\end{eqnarray}
where $\mathbf{1}$ is the indicator function.  Now, apply to this channel the uniform input distribution \distnarg[q]{X}, and the result is a desired input-output distribution $\distnarg[q]{X} \distnarg[q]{Y|X}$ that is uniform over all pairs $(X,Y)$ such that $X \neq Y$.  This distribution is studied as an example in \cite{witsenhausen76-common-information}, \cite{winter05-triples}, and \cite{cuff-permuter-cover10-coordination-capacity}.

We find the optimal rates $(R,R_0) \in {\cal S}$ by considering Markov distributions in ${\cal D}$.  As in the previous examples, the sparsity in the joint distribution \distnarg[q]{X,Y} simplifies this optimization.

For any $\distnarg{X,Y,U} \in {\cal D}$, the Markov property constrains that for each value $u$ in the support of $U$, the conditional distribution \distnarg{X,Y|U=u} is a product distribution $\distnarg{X|U=u} \distnarg{Y|U=u}$.  We categorize these into $(m-1)$ categories based on the support size of \distnarg{X|U=u}.  Call this support ${\cal X}_u$ with size $a_u = |{\cal X}_u|$. Then ${\cal X}_u \cap {\cal Y}_u = \emptyset$ to avoid any probability that $X = Y$, resulting in $|{\cal Y}_u| \leq m-a_u$.

\begin{figure}[ht]
  \begin{center}
    \begin{tikzpicture}[scale=2,
       dot/.style={draw,fill=black,circle,minimum size=0.5mm,inner sep=0pt},arw/.style={->,>=stealth},linestyle/.style={}]
       \draw[arw] (0,0) to (0,3.2) node[above] {$R_0$};
       \draw[arw] (0,0) to (2.2,0) node[right]{$R$};
       \draw (0,3) -- ++(-1/2pt,0) node[anchor=east,font=\small] {$3$};
       \draw (2,0) -- ++(0,-1/2pt) node[anchor=north,font=\small] {$2$};

       \draw[line width=3pt] (${ln(3/2)/ln(2)}*(1,0)+(0,3)$) -- (${ln(3/2)/ln(2)}*(1,0)+(0,1)$) -- (${ln(3/2)/ln(2)+1}*(1,0)$); 

       \foreach \m in {5} 
       {
          \pgfmathsetmacro{\start}{(\m+1)/2};
          \pgfmathsetmacro{\finish}{\m-2};
          \pgfmathsetmacro{\finishplus}{\m -1}
          \draw[line width=2pt] (${ln(\m/\finishplus)/ln(2)}*(1,0)+{ln((\m-1)/(\m-\finishplus))/ln(2)}*(0,1)$) -- (${ln(\m/\finishplus)/ln(2)}*(1,0)+(0,3)$);
          \foreach \a in {\start,...,\finish}
          {
          \draw[line width=2pt] (${ln(\m/\a)/ln(2)}*(1,0)+{ln((\m-1)/(\m-\a))/ln(2)}*(0,1)$) -- (${ln(\m/(\a+1))/ln(2)}*(1,0)+{ln((\m-1)/(\m-(\a+1)))/ln(2)}*(0,1)$);
          }
          \draw[line width=2pt] (${ln(\m/\start)/ln(2)}*(1,0)+{ln((\m-1)/(\m-\start))/ln(2)}*(0,1)$)--(${ln(\m/\start)/ln(2)+ln((\m-1)/(\m-\start))/ln(2)}*(1,0)$);
       }

      \foreach \m in {7} 
       {
          \pgfmathsetmacro{\start}{(\m+1)/2};
          \pgfmathsetmacro{\finish}{\m-2};
          \pgfmathsetmacro{\finishplus}{\m -1}
          \draw[line width=1pt] (${ln(\m/\finishplus)/ln(2)}*(1,0)+{ln((\m-1)/(\m-\finishplus))/ln(2)}*(0,1)$) -- (${ln(\m/\finishplus)/ln(2)}*(1,0)+(0,3)$);
          \foreach \a in {\start,...,\finish}
          {
          \draw[line width=1pt] (${ln(\m/\a)/ln(2)}*(1,0)+{ln((\m-1)/(\m-\a))/ln(2)}*(0,1)$) -- (${ln(\m/(\a+1))/ln(2)}*(1,0)+{ln((\m-1)/(\m-(\a+1)))/ln(2)}*(0,1)$);
          }
          \draw[line width=1pt] (${ln(\m/\start)/ln(2)}*(1,0)+{ln((\m-1)/(\m-\start))/ln(2)}*(0,1)$)--(${ln(\m/\start)/ln(2)+ln((\m-1)/(\m-\start))/ln(2)}*(1,0)$);
       }
      \end{tikzpicture}
    \caption{{\em Scatter Channel Rate Region:}  The boundaries of the achievable rate regions for synthesis of the scatter channel with symmetric inputs are shown for $|{\cal X}| = 3, 5$, and $7$ from thickest to thinnest.  Larger alphabets have greater benefit from common randomness.}
    \label{figure scatter channel region}
  \end{center}
\end{figure}
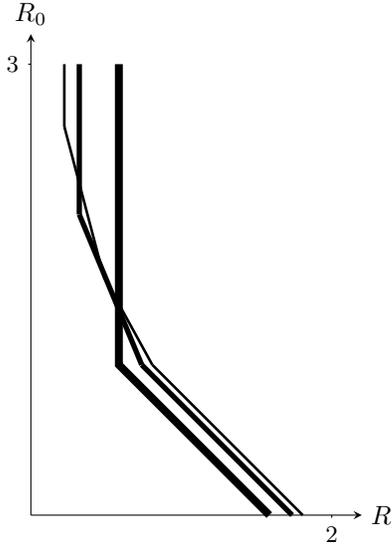

For each of the above categories, associated with $a_u = 1,...,m-1$, we have a trivial bound on conditional entropy:
\begin{eqnarray}
  H(X|U=u) & \leq & \log a_u, \label{eq:scatter entropy bound 1} \\
  H(X,Y|U=u) & \leq & \log a_u + \log(m-a_u). \label{eq:scatter entropy bound 2}
\end{eqnarray}
Thus, $(H(X|U), H(X,Y|U))$ must be in the convex hull of the union of two dimensional regions defined by \eqref{eq:scatter entropy bound 1} and \eqref{eq:scatter entropy bound 2} for each $a_u$.  On the other hand, the corner points of these regions can be achieved due to the symmetry of \distnarg[q]{X,Y}.  This is accomplished by constructing $m$-choose-$a_u$ conditional distributions $\distnarg{X|U=u_i} \distnarg{Y|U=u_i}$, one for each support structure consistent with $a_{u_i}=a_u$.   Let the conditional distributions be uniformly distributed over their supports.  Also, let \distnarg{U} be uniformly distributed over this set of $u_i$.  Therefore,
\begin{eqnarray}
\label{eq:S for scatter channel}
  \mkern-24mu {\cal S} & = & \mbox{Conv} \left( \left\{
  \begin{array}{rcl}
    (R,R_0) & \in & {\cal R}^2 : \\
    \exists \; a & \in & [m-1] \mbox{ s.t.} \\
    a & \geq & m/2, \\
    R & \geq & \log \left( \frac{m}{a} \right), \\
    R_0 + R & \geq & \log \left( \frac{m(m-1)}{a(m-a)} \right),
  \end{array}
  \right\} \right),
\end{eqnarray}
where $\mbox{Conv}(\cdot)$ indicated the convex hull.  This region is depicted in Fig.~\ref{figure scatter channel region} for $m = 3, 5$, and $7$.

{\em Common Information:}  The common information for this distribution was calculated in \cite{cuff-permuter-cover10-coordination-capacity} and can be obtained from the above region.  Let $\lceil m \rceil_2$ represent the value of $m$ rounded up to the nearest even number.
\begin{eqnarray}
C_{Q}(X;Y) & = & 2 \mbox{ bits } - \log \left( \frac{\lceil m \rceil_2}{\lceil m \rceil_2 - 1} \right).
\end{eqnarray}
Notice that this increases to $2$ bits as the alphabet size $m$ increases.

{\em Minimum Communication:}  In contract to the common information, the mutual information $I_Q (X;Y)$ decreases to zero as $m$ increases.
\begin{eqnarray}
I_{Q}(X;Y) & = & \log \left( \frac{m}{m-1} \right) \\
& \approx & \frac{1}{m} \log e.
\end{eqnarray}
The rate of common randomness needed to achieve this minimal communication rate is $R_0 \geq H_Q (Y \dag X) = H_Q (Y|X) = \log (m-1)$.

\subsection{Total Variation Distance}
\label{subsection total variation}

We use total variation to measure the distance between the induced input-output distribution and the desired input-output distribution.  Total variation between two distributions $\Pi$ and $\Gamma$ on a set ${\cal W}$ is defined in the following way:
\begin{eqnarray}
\label{eq:define tv}
\| \Pi - \Gamma \|_{TV} & \triangleq & \max_{S \subset {\cal W}} \; \left( \Pi(S) - \Gamma(S) \right) \\
& = & \max_{S \subset {\cal W}} \; \left( \Gamma(S) - \Pi(S) \right).
\end{eqnarray}
If ${\cal W}$ is countable and \distns[\pi]{w} and \distns[\gamma]{w} represent the probability mass functions associated with $\Pi$ and $\Gamma$, then
\begin{eqnarray}
\label{eq:tv is l1}
\| \distnarg[\pi]{} - \distnarg[\gamma]{} \|_{TV} & \triangleq & \frac{1}{2} \| \distnarg[\pi]{} - \distnarg[\gamma]{} \|_{1} \\
& = & \frac{1}{2} \sum_{w \in {\cal W}} | \pi(w) - \gamma(w) | \label{eq:tv is sum} \\
& = & \| \Pi - \Gamma \|_{TV}.
\end{eqnarray}

Total variation has properties that make it an attractive measure of tolerance.  First is statistical indistinguishability.  Consider a test that tries to detect a synthesized channel.  The performance of any binary hypothesis test is characterized by two parameters:  the probability of false positive ($\alpha$); and the probability of false negative ($\beta$).  Let $\Gamma$ be the null hypothesis (the channel is genuine) and $\Pi$ be the alternative hypothesis (the channel is synthetic).  If the two hypotheses yield identical distributions, then accurate detection is impossible, and $\alpha + \beta = 1$ (any value of $\beta = 1 - \alpha$ can be attained by adjusting the sensitivity of the test).  In general,
\begin{eqnarray}
\label{eq:alpha and beta bound}
  \alpha + \beta & \geq & 1 - \| \Pi - \Gamma \|_{TV}.
\end{eqnarray}
Therefore, if total variation is small, then reliable detection is not possible.  This is the objective of channel synthesis.

Another property of total variation is a bound related to expected values of bounded functions.
\begin{eqnarray}
\label{eq:tv expected value bound}
\left| \mathbf{E}_{\Pi} f(W) - \mathbf{E}_{\Gamma} f(W) \right| & \leq & 2 f_{max} \; \left\| \Pi - \Gamma \right\|_{TV},
\end{eqnarray}
where $f_{max} = \max_{w \in {\cal W}} |f(w)|$.  This bound implies continuity of $\mathbf{E} f(W)$, for bounded $f$, with respect to the distribution \distnarg{W}, using total variation as the distance metric.  If we know that the total variation is small between two distributions $\Pi$ and $\Gamma$ and we care about the expected value with respect to $\Pi$ of a bounded function, then we are free to instead analyze the expected value with respect to $\Gamma$, which is guaranteed to be nearly equivalent.  We use this technique when analyzing the payoff achieved in the game theoretic setting of \S\ref{subsection game theory} and in related secrecy work in \cite{cuff10-lossless-secrecy-Globecom}, \cite{cuff10-partial-secrecy-Allerton}, and \cite{schieler-cuff12-zero-rate-secrecy-ISIT}.

Other metrics of distance between probability distributions have been explored in related works, such as \cite{wyner75-common-information}, \cite{steinberg-verdu96-simulation-random-processes}, and \cite{bloch-laneman11-secrecy-from-resolvability-Arxiv}.  In particular, Kullback-Leibler divergence $d_{KL}(\Pi; \Gamma)$ makes an interesting choice of fidelity metric because it is an information-theoretic quantity closely related to other important quantities such as mutual information and, more importantly, because it has implications concerning the asymptotics of hypothesis testing, in the regime of highly reliable detection.  Wyner uses K-L divergence, normalized by the block-length, as his tolerance metric for generating correlated random variables in \cite{wyner75-common-information}.

Kullback-Leibler divergence is a stricter metric than total variation in general---Pinsker's inequality reveals that total variation converges to zero as K-L divergence approaches zero.  However, normalized K-L divergence forfeits this relationship.  Also, a reverse relationship holds for i.i.d. distributions.  That is, if $\Pi \ll \Gamma$ (i.e. $\Pi$ is absolutely continuous with respect to $\Gamma$), and $\Gamma$ is an i.i.d. discrete distribution of $n$ variables, then
\begin{eqnarray}
  d_{KL}(\Pi; \Gamma) & \in & O \left( \left( n + \log \frac{1}{\mbox{TV}} \right) \; \mbox{TV} \right), \label{eq:bound tv from divergence}
\end{eqnarray}
as $\mbox{TV} \triangleq \| \Pi - \Gamma \|_{TV}$ goes to zero and $n$ goes to infinity.\footnote{This statement uses \eqref{eq:tv expected value bound} and \cite[Theorem~17.3.3]{cover-thomas06-eit}.}  In particular, this means that an exponential decay of total variation with respect to $n$ produces an exponential decay in $d_{KL}(\Pi; \Gamma)$ with the same exponent.

It turns out that the rate region given in the main result (Theorem~\ref{theorem main result}) no longer holds if the total variation tolerance metric is replaced by Kullback-Leibler divergence in the direction used by Wyner in \cite{wyner75-common-information}.  That is, let $\Pi$ represent the induced distribution and let $\Gamma$ be the desired distribution.  Suppose achievability demands that $d_{KL}(\Pi; \Gamma)$ be made arbitrarily small (with or without normalization).  We again call upon the identity channel as a simple counterexample to Theorem~\ref{theorem main result}.  Notice that $R>H(X)$ is sufficient for the theorem, but $R>\log |{\cal X}|$ is necessary for exact lossless compression.  Therefore, any rate less than $\log |{\cal X}|$ will cause $d_{KL}(\Pi; \Gamma)=\infty$.  On the other hand, for any channel with $\dist[q]{Y|X} > 0$ for all $x$ and $y$, the divergence $d_{KL}(\Pi; \Gamma)$ goes to zero exponentially fast with the same exponent as total variation due to \eqref{eq:bound tv from divergence}.

For a careful comparison of inequalities involving total variation, K-L divergence, and normalized K-L divergence, see \cite{bloch-laneman11-secrecy-from-resolvability-Arxiv}.

\section{Extensions}
\label{section extensions}

\subsection{Broadcast Channel}
\label{subsection broadcast channel}

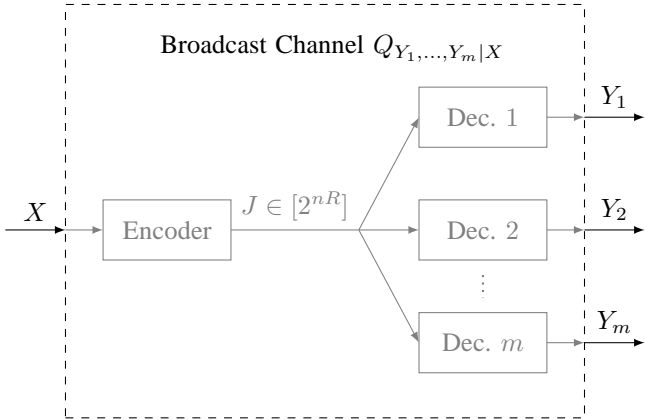
\begin{figure}[ht]
  \begin{center}
    \begin{tikzpicture}[node distance=2cm]
      \node (src) [coordinate] {};

      \begin{scope}[gray]
      \node (bbox) [coordinate,right=8mm of src] {};
      \node (enc) [node,right =5mm of bbox] {Encoder};
      \node (dummy) [coordinate, right=1.7cm of enc] {};
      \node (dec2) [node,right=8mm of dummy] {Dec. $2$};
      \node (dec1) [node,above=7mm of dec2] {Dec. $1$};
      \node (decm) [node,below=7mm of dec2] {Dec. $m$};
      \node (bbox2) [coordinate,right=5mm of dec2] {};
      \node (bbox1) [coordinate,right=5mm of dec1] {};
      \node (bboxm) [coordinate,right=5mm of decm] {};
      \end{scope}

      \node [rectangle,above =2.1cm of dummy,xshift=-.35cm] {Broadcast Channel $Q_{Y_1,\ldots,Y_m|X}$};

      \node (out2) [coordinate,right=8mm of bbox2] {};
      \node (out1) [coordinate,right=8mm of bbox1] {};
      \node (outm) [coordinate,right=8mm of bboxm] {};

      \draw[arw] (src) to node [midway,above] {$X$} (bbox);

      \begin{scope}[gray]
      \draw[arw] (bbox) to (enc);
      \draw (enc) to node [midway,above] {$J\in[2^{nR}]$} (dummy);
      \draw[arw] (dummy) to (dec1.west);
      \draw[arw] (dummy) to (dec2.west);
      \draw[arw] (dummy) to (decm.west);
      \draw[thick, dotted] ([yshift=-2mm] dec2.south) to ([yshift=2mm] decm.north);
      \draw[arw] (dec2) to (bbox2);
      \draw[arw] (dec1) to (bbox1);
      \draw[arw] (decm) to (bboxm);
      \end{scope}

      \draw[arw] (bbox2) to node [midway,above] {$Y_2$} (out2);
      \draw[arw] (bbox1) to node [midway,above] {$Y_1$} (out1);
      \draw[arw] (bboxm) to node [midway,above] {$Y_m$} (outm);
      \draw[dashed] ([yshift=-2.5cm] bbox) rectangle ([yshift=3cm] bbox2);
    \end{tikzpicture}
    \caption{{\em Broadcast Channel:}  This setting extends the main result to include the synthesis of a broadcast channel with separated decoders producing each channel output.  The decoders each receive a common communication message as well as common randomness.}
    \label{figure multiple decoders}
  \end{center}
\end{figure}

The main result of Theorem~\ref{theorem main result} can be readily extended to a situation with multiple separate decoders together synthesizing a memoryless broadcast channel \distnarg[q]{Y_1,...,Y_m|X}, each producing one of the channel output sequences after receiving a common transmission from the encoder as well as common randomness among all nodes.  This is depicted in Fig.~\ref{figure multiple decoders}.  The region of achievable rates for synthesis is given by
\begin{eqnarray}
\label{definition S broadcast}
  \mkern-18mu {\cal S}_{BC} & = & \left\{
  \begin{array}{rcl}
    (R,R_0) & \in & {\cal R}^2 : \\
    \exists \; \distnarg{} & \in & {\cal D}_{BC} \mbox{ such that} \\
    R & \geq & I(X;U), \\
    R_0 + R & \geq & I(X,Y_1,...,Y_m;U).
  \end{array}
  \right\},
\end{eqnarray}
where
\begin{eqnarray}
\label{definition D broadcast}
  \mkern-24mu {\cal D}_{BC} & = & \left\{ \!\!\!
  \begin{array}{rcl}
    \distnarg{X,Y_1,...,Y_m,U} & = & \distnarg{X,U} \prod_{i=1}^m \distnarg{Y_i|U} : \\
    \distnarg{X,Y_1,...,Y_m} & = & \distnarg[q]{X}\distnarg[q]{Y_1,...,Y_m|X}, \\
    |{\cal U}| & \leq & |{\cal X}| |{\cal Y}_1|...|{\cal Y}_m| + 1.
  \end{array}
  \!\!\! \right\}.
\end{eqnarray}

This region can be proven using the same steps as the main result.  Notice that ${\cal D}_{BC}$ contains a conditional independence constraint.

\subsection{Game Theory}
\label{subsection game theory}

Consider a zero-sum repeated game between two teams.  Team A consists of two players who on the $t$th iteration take actions $X_t \in {\cal X}$ and $Y_t \in {\cal Y}$.  The opponents on Team B take a combined action $Z_t \in {\cal Z}$.  The strategy sets ${\cal X}$, ${\cal Y}$, and ${\cal Z}$ are finite.  The payoff for Team A at each iteration is a time-invariant finite function $\pi(X_t, Y_t, Z_t)$:  As a zero-sum game, the payoff for Team B is $-\pi(X_t, Y_t, Z_t)$.  Each participant observes all actions from previous iterations, and each team wishes to maximize its time-averaged expected payoff.

Let Team A play conservatively by assuming the best strategy for Team B.  In the worst case (from the viewpoint of Team A), the expected payoff in the $t$th iteration is
\begin{eqnarray}
\label{payoff function}
\Pi_t & \triangleq & \min_{z(\cdot,\cdot)} {\mathbb E} \; \pi(X_t,Y_t,z(X^{t-1},Y^{t-1})).
\end{eqnarray}
Clearly \eqref{payoff function} could be maximized by finding an optimal mixed strategy $P_{X,Y}^*$ that maximizes $\min_{z \in {\cal Z}} {\mathbb E}_{P^*} \pi(X,Y,z)$ and choosing independent actions accordingly for each iteration.  This would correspond to the minimax strategy.

{\em Communication Constraint:}
Now consider an additional constraint on Team A.  The players on Team A have as their only means of coordinating their actions a secure channel of communication, limited to a rate of $R$ bits per game iteration.  Specifically, Player 1, who chooses the actions $X_t$, communicates at rate $R$ to Player 2, who chooses $Y_t$.\footnote{Communication and common randomness play the same role in this setting.}

We say a rate $R$ is achievable for payoff $\Pi$ if there exists a communication protocol that obeys a rate limit of $R$ and produces average expected payoff no less than $\Pi$.  That is, there exists a block-length $n$ and a random variable triple $(X^n,Y^n,U)$ that has the conditional independence structure $X_t - (U, X^{t-1}, Y^{t-1}) - Y_t$ for all $t$ and such that $|{\cal U}| \leq 2^{nR}$ and
\begin{eqnarray}
\label{equation game achievability}
\frac{1}{n} \sum_{t=1}^n \Pi_t & \geq & \Pi.
\end{eqnarray}
Let ${\cal G}$ be the closure of the set of achievable pairs $(R,\Pi)$.

We claim that optimality is obtained by producing i.i.d. actions with respect to a designed joint distribution.  Define,
\begin{eqnarray}
  \mkern-18mu {\cal G}_0 & \triangleq & \left\{
  \begin{array}{rcl}
    (R,\Pi) & \in & {\cal R}^2 : \\
    \exists \; \distnarg{X,Y} & & \mbox{such that} \\
    R & \geq & C(X;Y), \\
    \Pi & \leq & \min_{z \in {\cal Z}} {\bf E} \; \pi(X,Y,z).
    \end{array}
  \right\},
\end{eqnarray}
where $C(X;Y)$ is the common information defined in \eqref{definition common information}.

\begin{lemma}[Optimal cooperative strategy]
\label{lemma game theory}
\begin{eqnarray}
{\cal G} & = & \mbox{Conv} ({\cal G}_0).
\end{eqnarray}
\end{lemma}

{\em Comments:}  Notice that Lemma~\ref{lemma game theory} involves a convexification of ${\cal G}_0$.  This means that it may be optimal to split time between two different efficient strategies---one that operates at a low communication rate and one that operates at a high communication rate---in order to satisfy the average rate constraint while competing effectively in the game.

Variants of this problem have been considered in \cite{cuff09-bayesian-games-Allerton}, \cite{cuff10-lossless-secrecy-Globecom}, \cite{cuff10-partial-secrecy-Allerton}, and \cite{schieler-cuff12-zero-rate-secrecy-ISIT}.  The difference in those works is that $X^n$ (the actions of Player 1 in this setting) are instead observed states of nature.  Their distribution is not designed by Team A.  The job of Player 1 is to compress and communicate the observed sequence efficiently to Player 2.  If the communication occurs over a public channel, with use of common randomness to conceal the communication, then the optimal solution is exactly characterized in \cite{cuff10-partial-secrecy-Allerton} and is integrally related to the ability to synthesize a memoryless channel.  However, communication over a private channel, as in the present setting, is addressed in \cite{cuff09-bayesian-games-Allerton} and still open.

The proof of Lemma~\ref{lemma game theory} is in the appendix.

\subsection{Public Channel}
\label{subsection public channel}

What if the communication used for distributed channel synthesis occurs over a public channel and we wish for the synthesis to be immune to statistical tests that utilize the public message $J$?  We require $X^n$ and $Y^n$ to pass as the input and output of a memoryless channel and $J$ to appear unrelated to $X^n$ and $Y^n$.  That is, for rates $(R,R_0)$ to be achievable, there must exist a sequence of $(R,R_0)$ channel synthesis codes such that the induced distribution \distnarg{X^n,Y^n,J} satisfies
\begin{eqnarray}
  \lim_{n \to \infty} \left\| \distnarg{X^n,Y^n,J} - \distnarg{J} \prod \distnarg[q]{X} \distnarg[q]{Y|X} \right\|_{TV} & = & 0.
\end{eqnarray}

This setting falls into the context of secrecy, related to \cite{cuff10-lossless-secrecy-Globecom}, \cite{cuff10-partial-secrecy-Allerton}, and \cite{satpathy-cuff13-cascade-ISIT}.  Common randomness can be used as a secret key to encrypt the public communication.  We find that this straightforward adaptation to distributed channel synthesis, where extra common randomness is used as a one-time-pad on the public communication, produces the optimal rate pairs $(R,R_0)$.  The closure of the set of achievable rate pairs is given by
\begin{eqnarray}
\label{definition S public}
  \mkern-24mu {\cal S}_{PC} & = & \left\{ \!\!
  \begin{array}{rcl}
    (R,R_0) \in {\cal R}^2 & : & \exists \; \distnarg{X,Y,U} \in {\cal D} \mbox{ s.t.} \\
    R & \geq & I(X;U), \\
    R_0 & \geq & I(X,Y;U).
  \end{array}
  \!\! \right\},
\end{eqnarray}
where ${\cal D}$ is defined in \eqref{definition D}.

Surprisingly, the common randomness rate requirement $R_0$ is greater than the communication rate requirement $R$ in the case of public communication.  The common randomness rate can be reduced to the common information $C_Q(X;Y)$, and the communication rate can be reduced to $I_Q(X;Y)$, but the two extremes cannot be achieved simultaneously in general.

Proof of this result is in the appendix.

\subsection{Limited Duration Fidelity}
\label{subsection limited memory}

Consider a relaxed objective for channel synthesis.  Suppose the objective is to synthesize a memoryless channel with high enough fidelity that it would pass any statistical test with limited memory of length $B$.  That is, for any $\epsilon>0$ we desire an encoding such that
\begin{eqnarray}
  \left\| \distnarg{X_{t-B}^{t},Y_{t-B}^{t}} - \prod \distnarg[q]{X} \distnarg[q]{Y|X} \right\|_{TV} & \leq & \epsilon \quad \forall t
\end{eqnarray}
where $B$ may be much smaller than the encoding block $n$.

The region of interest for a sharp rate requirement occurs when $B$ grows linearly with the encoding block-length:  $B = bn$.  In this case, the region of achievable rate pairs $(R,R_0)$ contains the following region:
\begin{eqnarray}
\label{definition S limited memory}
  \mkern-24mu {\cal S}_{LM} & = & \left\{ \!\!
  \begin{array}{rcl}
    (R,R_0) \in {\cal R}^2 & : & \exists \; \distnarg{X,Y,U} \in {\cal D} \mbox{ s.t.} \\
    R & \geq & I(X;U), \\
    R_0 + R & \geq & b I(X,Y;U).
  \end{array}
  \!\! \right\},
\end{eqnarray}
where ${\cal D}$ is defined in \eqref{definition D}.

In particular this means that for finite memory $B$ not growing with $n$, no common randomness is required, and the communication rate must only exceed $R \geq I_Q(X;Y)$.  Notice that the sum-rate bound in \eqref{definition S limited memory} is dominated by the communication rate bound in \eqref{definition S limited memory} when $U=Y$ and $n$ is large enough that $I_Q(X;Y) > \frac{B}{n} H_Q(Y)$.

See the appendix for the proof.

\subsection{Local Randomness}
\label{subsection local randomness}

The optimal encoder design for distributed channel synthesis, presented in \S\ref{section achievability}, calls for randomization at the encoder and decoder.  The randomization at the encoder is insignificant and perhaps even avoidable altogether.  It is easy to show, for example, that $H_P(J|X^n,K)$ scales no more than linearly with $n$ at a rate close to the arbitrarily small excess rate $R - I(X;U)$, where $U$ is the auxiliary random variable in the region ${\cal S}$ of \eqref{definition S}.  On the other hand, the private randomization required by the decoder is much larger.  The decoder of \S\ref{section achievability} locally synthesizes a memoryless channel according to \distnarg{Y|U} and applies the input $u^n(j,k)$ from the codebook to this synthesized channel.

\begin{figure}[ht]
  \begin{center}
    \begin{tikzpicture}[node distance=2cm]
     \node (src)   [coordinate] {};
     \node (enc)   [node,minimum width=2cm,right=1cm of src] {$F_{J|X^n K}$};
     \node (dummy) [coordinate,right=1cm of enc] {};
     \node (rnd) [rectangle,above=1cm of dummy] {$K\in [2^{nR_0}]$};
     \node (dec)    [node,minimum width=2cm,right=2cm of enc] {$G(J,K,L)$};
     \node (dummy2) [coordinate] at ([xshift=5mm] dec.center) {};
     \node (side) [rectangle, above=1cm of dummy2] {$L\in[2^{nR_L}]$};
     \node (out)  [coordinate,right=1cm of dec] {};

     \draw[arw] (src) to node [midway,above] {$X^n$} (enc);
     \draw[arw] (enc) to node [midway,above] {$J\in[2^{nR}]$} (dec);
     \draw[arw] (rnd) to [out=180,in=90] ([xshift=3mm] enc.north);
     \draw[arw] (rnd) to [out=0,in=90] ([xshift=-3mm] dec.north);
     \draw[arw] (side) to ([xshift=5mm] dec.north);
     \draw[arw] (dec) to node [midway,above] {$Y^n$} (out);
    \end{tikzpicture}
    \caption{{\em Local Randomness:}  In this extension to the main result, the decoder is deterministic but makes use of rate-limited local randomness.}
    \label{figure local randomness}
  \end{center}
\end{figure}
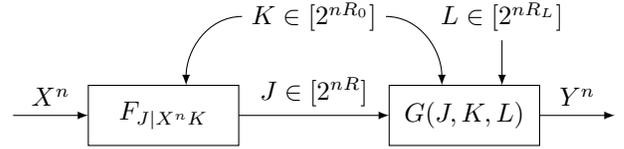

Here we quantify explicitly the amount of local randomness required by the decoder, similar to \cite{bloch-kliewer12-constrained-randomness-Arxiv} and \cite{steinberg-verdu94-channel-simulation}.  Let $R_L$ be the rate of random bits $L \in [2^{n R_L}]$ available to the decoder only, and define the decoder as a deterministic function
\begin{eqnarray}
  G & : & {\cal J} \times {\cal K} \times {\cal L} \to {\cal Y}^n.
\end{eqnarray}
This is depicted in Fig.~\ref{figure local randomness}.

We now aim to characterize the set of rate triples $(R,R_0,R_L)$ that can synthesize a memoryless channel \distnarg[q]{Y|X} with input distribution \distnarg[q]{X}, and we claim that the closure of this set is given by
\begin{eqnarray}
\label{definition S local randomness}
  {\cal S}_{LR} & = & \left\{
  \begin{array}{rcl}
    (R,R_0,R_L) & \in & {\cal R}^3 : \\
    \exists \; \distnarg{X,Y,U} & \in & {\cal D} \mbox{ such that} \\
    R & \geq & I(X;U), \\
    R_0 + R & \geq & I(X,Y;U), \\
    R_L & \geq & H(Y|U).
  \end{array}
  \right\},
\end{eqnarray}

The total amount of randomness flowing into our synthetic channel (ignoring the minimally random encoder), when all inequalities in ${\cal S}_{LR}$ of \eqref{definition S local randomness} are exercised with equality, is $R_0 + R_L = I(X,Y;U) - I(X;U) + H(Y|U) = H(Y|X)$.  To our delight, distributed channel synthesis is efficient even compared to the local synthesis in \cite{steinberg-verdu94-channel-simulation} and in Corollary~\ref{cor:cloud local synthesis}.

This proof can be found in the appendix.

\subsection{Common Information Duality}
\label{subsection common information duality}

Two notions of common information were introduced at nearly the same time in the literature.  One by G\'{a}cs and K\"{o}rner \cite{gacs-korner73-common-information} is defined as
\begin{eqnarray}
  C_{G-K}(X;Y) & \triangleq & \max_{f(\cdot) \; : \; H(f(X)|Y)=0} H(f(X)).
\end{eqnarray}
The other common information by Wyner in \cite{wyner75-common-information} is stated in \eqref{definition common information}.  For this discussion, we refer to Wyner's common information as $C_W(X;Y)$.

Attention has been drawn in the literature to dual properties of these two quantities.  For example,
\begin{eqnarray}
  C_W(X;Y) & \geq & I(X;Y) \label{eq:CI w MI} \\
  C_{G-K}(X;Y) & \leq & I(X;Y) \label{eq:CI gk MI}
\end{eqnarray}
Also, both can be viewed as extreme points for the common message rate in the Gray-Wyner network \cite{gray-wyner74-source-coding}.  In this network, correlated sources are encoded jointly using three messages and decoded separately, each decoder receiving only two of the messages.  The message received by both is the common message.  If we imagine the three messages traveling down a cable to a midway point (Segment~1) and then splitting into separate cables to travel to the separate decoders (Segment~2), with the common message duplicated at the juncture, then a simple duality can be stated.  When the sum rate of the first segment is efficient, the common message rate is at least $C_W(X;Y)$.  When the sum rate of the second segment is efficient, the common message rate is no more than $C_{G-K}(X;Y)$.  Furthermore, the first case yields inefficiency in the second segment equal to $C_W(X;Y) - I(X;Y)$, and the second case yields inefficiency in the first segment equal to $I(X;Y) - C_{G-K}(X;Y)$.  Thus, equality holds in both \eqref{eq:CI w MI} and \eqref{eq:CI gk MI} or in neither.

Here we emphasize another duality, using the present results to enrich the operational symmetry.

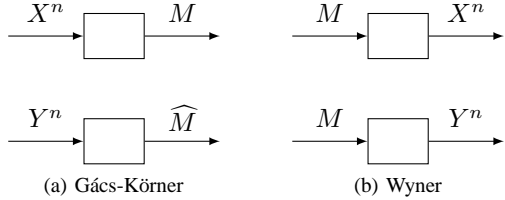
\begin{figure}[ht]
  \begin{center}
    \subfloat[G\'{a}cs-K\"{o}rner]{
    \begin{tikzpicture}[minimum height=6mm, minimum width=8mm]
      \node (in1) [coordinate] at (0,0) {};
      \node (ch1) [rectangle,draw,right=1cm of in1] {};
      \node (out1) [coordinate,right =1cm of ch1] {};
      \node (in2) [coordinate,below=1.4cm of in1] at (0,0) {};
      \node (ch2) [rectangle,draw,right=1cm of in2] {};
      \node (out2) [coordinate,right =1cm of ch2] {};

      \draw[arw] (in1) to node [midway,above] {$X^n$} (ch1);
      \draw[arw] (in2) to node [midway,above] {$Y^n$} (ch2);
      \draw[arw] (ch1) to node [midway,above] {$M$} (out1);
      \draw[arw] (ch2) to node [midway,above] {$\widehat{M}$} (out2);
    \end{tikzpicture}
    \label{subfigure common information gk}
    }
    \qquad
    \subfloat[Wyner]{
    \begin{tikzpicture}[minimum height=6mm, minimum width=8mm]
      \node (in1) [coordinate] at (0,0) {};
      \node (ch1) [rectangle,draw,right=1cm of in1] {};
      \node (out1) [coordinate,right =1cm of ch1] {};
      \node (in2) [coordinate,below=1.4cm of in1] at (0,0) {};
      \node (ch2) [rectangle,draw,right=1cm of in2] {};
      \node (out2) [coordinate,right =1cm of ch2] {};

      \draw[arw] (in1) to node [midway,above] {$M$} (ch1);
      \draw[arw] (in2) to node [midway,above] {$M$} (ch2);
      \draw[arw] (ch1) to node [midway,above] {$X^n$} (out1);
      \draw[arw] (ch2) to node [midway,above] {$Y^n$} (out2);
    \end{tikzpicture}
    \label{subfigure common information w}
    }
    \caption{{\em Operational duality of common information.}}
    \label{figure common information}
  \end{center}
\end{figure}

Fig.~\ref{figure common information} shows two complementary settings.  In the first, i.i.d. observations of correlated random variables are used by separate, independent nodes to generate the same random bits (with high probability).  The rate with which random bits can be generated is $C_{G-K}(X;Y)$.  In the second, equal random bits are provided to two independent nodes which must produce a correlated i.i.d. sequence (with high fidelity).  The required rate of random bits is $C_W(X;Y)$.  These results come directly from the original work in \cite{gacs-korner73-common-information} and \cite{wyner75-common-information}.

\begin{figure}[ht]
  \begin{center}
    \subfloat[Key Agreement]{
    \begin{tikzpicture}[minimum height=6mm, minimum width=8mm]
      \node (in1) [coordinate] at (0,0) {};
      \node (ch1) [rectangle,draw,right=1cm of in1] {};
      \node (out1) [coordinate,right =1cm of ch1] {};
      \node (in2) [coordinate,below=1.4cm of in1] at (0,0) {};
      \node (ch2) [rectangle,draw,right=1cm of in2] {};
      \node (out2) [coordinate,right =1cm of ch2] {};

      \upratearrow{ch2.north}{ch1.south}{1.5pt}{3pt}{$R_0$}{2pt}
      \draw[arw] (in1) to node [midway,above] {$X^n$} (ch1);
      \draw[arw] (in2) to node [midway,above] {$Y^n$} (ch2);
      \draw[arw] (ch1) to node [midway,above] {$M$} (out1);
      \draw[arw] (ch2) to node [midway,above] {$\widehat{M}$} (out2);
    \end{tikzpicture}
    \label{subfigure common information gk with comm}
    }
    \qquad
    \subfloat[Channel Synthesis]{
    \begin{tikzpicture}[minimum height=6mm, minimum width=8mm]
      \node (in1) [coordinate] at (0,0) {};
      \node (ch1) [rectangle,draw,right=1cm of in1] {};
      \node (out1) [coordinate,right =1cm of ch1] {};
      \node (in2) [coordinate,below=1.4cm of in1] at (0,0) {};
      \node (ch2) [rectangle,draw,right=1cm of in2] {};
      \node (out2) [coordinate,right =1cm of ch2] {};

      \upratearrow{ch2.north}{ch1.south}{1.5pt}{3pt}{$R_0$}{2pt}
      \draw[arw] (in1) to node [midway,above] {$M$} (ch1);
      \draw[arw] (in2) to node [midway,above] {$M$} (ch2);
      \draw[arw] (ch1) to node [midway,above] {$X^n$} (out1);
      \draw[arw] (ch2) to node [midway,above] {$Y^n$} (out2);
    \end{tikzpicture}
    \label{subfigure common information w with comm}
    }
    \caption{{\em Operational duality with communication:}  The rates of randomness $M$ for both situations relax to mutual information when communication is allowed (independent of the receiver output).}
    \label{figure common information with comm}
  \end{center}
\end{figure}
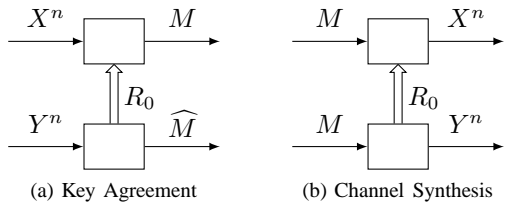

Now we add a communication link between the two encoders with a somewhat peculiar constraint.  The communication is required to be independent of the output of the receiving node (nearly independent as measured by total variation).  This alteration is depicted in Fig.~\ref{figure common information with comm}.

The setting of Fig.~\ref{subfigure common information gk with comm} has been studied for the purpose of secret key generation in \cite{maurer93-secret-key-agreement}, \cite{ahlswede-csiszar93-secret-key-agreement-part-1}, \cite{ahlswede-csiszar98-secret-key-agreement-part-2}, \cite{csiszar-narayan00-secret-key-agreement}, and \cite{maurer-wolf00-strong-secret-key-agreement}.  It is shown that with a high enough rate of communication, namely $H(Y|X)$, the rate of extraction of random bits in agreement increases to the mutual information $I(X;Y)$.

The dual setting of Fig.~\ref{subfigure common information w with comm} is solved by Theorem~\ref{theorem main result}.  With a high enough communication rate, the required rate of random bits reduces to the mutual information $I(X;Y)$.  Furthermore, a communication rate of $H(Y|X)$ is sufficient in this setting as well (and necessary for most distributions).

To see how this follows from Theorem~\ref{theorem main result}, consider the equivalent description of the distributed channel synthesis problem given in the beginning of \S\ref{section achievability}.  That description applies exactly to this situation as well.

For added curiosity, we can state the corner points $(R,R_0)$ of the characterizations of the achievable rate region for the two settings in a way that suggests a deeper relationship.  Both rate regions can be described as the union of simple regions, each defined by the choice of an auxiliary random variable.  In the case of Fig.~\ref{subfigure common information gk with comm}, the simple regions are rectangles defined by an upper bound on $R$ and a lower bound on $R_0$.  In the case of Fig.~\ref{subfigure common information w with comm}, the simple regions are pentagons defined by a lower bound on $R$ and a lower bound on the sum rate $(R+R_0)$.  In both cases, we now specify the corner points that define the regions.

For the setting of Fig.~\ref{subfigure common information gk with comm}, the corner points of the rate region, identified in \cite{ahlswede-csiszar98-secret-key-agreement-part-2}, are $(R,R_0)$ such that
\begin{eqnarray}
  R & = & I(X;U), \\
  R_0 & = & I(Y;U|X),
\end{eqnarray}
for some $U$ such that $X - Y - U$ forms a Markov chain.

For the setting of Fig.~\ref{subfigure common information w with comm}, the corner points of the rate region, identified by Theorem~\ref{theorem main result}, are $(R,R_0)$ such that
\begin{eqnarray}
  R & = & I(X;U), \\
  R_0 & = & I(Y;U|X),
\end{eqnarray}
for some $U$ such that $X - U - Y$ forms a Markov chain.

\section{Soft Covering Lemma}
\label{section:cloud}

\subsection{Discussion}

Our achievability proof centers around a {\em soft covering lemma}\footnote{In \cite{cuff09-dissertation} we referred to this as the ``cloud mixing'' lemma.} that is conceptually rooted in \cite[Theorem~6.3]{wyner75-common-information} by Wyner and further studied in \cite{han-verdu93-output-statistics} and elsewhere.  In this section we state the simple form of the lemma needed for our proof, accompanied by a new exponential bound.  We then discuss this principle in greater depth, including proofs, in \S\ref{section:soft covering continued}.

\begin{figure}[ht]
\begin{center}
\tikzstyle{arw}=[->,>=latex]

\pgfmathsetseed{3}
\pgfmathsetmacro{\major}{1+0.2*rand}
\pgfmathsetmacro{\minor}{1+0.2*rand}
\pgfmathsetmacro{\angle}{(rand+1)*180}

\begin{tikzpicture}
 \pgfmathsetmacro{\major}{1+0.2*rand}
 \pgfmathsetmacro{\minor}{1+0.2*rand}
 \pgfmathsetmacro{\angle}{(rand+1)*180}
 \draw (0,0) +($.2*(rand,rand)$) [rotate around = {\angle:+(0,0)}] ellipse (\major cm and \minor cm) node {\distnarg[\Phi]{V^n|U^n(7)}};

 \foreach \i in {1,...,6}
 {
 \pgfmathsetmacro{\major}{1+0.2*rand};
 \pgfmathsetmacro{\minor}{1+0.2*rand};
 \pgfmathsetmacro{\angle}{(rand+1)*180};
 \draw ($(60*\i:1.8)$) +($.2*(rand,rand)$) [rotate around = {\angle:+(0,0)}] ellipse (\major cm and \minor cm) node {\distnarg[\Phi]{V^n|U^n(\i)}};
 }

\node (space) [ellipse,minimum height=5.6 cm,minimum width=5.2 cm,draw,dashed] at (0,0) {};
\node (dummy) [coordinate, right=3mm of space] {};
\node (lab) [above=17mm of dummy] {$\prod \distnarg[\Phi]{V}$};
\draw [arw] (lab) to (space);
\end{tikzpicture}
\caption{{\em Soft Covering:}  A sparse collection (codebook) of conditional distributions \distnarg[\Phi]{V^n|U^n=u^n(i)} is averaged together to approximate a marginal distribution $\prod \distnarg[\Phi]{V}$.  For an i.i.d. codebook distribution and a memoryless channel \distnarg[\Phi]{V^n|U^n}, an exponentially large codebook of rate $R > I(U;V)$ is sufficient.}
\label{figure cloud mixing diagram}
\end{center}
\end{figure}
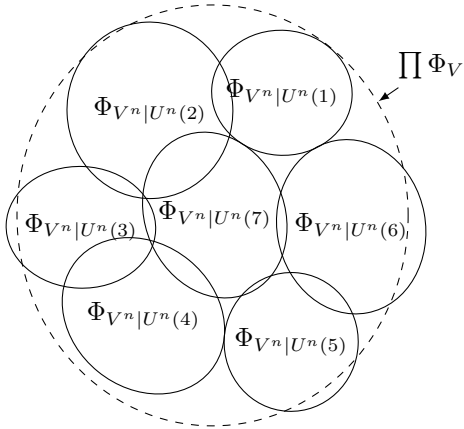

The lemma pertains to the distribution induced by selecting uniformly at random from a random codebook and passing the codeword through a memoryless channel.  If the size of the codebook is large enough, then the resulting distribution on the output of the channel, illustrated in Fig.~\ref{figure cloud mixing diagram}, will be well approximated by an i.i.d. distribution.  Not surprisingly, the rate of the codebook sufficient to observe this phenomenon is the mutual information associated with the codebook distribution and the channel.  Wyner used this observation in his achievability proof for common information.  This lemma will also play a key role in our achievability proof for distributed channel synthesis, providing us with a simple analysis.

This phenomenon has a close relative, often referred to simply as covering.  Given a joint distribution \distnarg{U,V}, in the limit of large block-length, a random codebook of size $2^{nR}$ of $u^n$ sequences, with $R > I(U;V)$, covers the i.i.d. distribution associated with \distnarg{V} in the sense that for a random $V^n$ there will exist with high probability a sequence in the codebook such that $(u^n,V^n)$ are jointly typical per the definition (10.107) in \cite{cover-thomas06-eit}.  That is, the union of conditionally typical sets in ${\cal V}^n$ induced by the codebook will contain all of the probability of the i.i.d. distribution in the limit.  The soft covering lemma is a strengthening of this statement.  Instead of looking at the union of conditionally typical sets, it states that the average of conditional distributions from each codeword in the codebook will match the i.i.d. distribution to arbitrarily high fidelity.

In \cite{han-verdu93-output-statistics}, Han and Verd\'{u} study this soft covering phenomenon in depth.  Notably, they examine the converse statement, characterizing the necessary codebook rates, in addition to sufficient rates.  This they refer to as the ``resolvability'' of a channel.  Furthermore, their work looks beyond memoryless channels. Also, they consider total variation as a metric for the fidelity of the distribution (as we do here), in addition to normalized K-L divergence, which was the metric that Wyner used.

Other work has also continued the study of this phenomenon.  Hayashi's derivation in \cite{hayashi06-resolvability} provided the tightest previous bound in the literature for memoryless channels, upon which this work improves.  Also, \cite{ahlswede-winter02-quantum-id-converse} and Chapter~16 of \cite{wilde11-quantum-textbook} contain similar lemmas in a broader quantum context, both referring to the tool as a ``covering lemma.''  The ``sampling lemma'' in \cite{winter05-triples} makes a particularly strong claim that the soft covering phenomenon occurs even under the stricter fidelity metric of K-L divergence (not normalized by the block-length).

Recent work in \cite{yassaee-aref-gohari12-random-binning-ISIT} and \cite{schieler13-coodination-codes} has developed alternative constructions and analysis tools for obtaining similar properties to what soft covering provides, partially motivated by our work in \cite{cuff08-channel-synthesis-ISIT} and \cite{cuff-permuter-cover10-coordination-capacity}.

\subsection{Soft Covering Lemma Statement}
\label{subsection:cloud memoryless}

The simplest statement of the soft covering principle, and all that we need for the proof of distributed channel synthesis, involves a memoryless channel with memoryless input.  Let \distnarg[\Phi]{U} be a distribution on ${\cal U}$ that induces a distribution \distnarg[\Phi]{V} when applied to the channel \distnarg[\Phi]{V|U}.  For $n$ channel uses, the corresponding input-output joint distribution is then
\begin{eqnarray}
\distnarg[\Phi]{U^n,V^n} & = &
\begin{array}{cc}
\mbox{Input} & \mbox{Channel} \\
\left( \prod \distnarg[\Phi]{U} \right) & \left( \prod \distnarg[\Phi]{V|U} \right)
\end{array} \\
& = & \prod \distnarg[\Phi]{U,V},
\end{eqnarray}
yielding the {\em desired output distribution}
\begin{eqnarray}
\label{eq:memoryless mixing desired output}
\dist[q]{V^n} & \triangleq & \sum_{u^n} \dist[\Phi]{U^n,V^n} \\
& = & \prod_{t=1}^{n} \distalt[\Phi]{V}{v_t}.
\end{eqnarray}

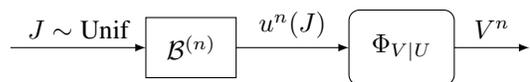
\begin{figure}[ht]
\begin{center}
\begin{tikzpicture}[node distance=2cm]
 \node (src)   [coordinate] {};
 \node (enc)   [node,minimum width=12mm,right=18mm of src] {${\cal B}^{(n)}$};
 \node (ch)    [node,minimum width=14mm,minimum height=1cm,rounded corners,right=15mm of enc] {\distnarg[\Phi]{V|U}};
 \node (outV)  [coordinate,right=1cm of ch] {};

 \draw[arw] (src) to node [midway,above] {$J \sim \text{Unif}$} (enc);
 \draw[arw] (enc) to node [midway,above] {$u^n(J)$} (ch);
 \draw[arw] (ch) to node [midway,above] {$V^n$} (outV);
\end{tikzpicture}
\caption{{\em Soft covering - Memoryless channel:}  An i.i.d. output distribution $\prod \distnarg[\Phi]{V}$ is synthesized by randomly selecting a codeword from a codebook of $u^n$ sequences and passing it through a memoryless channel.  Lemma~\ref{lemma:cloud memoryless} gives a sufficient codebook construction.}
\label{figure cloud memoryless}
\end{center}
\end{figure}

The lemma, which follows, states that we can nearly produce the desired output distribution by applying a uniform distribution to a collection ${\cal B}^{(n)}$ of $2^{n (I(U;V) + \epsilon)}$ randomly generated channel input sequences, as depicted in Fig.~\ref{figure cloud memoryless}.  The criterion for nearly producing the desired output distribution is that the {\em induced output distribution}
\begin{eqnarray}
\label{eq:mixing induced output}
\dist{V^n} & \triangleq & \frac{1}{|{\cal B}^{(n)}|} \sum_{j=1}^{|{\cal B}^{(n)}|} \distalt[\Phi]{V^n|U^n}{v^n|u^n(j)}
\end{eqnarray}
has vanishing total variation from the desired output distribution as $n$ increases.

\begin{lemma}[Soft Covering - Memoryless channel]
\label{lemma:cloud memoryless}
Let ${\cal B}^{(n)}$ be a randomly generated collection of $2^{nR}$ sequences in ${\cal U}^n$, each drawn independently and i.i.d. from the codebook distribution \distnarg[\Phi]{U}.  A memoryless channel specified by \distnarg[\Phi]{V|U} induces an output distribution defined in \eqref{eq:mixing induced output}.  This output distribution is random because ${\cal B}^{(n)}$ is random.

If $R > I_{\Phi}(U;V)$ then the expected value of the total variation between the induced output distribution and the desired output distribution defined in \eqref{eq:memoryless mixing desired output} vanishes with $n$.  That is,
\begin{eqnarray}
\label{eq:cloud mixing memoryless part 1}
\mkern-24mu R > I_{\Phi}(U;V) & \Longrightarrow & \lim_{n \to \infty} {\bf E} \left\| \distnarg{V^n} - \distnarg[q]{V^n} \right\|_{TV} = 0.
\end{eqnarray}
Furthermore, the expected total variation vanishes exponentially fast:
\begin{eqnarray}
\label{eq:cloud mixing memoryless part 2}
{\bf E} \left\| \distnarg{V^n} - \distnarg[q]{V^n} \right\|_{TV} & \leq & \frac{3}{2} \exp (- \gamma n),
\end{eqnarray}
where $\gamma$ is given in \eqref{eq:exponent digital rate} of \S\ref{subsection exponential convergence}.
\end{lemma}

Lemma~\ref{lemma:cloud memoryless} can be derived as a corollary of Theorem~\ref{lemma:cloud general source and channel}, which is stated in \S\ref{section:soft covering continued} as a generalization of the soft covering principle, along with a variety of corollaries, the proof, and analysis of the total variation exponent.

\section{Achievability}
\label{section achievability}

\subsection{Synopsis}

In this section we prove ${\cal C} \supset {\cal S}$.  That is to say, for any rate pair $(R,R_0)$ in the interior of the rate region specified by ${\cal S}$, $(R,R_0)$ is achievable for synthesizing the memoryless channel \distnarg[q]{Y|X} with input distribution \distnarg[q]{X}.  The definition of achievability in Definition~\ref{definition achievability} concerns the existence of $(R,R_0,n)$ channel synthesis codes.  However, the same achievability criterion can be stated simply in terms of the existence of a joint distribution satisfying certain properties, removing the emphasis from the usual causal description of how an encoder or decoder takes an input and returns an output.  This method can be used to redefine any of the familiar communication problems in information theory, but we find it particularly useful in this case.

Consider the induced joint distribution \distnarg{X^n,Y^n,J,K} defined in Definition~\ref{definition induced distribution}.  The rates $(R,R_0)$ are achievable if for any $\epsilon>0$ there exists an $N$ such that for all block lengths $n>N$ there exists an induced joint distribution \distnarg{X^n,Y^n,J,K} satisfying the following properties:
\begin{enumerate}
\item $X^n - (J,K) - Y^n$ form a Markov chain.
\item $X^n$ and $K$ are independent.
\item $X^n$ is i.i.d. $\sim \distnarg[q]{X}$.
\item $|{\cal J}| = 2^{nR}$.
\item $|{\cal K}| = 2^{nR_0}$. \footnote{The result does not change if $K$ is required to be uniformly distributed per the original problem statement.}
\item $\left\| \distnarg{X^n,Y^n} - \prod \distnarg[q]{X} \distnarg[q]{Y|X} \right\|_{TV} < \epsilon$.
\end{enumerate}
This is simply an exhaustive list of all of the constraints imposed by the definitions of channel synthesis codes and the induced joint distribution, with the addition of 6), the synthesis requirement.

Our approach will be to construct a joint distribution \distnarg[\Upsilon]{X^n,Y^n,J,K} that satisfies 1), 4), and 5) by construction.  We will then use the soft covering lemma of \S\ref{section:cloud} to show that 6) is satisfied while 2) and 3) are nearly satisfied.  Fortunately, due to some basic properties of total variation, we can augment the joint distribution to exactly satisfy 2) and 3) while not destroying the other properties.

The key idea for developing this proof is to relax some of the strict requirements (properties 2) and 3)), knowing that this relaxation can be corrected at the end.  By doing so, we reveal a large degree of symmetry in the problem statement.  Rather than design the joint distribution from left to right (referring to the Markov chain in property 1)), we design from the middle outward.

The consequence of this technique is that we design the encoder in reverse.  The result is best described as a {\em likelihood encoder} (see \S\ref{subsection encoder comments}), which is stochastic.  Similarity between the behavior of this encoder and other encoders used for source coding is analyzed in \cite{schieler13-coodination-codes}.  Also, an alternative proof construction based on random binning, which yields similar behavior to the likelihood encoder, is proposed in \cite{yassaee-aref-gohari12-random-binning-ISIT}.

\subsection{Construction}

Begin by finding $\distnarg[q]{X,Y,U} \in {\cal D}$ defined in \eqref{definition D} such that $R > I_{\distnarg[q]{}}(X;U)$ and $R_0 + R > I_{\distnarg[q]{}}(X,Y;U)$.  Our reuse of the label $Q$ is intentional.  By the definition of ${\cal D}$, the marginal distribution of \distnarg[q]{X,Y,U} must coincide with the desired input-output distribution specified by $\distnarg[q]{X} \distnarg[q]{Y|X}$.

\begin{figure}[ht]
\begin{center}
\begin{tikzpicture}[node distance=2cm]
  \node (level1) [coordinate] {};
  \node (level2) [coordinate, below=1cm of level1] {};
  \node (level1c) [coordinate, above=3mm of level1] {};
  \node (level2c) [coordinate, below=3mm of level2] {};
  \node (levelm) [coordinate, below=5mm of level1] {};
  \node (enc) [node, minimum width=12mm, minimum height=18mm, right=18mm of levelm] {${\cal B}^{(n)}$};
  \node (split)   [coordinate, right=2cm of enc] {};
  \node (chh) [coordinate, right=1.5cm of split] {};
  \node (ch1) [node, minimum width=14mm,minimum height=1cm, rounded corners] at (level1c -| chh) {\distnarg[q]{X|U}};
  \node (ch2) [node, minimum width=14mm,minimum height=1cm, rounded corners] at (level2c -| chh) {\distnarg[q]{Y|U}};
  \node (out1)  [coordinate, right=1cm of ch1] {};
  \node (out2)  [coordinate, right=1cm of ch2] {};

\draw[arw] (level1) to node [midway, above] {$J \sim \text{Unif}$} (level1 -| enc.west);
\draw[arw] (level2) to node [midway, above] {$K \sim \text{Unif}$} (level2 -| enc.west);
\draw[arw] (ch1) to node [midway, above] {$X^n$} (out1);
\draw[arw] (ch2) to node [midway, above] {$Y^n$} (out2);
\draw (enc) to node [midway, above] {$u^n(J,K)$} (split);
\draw (split) to (level1c -| split);
\draw (split) to (level2c -| split);
\draw[arw] (level1c -| split) to (ch1);
\draw[arw] (level2c -| split) to (ch2);
\end{tikzpicture}
\caption{{\em Codec Construction:}  The first step in deriving an efficient encoder and decoder is to construct a joint distribution \distnarg[\Upsilon]{X^n,Y^n,J,K} that nearly satisfies the six conditions for achievability.  This is done by constructing a randomly generated codebook ${\cal B}^{(n)}$ of sequences $u^n(j,k)$.  Independent and uniformly distributed indices $J$ and $K$ select from the codebook the input to a memoryless broadcast channel specified by $\distnarg[q]{X|U} \distnarg[q]{Y|U}$.   If the cardinalities of $J$ and $K$ are large enough, $X^n$ and $Y^n$ can be shown using the soft covering lemma to be nearly i.i.d. according to the desired distribution, with $X^n$ nearly independent of $K$.}
\label{figure encoder construction}
\end{center}
\end{figure}
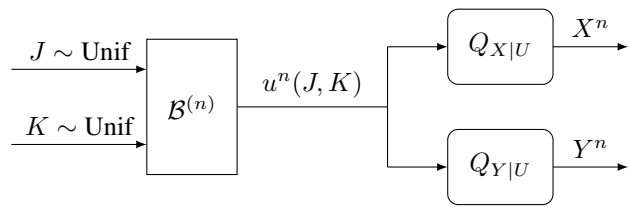

Using the standard practice of random codebook construction to prove the existence of a good codebook, generate a codebook ${\cal B}^{(n)}$ of $u^n$ sequences indexed by $j \in [2^{nR}]$ and $k \in [2^{nR_0}]$ independently according to $\prod_{t=1}^n \distalt[q]{U}{u_t}$.  Construct a joint distribution as depicted in Fig.~\ref{figure encoder construction} and as follows.  Define \distnarg[\Upsilon]{X^n,Y^n,J,K} such that $J$ and $K$ are uniformly distributed over their supports and $X^n$ and $Y^n$ are the result of the codeword $u^n(J,K)$ passed through the memoryless channel defined by \distnarg[q]{X,Y|U}:
\begin{eqnarray}
  & & \mkern-54mu \dist[\Upsilon]{X^n,Y^n,J,K} \nonumber \\
  & \triangleq & \frac{1}{2^{n (R_0 + R)}} \left( \prod_{t=1}^n \distalt[q]{X,Y|U}{x_t,y_t|u_t(j,k)} \right).
\end{eqnarray}
Notice that the channel \distnarg[q]{X,Y|U} separates into $\distnarg[q]{X|U} \distnarg[q]{Y|U}$, as shown in Fig.~\ref{figure encoder construction}, because of the Markov chain property of all distributions in ${\cal D}$.

It is clear that \distnarg[\Upsilon]{X^n,Y^n,J,K} satisfies properties 1), 4), and 5) by construction.  Our next step is to construct \distnarg{X^n,Y^n,J,K} from \distnarg[\Upsilon]{X^n,Y^n,J,K} in a way that satisfies properties 2) and 3).  Define \distnarg{X^n,Y^n,J,K} as follows:
\begin{eqnarray}
  \distnarg{X^n,Y^n,J,K} & = & \frac{1}{2^{nR_0}} \left( \prod \distnarg[q]{X} \right) \distnarg[\Upsilon]{Y^n,J|X^n,K}.
\end{eqnarray}
The conditional distribution \distnarg[\Upsilon]{Y^n,J|X^n,K} is derived from \distnarg[\Upsilon]{X^n,Y^n,J,K} and well defined for all values of $(X^n,K)$ with positive probability.  For others values we can simply assign the uniform distribution over ${\cal Y}^n \times {\cal J}$.

Notice that \distnarg{X^n,Y^n,J,K} satisfies property 1) because $\distnarg{Y^n|X^n,J,K} = \distnarg[\Upsilon]{Y^n|X^n,J,K} = \distnarg[\Upsilon]{Y^n|J,K}$.  Thus, \distnarg[\Upsilon]{J|X^n,K} defines the likelihood encoder, and \distnarg[\Upsilon]{Y^n|J,K} is the decoder.  The distribution \distnarg{X^n,Y^n,J,K} satisfies 2), 3), 4), and 5) as well by construction. Only property 6) is left to be verified.

\subsection{Synthesis Analysis}

Recall that \distnarg[\Upsilon]{X^n,Y^n,J,K} and \distnarg{X^n,Y^n,J,K} are random because the codebook ${\cal B}^{(n)}$ is random.  We now call on the soft covering lemma (Lemma~\ref{lemma:cloud memoryless}) twice.  First, we have a straightforward conclusion.  Since $R_0 + R > I_{\distnarg[q]{}}(X,Y;U)$,
\begin{eqnarray}
  \lim_{n \to \infty} \mathbf{E} \left\| \distnarg[\Upsilon]{X^n,Y^n} - \prod \distnarg[q]{X} \distnarg[q]{Y|X} \right\|_{TV} & = & 0. \label{eq:ach objective bound}
\end{eqnarray}

The second use of Lemma~\ref{lemma:cloud memoryless} is a little more subtle and similar to a technique used by Winter in \cite{winter05-triples} and by Bloch and Laneman in \cite{bloch-laneman11-secrecy-from-resolvability-Arxiv}.  Notice that for any fixed $k$, the collection $\{u^n(j,k)\}_j$ is a collection of $2^{nR}$ randomly generated codewords.  If we consider only the memoryless channel specified by \distnarg[q]{X|U} with channel output $X^n$, then $R > I_{\distnarg[q]{}}(U;X)$ satisfies the condition of the lemma. Therefore, for any $k$,
\begin{eqnarray}
  \mathbf{E} \left\| \distnarg[\Upsilon]{X^n|K=k} - \prod \distnarg[q]{X} \right\|_{TV} & < & \epsilon_n \; \to \; 0
\end{eqnarray}
as $n \to \infty$.  The expression on the left-hand side is constant over all values of $k$ for any fixed $n$ because of the symmetric nature of an i.i.d. codebook.

By the definition of total variation in \eqref{eq:tv is sum},
\begin{eqnarray}
  & & \mkern-54mu \mathbf{E} \left\| \distnarg[\Upsilon]{X^n,K} - \frac{1}{2^{nR_0}} \prod \distnarg[q]{X} \right\|_{TV} \nonumber \\
  & = & \mathbf{E} \frac{1}{2} \sum_{x^n,k} \left| \dist[\Upsilon]{X^n,K} - \frac{1}{2^{nR_0}} \prod_{t=1}^n \distalt[q]{X}{x_t} \right| \nonumber \\
  & = & \frac{1}{2^{nR_0}} \mathbf{E} \frac{1}{2} \sum_{x^n,k} \left| \dist[\Upsilon]{X^n|K} - \prod_{t=1}^n \distalt[q]{X}{x_t} \right| \nonumber \\
  & = & \frac{1}{2^{nR_0}} \sum_k \mathbf{E} \left\| \distnarg[\Upsilon]{X^n|K=k} - \prod \distnarg[q]{X} \right\|_{TV} \nonumber \\
  & \leq & \frac{1}{2^{nR_0}} \sum_k \epsilon_n \nonumber \\
  & = & \epsilon_n \nonumber \\
  & \to & 0.  \label{eq:ach constraint bound}
\end{eqnarray}
Thus, \distnarg[\Upsilon]{X^n,Y^n,J,K} satisfies property 6) by \eqref{eq:ach objective bound}, and it nearly satisfies properties 2) and 3) by \eqref{eq:ach constraint bound}.  We next invoke two properties of total variation.

\begin{lemma}[Total Variation of Marginal Distribution]
\label{lemma:tv marginal}
Total variation cannot be larger between marginal distributions than between encompassing joint distributions.  That is,
\begin{eqnarray}
\label{eq:tv marginal}
  \left\| \distnarg[\Pi]{W} - \distnarg[\Gamma]{W} \right\|_{TV} & \leq & \left\| \distnarg[\Pi]{W,Z} - \distnarg[\Gamma]{W,Z} \right\|_{TV}.
\end{eqnarray}
\end{lemma}

\begin{IEEEproof}
Referring to the definition of total variation given in \eqref{eq:define tv}, the left-hand side of \eqref{eq:tv marginal} is a maximization over a smaller set than the right-hand side.
\end{IEEEproof}

\begin{lemma}[Total Variation with Common Channel]
\label{lemma:tv channel}
When two random variables are passed through the same channel, the total variation between the resulting input-output joint distributions is the same as the total variation between the input distributions.  That is,
\begin{eqnarray}
\label{eq:tv channel}
  \mkern-18mu \left\| \distnarg[\Pi]{W} \distnarg[\Pi]{Z|W} - \distnarg[\Gamma]{W} \distnarg[\Pi]{Z|W} \right\|_{TV} & = & \left\| \distnarg[\Pi]{W} - \distnarg[\Gamma]{W} \right\|_{TV}.
\end{eqnarray}
\end{lemma}

\begin{IEEEproof}
  Referring to the equivalent definition for total variation given in \eqref{eq:tv is l1}, the non-negative \distnarg[\Pi]{Z|W} term from the left-hand side of \eqref{eq:tv channel} factors out of the absolute value and sums to one.
\end{IEEEproof}

We continue with the final steps of the analysis of \distnarg{X^n,Y^n} using the triangle inequality:
\begin{eqnarray}
  & & \mkern-54mu \left\| \distnarg{X^n,Y^n} - \prod \distnarg[q]{X} \distnarg[q]{Y|X} \right\|_{TV} \nonumber \\
  & \leq & \left\| \distnarg{X^n,Y^n} - \distnarg[\Upsilon]{X^n,Y^n} \right\|_{TV} \nonumber \\
   & & \quad \quad \quad + \; \left\| \distnarg[\Upsilon]{X^n,Y^n} - \prod \distnarg[q]{X} \distnarg[q]{Y|X} \right\|_{TV} \nonumber \\
  & \stackrel{(a)}{\leq} & \left\| \distnarg{X^n,Y^n,J,K} - \distnarg[\Upsilon]{X^n,Y^n,J,K} \right\|_{TV} \nonumber \\
   & & \quad \quad \quad + \; \left\| \distnarg[\Upsilon]{X^n,Y^n} - \prod \distnarg[q]{X} \distnarg[q]{Y|X} \right\|_{TV} \nonumber \\
  & \stackrel{(b)}{=} & \left\| \distnarg{X^n,K} - \distnarg[\Upsilon]{X^n,K} \right\|_{TV} \nonumber \\
   & & \quad \quad \quad + \; \left\| \distnarg[\Upsilon]{X^n,Y^n} - \prod \distnarg[q]{X} \distnarg[q]{Y|X} \right\|_{TV} \nonumber \\
  & = & \left\| \frac{1}{2^{nR_0}} \left( \prod \distnarg[q]{X} \right) - \distnarg[\Upsilon]{X^n,K} \right\|_{TV} \nonumber \\
   & & \quad \quad \quad + \; \left\| \distnarg[\Upsilon]{X^n,Y^n} - \prod \distnarg[q]{X} \distnarg[q]{Y|X} \right\|_{TV}.  \label{eq:ach summary bound}
\end{eqnarray}
Both terms vanish as $n \to \infty$ because of \eqref{eq:ach constraint bound} and \eqref{eq:ach objective bound}.  Inequality (a) comes from Lemma~\ref{lemma:tv marginal}, and (b) comes from Lemma~\ref{lemma:tv channel}.

Therefore, for $n$ large enough, there exists a distribution satisfying all properties for achievability.  Furthermore, the soft covering lemma asserts that each of the total variation terms in the bound vanishes exponentially quickly.  \hfill $\blacksquare$

\subsection{Comments} \label{subsection encoder comments}
We now summarize the behavior of the optimal encoder and decoder constructed in this section: \\

\noindent
{\em Likelihood Encoder:}  The encoder inspects the codebook ${\cal B}^{(n)}$ of $u^n$ sequences indexed by $j$ and $k$ and considers only the subset where $k$ is equal to the common randomness observed.  In other words, the common randomness selects a sub-codebook.  The encoder then considers each $u^n$ sequence in the sub-codebook and selects one randomly with probability proportional to its likelihood associated with the memoryless channel \distnarg[q]{X|U} and the observed source sequence $X^n$.

It may happen that every codeword has a positive probability of being selected by the encoder; however, most of the probability will be concentrated on those codewords that are jointly typical with $X^n$.  Still, there are many jointly typical sequences to choose from randomly.  An interesting endeavor would be to design a deterministic encoder, if possible, that successfully operates throughout the region where both inequalities of Theorem~\ref{theorem main result} are active. \\

\noindent
{\em Decoder:}  The decoder identifies a codeword $u^n$ given by the codebook ${\cal B}^{(n)}$, the common randomness $K$, and the message $J$.  He then locally synthesizes a memoryless channel according to \distnarg[q]{Y|U} to produce $Y^n$ from $u^n$.

In the decoder's case, a specific amount of randomization ($H(Y|U)$ per channel use) is fundamental to the design and unavoidable according to \S\ref{subsection local randomness}.

\section{Converse}
\label{section converse}

In this section we prove ${\cal C} \subset {\cal S}$.  That is, any achievable rate pair $(R,R_0)$ for synthesizing the memoryless channel \distnarg[q]{Y|X} with input distribution \distnarg[q]{X} must fall in ${\cal S}$.\footnote{The set ${\cal S}$ is a closed set.}

\subsection{Cardinality Bound}
\label{subsection cardinality bound}

The cardinality bound on the auxiliary random variable $U$ in the definition of ${\cal D}$ in \eqref{definition D} not only makes the region computable but is an essential step in the converse, as will be apparent in \S\ref{subsection lower semi continuity}.

\begin{lemma}[Cardinality Bound]
\label{lemma cardinality bound}
For any discrete random variables $(X,Y,W) \sim \distnarg[\Pi]{X,Y,W}$ forming a Markov chain $X - W - Y$, there exists a distribution \distnarg[\Gamma]{X,Y,U} forming a Markov chain $X - U - Y$ such that
\begin{eqnarray}
  |{\cal U}| & \leq & |{\cal X}| |{\cal Y}| + 1, \\
  \distnarg[\Gamma]{X,Y} & = & \distnarg[\Pi]{X,Y}, \\
  I_{\Gamma}(X;U) & = & I_{\Pi}(X;W), \\
  I_{\Gamma}(X,Y;U) & = & I_{\Pi}(X,Y;W).
\end{eqnarray}
\end{lemma}

\begin{IEEEproof}
Consider the set of points ${\cal A} \in {\cal R}^{|{\cal X}| |{\cal Y}| + 2}$ such that the first $|{\cal X}| |{\cal Y}|$ coordinates represent the mass values of a product (independent) distribution $\distnarg{X} \distnarg{Y}$ and the last two coordinates are $H_{P}(X)$ and $H_{P}(X,Y)$.  This is a connected and compact set because each coordinate is a continuous function on the connected and compact set of all product distributions.

Recall that \distnarg[\Pi]{X,Y,W} is the distribution of the Markov chain $X - W - Y$ in question.  Consider the point $\pi \in {\cal R}^{|{\cal X}| |{\cal Y}| + 2}$ where \distnarg[\Pi]{X,Y} specifies the first $|{\cal X}| |{\cal Y}|$ coordinates and $H_{\Pi}(X|W)$ and $H_{\Pi}(X,Y|W)$ the last two.  Notice that $\pi$ is in the convex hull of ${\cal A}$.  It is a convex combination of points, each represented by a particular value of $w$, with convex weight equal to \dist[\Pi]{W}.  The constituent product distributions are the distributions of $(X,Y)$ conditioned on $W=w$.

Notice that the connected and compact set ${\cal A}$ is in fact contained in a $(|{\cal X}| |{\cal Y}| + 1)$-dimensional subspace of ${\cal R}^{|{\cal X}| |{\cal Y}| + 2}$ because of the linear constraint that a probability mass function sum to one.  Thus, the Carath\'{e}odory theorem for a connected set\footnote{This theorem is often referred to as the Carath\'{e}odory-Fenchel-Eggleston theorem.} states that $\pi$ is a convex combination of $(|{\cal X}| |{\cal Y}| + 1)$ points in ${\cal A}$ (see original publications:  \cite{caratheodory11-convex-set} for compact sets, \cite{steinitz13-caratheodory-bound} for general sets, \cite{eggleston63-caratheodory-bound} for connected sets; application to cardinality bounds of auxiliary variables:  \cite{salehi78-cardinality-bound}, \cite[Lemma~15.4]{csiszar-korner11-information-theory-book}, \cite{elgamal-kim11-nit}).  Associate each point with a value $u$.  We use these points to construct the distribution \distnarg[\Gamma]{X,Y,U}.  The convex weight of the points becomes \dist[\Gamma]{U}, and the associated product distributions are the conditional distributions \distnarg[\Gamma]{X,Y|U=u}, yielding the desired Markov chain property $X - U - Y$.  Notice that the joint distribution of $(X,Y)$ and the conditional entropies are preserved by the construction of $\pi$.
\end{IEEEproof}

\subsection{Entropy bounds}
\label{subsection entropy bounds}

A few preliminary bounds are needed to show that sequences that are nearly i.i.d. in total variation will have information properties close to their i.i.d. counterparts.

\begin{lemma}[Total Variation of Random Sample]
\label{lemma:tv random time index}
The total variation between the distributions of two random sequences is an upper bound on the total variation between the distributions of the variables in the sequences at a random time index (independent of the sequences).

Let $T \in \{1,...,n\}$ be a random time index distributed according to \distnarg[\Pi]{T}.  Also let \distnarg[\Pi]{W^n} and \distnarg[\Gamma]{W^n} be distributions independent of $T$, so that $\distnarg[\Pi]{W^n,T} = \distnarg[\Pi]{W^n} \distnarg[\Pi]{T}$ and $\distnarg[\Gamma]{W^n,T} = \distnarg[\Gamma]{W^n} \distnarg[\Pi]{T}$.  Then,
\begin{eqnarray}
  \left\| \distnarg[\Pi]{W_T} - \distnarg[\Gamma]{W_T} \right\|_{TV} & \leq & \left\| \distnarg[\Pi]{W^n} - \distnarg[\Gamma]{W^n} \right\|_{TV}.
\end{eqnarray}
\end{lemma}

\begin{IEEEproof}
  Notice that the channel $\distalt[\Pi]{W_T|W^n}{a|w^n} = \sum_{t=1}^n \dist[\Pi]{T} \mathbf{1} \left( a = w_t \right)$ defines the process that takes $W^n$ and selects a random time index according to \distnarg[\Pi]{T}.  This Lemma simply requires that output distributions from a common channel are as close as input distributions in total variation---a consequence of Lemma~\ref{lemma:tv marginal} and Lemma~\ref{lemma:tv channel}.
\end{IEEEproof}

Now we build on the fact that for finite alphabets we can upper-bound the difference in entropy in terms of total variation \cite[Theorem~17.3.3]{cover-thomas06-eit}.

\begin{lemma}[Entropy and Timing Information of Nearly i.i.d. Sequences]
\label{lemma entropy and information bound}
For any discrete random sequence $W^n \sim \distnarg[\Pi]{W^n}$ where $W_t \in {\cal W}$ for all $t \in \{1,...,n\}$, if there exists a distribution \distnarg[\Gamma]{W} on the alphabet ${\cal W}$ such that
\begin{eqnarray}
  \left\| \distnarg[\Pi]{W^n} - \prod \distnarg[\Gamma]{W} \right\|_{TV} & \leq & \epsilon \; < \; 1/4,
\end{eqnarray}
then
\begin{eqnarray}
\label{eq:entropy bound}
  \frac{1}{n} \sum_{t=1}^n I_{\Pi}(W_t;W^{t-1}) & \leq & 4 \epsilon \left( \log |{\cal W}| + \log \frac{1}{\epsilon} \right),
\end{eqnarray}
and for any random variable $T \in \{1,...,n\}$ independent of $W^n$,
\begin{eqnarray}
\label{eq:information bound}
  I_{\Pi}(W_T;T) & \leq & 4 \epsilon \left( \log |{\cal W}| + \log \frac{1}{\epsilon} \right).
\end{eqnarray}
\end{lemma}

\begin{IEEEproof}
We start by applying Lemma~\ref{lemma:tv random time index} for the arbitrary random time index $T$ referred to in the Lemma as well as for each individual deterministic time index (each a special case of \distnarg[\Pi]{T}).  Then by Theorem~17.3.3 of \cite{cover-thomas06-eit},
\begin{eqnarray}
\left| H_{\Pi}(W^n) - H_{\Gamma}(W^n) \right| & \leq & 2 \epsilon \log \left( \frac{|{\cal W}|^n}{\epsilon} \right), \label{eq:entropy sequence 1} \\
\left| H_{\Pi}(W_T) - H_{\Gamma}(W) \right| & \leq & 2 \epsilon \log \left( \frac{|{\cal W}|}{\epsilon} \right), \label{eq:entropy sequence 2} \\
\left| H_{\Pi}(W_t) - H_{\Gamma}(W) \right| & \leq & 2 \epsilon \log \left( \frac{|{\cal W}|}{\epsilon} \right), \label{eq:entropy sequence 3}
\end{eqnarray}
for all $t \in \{1,...,n\}$.

As with any i.i.d. distribution, $H_{\Gamma}(W^n) = \sum_{t=1}^n H_{\Gamma}(W_t)$.  Therefore, the triangle inequality yields,
\begin{eqnarray}
  & & \mkern-54mu \frac{1}{n} \sum_{t=1}^n I_{\Pi}(W_t;W^{t-1}) \nonumber \\
  & = & \frac{1}{n} \left( \left( \sum_{t=1}^n H_{\Pi}(W_t) \right) - H_{\Pi}(W^n) \right) \nonumber \\
  & \leq & \frac{1}{n} \left| H_{\Pi}(W^n) - H_{\Gamma}(W^n) \right| \nonumber \\
  & & \quad \quad \quad + \; \frac{1}{n} \sum_{t=1}^n \left| H_{\Pi}(W_t) - H_{\Gamma}(W) \right| \nonumber \\
  & \stackrel{(a)}{\leq} & \frac{2}{n} \epsilon \log \left( \frac{|{\cal W}|^n}{\epsilon} \right) + 2 \epsilon \log \left( \frac{|{\cal W}|}{\epsilon} \right) \nonumber \\
  & = & 4 \epsilon \log |{\cal W}| + \frac{n+1}{n} 2 \epsilon \log \frac{1}{\epsilon} \nonumber \\
  & \leq & 4 \epsilon \left( \log |{\cal W}| + \log \frac{1}{\epsilon} \right),
\end{eqnarray}
where (a) refers to \eqref{eq:entropy sequence 1} and \eqref{eq:entropy sequence 3}.

Furthermore, denoting the distribution of $T$ as \distnarg[\Pi]{T}, notice that
\begin{eqnarray}
  H_{\Pi}(W_T|T) & = & \sum_{t=1}^n \dist[\Pi]{T} H_{\Pi}(W_t)
\end{eqnarray}
because of the independence of $T$ and $W^n$.  Therefore,
\begin{eqnarray}
  & & \mkern-54mu \left| H_{\Pi}(W_T|T) - H_{\Gamma}(W) \right| \nonumber \\
  & = & \left| \sum_{t=1}^n \dist[\Pi]{T} \left( H_{\Pi}(W_t) - H_{\Gamma}(W) \right) \right| \nonumber \\
  & \leq & \sum_{t=1}^n \dist[\Pi]{T} \left| H_{\Pi}(W_t) - H_{\Gamma}(W) \right| \nonumber \\
  & \stackrel{(a)}{\leq} & \sum_{t=1}^n \dist[\Pi]{T} \; 2 \epsilon \log \left( \frac{|{\cal W}|}{\epsilon} \right) \nonumber \\
  & = & 2 \epsilon \log \left( \frac{|{\cal W}|}{\epsilon} \right),
\end{eqnarray}
where (a) refers to \eqref{eq:entropy sequence 3}.  Combining this with \eqref{eq:entropy sequence 2} gives
\begin{eqnarray}
  I_{\Pi}(W_T;T) & \leq & 4 \epsilon \left( \log |{\cal W}| + \log \frac{1}{\epsilon} \right),
\end{eqnarray}
by way of the triangle inequality.
\end{IEEEproof}

\subsection{Epsilon Rate Region}
\label{subsection epsilon rate region}

Now we use information theoretic inequalities and lemmas~\ref{lemma entropy and information bound} and \ref{lemma cardinality bound} to nearly complete the proof.  We define a region ${\cal S}_{\epsilon}$ for $\epsilon>0$ that gracefully expands the region ${\cal S}$ of the main result.  Then we show that an achievable rate pair $(R,R_0)$ is in ${\cal S}_{\epsilon}$.

Let the {\em epsilon rate region} be defined as
\begin{eqnarray}
\label{definition S epsilon}
  \mkern-18mu {\cal S}_{\epsilon} & \triangleq & \left\{
  \begin{array}{rcl}
    (R,R_0) & \in & {\cal R}^2 : \\
    \exists \; \distnarg{X,Y,U} & \in & {\cal D}_{\epsilon} \mbox{ such that} \\
    R & \geq & I(X;U), \\
    R_0 + R & \geq & I(X,Y;U) - 2g(\epsilon).
  \end{array}
  \right \},
\end{eqnarray}
where
\begin{eqnarray}
\label{definition D epsilon}
  {\cal D}_{\epsilon} & \triangleq & \left\{
  \begin{array}{l}
    \distnarg{X,Y,U} : \\
    \left\| \distnarg{X,Y} - \distnarg[q]{X} \distnarg[q]{Y|X} \right\|_{TV} \leq \epsilon, \\
    X - U - Y \mbox{ Markov}, \\
    |{\cal U}| \leq |{\cal X}| |{\cal Y}| + 1.
  \end{array}
  \right\},
\end{eqnarray}
and
\begin{eqnarray}
\label{definition g epsilon}
  g(\epsilon) & \triangleq & 4 \epsilon \left( \log|{\cal X}| + \log|{\cal Y}| + \log \frac{1}{\epsilon} \right).
\end{eqnarray}

\begin{lemma}[Epsilon Rate Region]
\label{lemma epsilon rate region}
If the rate pair $(R, R_0)$ is achievable for channel \distnarg[q]{Y|X} and source \distnarg[q]{X}, then
\begin{eqnarray}
  (R, R_0) & \in & {\cal S}_{\epsilon} \quad \forall \epsilon > 0.
\end{eqnarray}
\end{lemma}

\begin{IEEEproof}
Since ${\cal S}_{\epsilon}$ shrinks with $\epsilon$, let us only consider $\epsilon < 1/4$.  Let $(R, R_0)$ be achievable.  Then there exists an $(R,R_0,n)$ channel synthesis code such that
\begin{eqnarray}
  \left\| \distnarg{X^n,Y^n} - \prod \distnarg[q]{X} \distnarg[q]{Y|X} \right\|_{TV} & < & \epsilon.
\end{eqnarray}

Let the random variable $T$ be uniformly distributed over the set $\{1,...,n\}$ and independent of the induced joint distribution \distnarg{X^n,Y^n,J,K}.  The variable $T$ will serve as a random time index.  The variable $X_T$ is independent of $T$ because $X^n$ is an i.i.d. source sequence (see \cite{cuff-permuter-cover10-coordination-capacity}, Property 1).  However, $Y_T$ need not be independent of $T$.

We lower bound $R$ by,
\begin{eqnarray}
  n R & \geq & H_P(J) \nonumber \\
  & \geq & H_P(J|K) \nonumber \\
  & \geq & I_P(X^n;J|K) \nonumber \\
  & \stackrel{(a)}{=} & I_P(X^n;J,K) \nonumber \\
  & = & \sum_{t=1}^n I_P(X_t;J,K|X^{t-1}) \nonumber \\
  & \stackrel{(b)}{=} & \sum_{t=1}^n I_P(X_t;J,K,X^{t-1}) \nonumber \\
  & \geq & \sum_{t=1}^n I_P(X_t;J,K) \nonumber \\
  & = & n I_P(X_T;J,K|T) \nonumber \\
  & \stackrel{(c)}{=} & n I_P(X_T;J,K,T), \label{eq:converse rate bound}
\end{eqnarray}
where (a) comes from the problem statement and (b) and (c) are due to the i.i.d. nature of $X^n$.

Similarly, we lower bound the sum rate by,
\begin{eqnarray}
  & & \mkern-54mu n (R_0 + R) \nonumber \\
  & \geq & H_P(J,K) \nonumber \\
  & \geq & I_P(X^n,Y^n;J,K) \nonumber \\
  & = & \sum_{t=1}^n I_P(X_t,Y_t;J,K|X^{t-1},Y^{t-1}) \nonumber \\
  & = & \sum_{t=1}^n I_P(X_t,Y_t;J,K,X^{t-1},Y^{t-1}) \nonumber \\
  & & \quad \quad \quad - \; \sum_{t=1}^n I_P(X_t,Y_t;X^{t-1},Y^{t-1}) \nonumber \\
  & \stackrel{(a)}{\geq} & \sum_{t=1}^n I_P(X_t,Y_t;J,K,X^{t-1},Y^{t-1}) - n g(\epsilon) \nonumber \\
  & \geq & \sum_{t=1}^n I_P(X_t,Y_t;J,K) - n g(\epsilon) \nonumber \\
  & = & n I_P(X_T,Y_T;J,K|T) - n g(\epsilon) \nonumber \\
  & = & n I_P(X_T,Y_T;J,K,T) - n I_P(X_T,Y_T;T) - n g(\epsilon) \nonumber \\
  & \stackrel{(b)}{\geq} & n I_P(X_T,Y_T;J,K,T) - 2n g(\epsilon), \label{eq:converse sum rate bound}
\end{eqnarray}
where (a) and (b) are both consequences of Lemma~\ref{lemma entropy and information bound}, and $g(\epsilon)$ is defined in (\ref{definition g epsilon}).

Notice the Markov chain given by $X_T - (J,K,T) - Y_T$.  This comes about because the entire sequences $X^n$ and $Y^n$ are conditionally independent given $J$ and $K$, according to the problem statement, so in particular conditional independence holds for $X_T$ and $Y_T$ for any specific value of $T=t$.  Therefore, by Lemma~\ref{lemma cardinality bound} we can find a \distnarg[\Gamma]{X,Y,U} such that
\begin{eqnarray}
  |{\cal U}| & \leq & |{\cal X}| |{\cal Y}| + 1, \\
  \distnarg[\Gamma]{X,Y} & = & \distnarg{X_T,Y_T}, \\
  I_{\Gamma}(X;U) & = & I_P(X_T;J,K,T) \\
  I_{\Gamma}(X,Y;U) & = & I_P(X_T,Y_T;J,K,T).
\end{eqnarray}

We see from \eqref{eq:converse rate bound} and \eqref{eq:converse sum rate bound} that \distnarg[\Gamma]{X,Y,U} satisfies the inequalities in \eqref{definition S epsilon}.  What remains is to verify that $\distnarg[\Gamma]{X,Y,U} \in {\cal D}$.  This is indeed confirmed by applying Lemma~\ref{lemma:tv random time index}:
\begin{eqnarray}
  \left\| \distnarg[\Gamma]{X,Y} - \distnarg[q]{X} \distnarg[q]{Y|X} \right\|_{TV} & = & \left\| \distnarg{X_T,Y_T} - \distnarg[q]{X} \distnarg[q]{Y|X} \right\|_{TV} \nonumber \\
  & \leq & \left\| \distnarg{X^n,Y^n} - \prod \distnarg[q]{X} \distnarg[q]{Y|X} \right\|_{TV} \nonumber \\
  & < & \epsilon.
\end{eqnarray}
\end{IEEEproof}

\subsection{Continuity of ${\cal S}_{\epsilon}$ at Zero}
\label{subsection lower semi continuity}

The final step in the proof is to show that the intersection of all ${\cal S}_{\epsilon}$ with $\epsilon>0$ is equal to ${\cal S}$, a closed set.  This may seem like unnecessary detail.  It may seem obvious because of how ${\cal S}_{\epsilon}$ was deliberately designed, namely ${\cal S}_{0} = {\cal S}$, and the non-strict inequalities in the definition of ${\cal S}$ seem to make it a closed set.

There are a few subtle points to consider.  Yes, ${\cal S}$ is closed, but this assertion relies on the cardinality bound of ${\cal U}$.  Also, notice that ${\cal S}_{\epsilon}$ allows not only a relaxation in the sum rate but also a relaxation in the set of distributions ${\cal D}_{\epsilon}$.  We must show that a distribution near the desired input-output distribution does not have a significantly larger achievable rate region, as bounded by ${\cal S}_{\epsilon}$.  Notice that in other work, such as \cite{cuff-permuter-cover10-coordination-capacity}, this complication is avoided by defining the achievable region as the closure of the set of achievable rates and distributions.  In the present work, we define the achievable region as the closure of the set of rates for a given distribution---a more precise characterization of the achievable set---which requires this additional precision in the proof.

\begin{lemma}[Continuity of ${\cal S}_{\epsilon}$ at Zero]
\label{lemma continuity}
  The epsilon rate regions ${\cal S}_{\epsilon}$ decrease to the closed set ${\cal S}$ as $\epsilon$ decreases to zero:
  \begin{eqnarray}
    \bigcap_{\epsilon > 0} {\cal S}_{\epsilon} & = & {\cal S}.
  \end{eqnarray}
\end{lemma}

\begin{IEEEproof}
One direction of equality is trivial because ${\cal S}_{\epsilon}$ shrinks as $\epsilon$ shrinks and ${\cal S}_0 = {\cal S}$:
\begin{eqnarray}
  \bigcap_{\epsilon > 0} {\cal S}_{\epsilon} & \supset & {\cal S}.
\end{eqnarray}

Notice that $\lim_{\epsilon \to 0} g(\epsilon) = 0$.

First we take care of the easy part.  Define ${\cal S}_{\epsilon}'$ to remove the relaxation in the sum rate:
\begin{eqnarray}
\label{definition S epsilon prime}
  \mkern-18mu {\cal S}_{\epsilon}' & \triangleq & \left\{ \!\!
  \begin{array}{rcl}
    (R,R_0) \in {\cal R}^2 & : & \exists \; \distnarg{X,Y,U} \in {\cal D}_{\epsilon} \mbox{ s.t.} \\
    R & \geq & I(X;U), \\
    R_0 + R & \geq & I(X,Y;U).
  \end{array}
  \!\! \right \},
\end{eqnarray}
using the same definition for ${\cal D}_{\epsilon}$ as in \eqref{definition D epsilon}.  Notice that
\begin{eqnarray}
  \bigcap_{\epsilon > 0} {\cal S}_{\epsilon} & \subset & \mbox{Closure} \left( \bigcap_{\epsilon > 0} {\cal S}_{\epsilon}' \right).
\end{eqnarray}
This can be verified by contradiction.  Suppose $(a,b)$ is in the left-hand side but not the right-hand side.  Find the smallest $b^*$ such that $(a,b^*)$ is in the right-hand side.  Then $b^*>b$.  Choose $\epsilon$ small enough to exclude $(a, (b^*+b)/2)$ from ${\cal S}_{\epsilon}'$ and so that $g(\epsilon) < (b^*-b)/2$.  Thus, a contradiction.

Now define the function $f:\Delta^{|{\cal X}||{\cal Y}||{\cal U}|-1} \to {\cal R}^2$ as follows:
\begin{eqnarray}
  f \left( \distnarg{X,Y,U} \right) & = & \left( I(X;U), I(X,Y;U) \right).
\end{eqnarray}
The images $f(D)$ and $f({\cal D}_{\epsilon})$ characterize the rate regions ${\cal S}$ and ${\cal S}_{\epsilon}'$. That is, the Pareto optimal points in the images and the respective rate regions are the same.  Had the rate regions ${\cal S}$ and ${\cal S}_{\epsilon}'$ been defined with equality for the rate constraints rather than inequality, then they would precisely equal the images $f(D)$ and $f({\cal D}_{\epsilon})$.

Notice that
\begin{eqnarray}
  \bigcap_{\epsilon > 0} f({\cal D}_{\epsilon}) & = & f(D),
\end{eqnarray}
because $\bigcap_{\epsilon > 0} {\cal D}_{\epsilon} = {\cal D}$, the sets ${\cal D}_{\epsilon}$ are decreasing subsets (as $\epsilon$ decreases) of the compact probability simplex (due to the cardinality bound), and $f$ is a continuous function.  This implies,
\begin{eqnarray}
  \bigcap_{\epsilon > 0} {\cal S}_{\epsilon}' & = & {\cal S}.
\end{eqnarray}

Finally, ${\cal S}$ is closed due to $f$ continuous and ${\cal D}$ compact.
\end{IEEEproof}

\section{Soft Covering Generalization and Analysis}
\label{section:soft covering continued}

In this section we present a variety of distribution matching results built from the soft covering principle, provide a simple proof, and investigate error exponents for memoryless sources and channels.  We begin with a broad theorem for a general source and channel, from which a subtle improvement to Hayashi's result \cite[Lemma~2]{hayashi06-resolvability} is derived.  We then illustrate a variety of implications of the theorem.

All statements in this section apply to general distributions, with Radon-Nikodym derivatives substituted where appropriate, although probability mass functions are used for notational simplicity.

\subsection{Soft Covering - General Source and Channel}
\label{subsection:cloud general}

\begin{figure}[ht]
\begin{center}
\begin{tikzpicture}[node distance=2cm]
 \node (src1)   [coordinate] {};
 \node (src2)   [coordinate,below=1cm of src1] {};
 \node (enc)    [node,minimum width=12mm,right=15mm of src2] {$f(\cdot)$};
 \node (ch)     [node,minimum width=14mm,minimum height=1cm,rounded corners,right=15mm of enc] {\distnarg[\Phi]{V|W,U}};
 \node (split1a) [coordinate] at (src1 -| enc.center) {};
 \node (split1b) [coordinate] at (src1 -| ch.center) {};
 \node (outline)  [coordinate,right=1cm of ch] {};
 \node (out1)   [coordinate] at (src1 -| outline) {};
 \node (out2)   [coordinate] at (src2 -| outline) {};

 \draw (src1) to node [midway,above] {$W \sim \distnarg[\Phi]{W}$} (src1-|enc.center);
 \draw (src1-|enc.center) to (src1-|ch.center);
 \draw[arw] (src1-|enc.center) to (enc);
 \draw[arw] (src1-|ch.center) to (ch);
 \draw[arw] (enc) to node [midway,above] {$U$} (ch);
 \draw[arw] (ch) to node [midway,above] {$V$} (out2);
\end{tikzpicture}
\caption{Theorem~\ref{lemma:cloud general source and channel} is a statement about the soft covering principle for a general source and channel.  Here a random variable $W$ is the input to a deterministic encoder which produces an output $U$.  A channel then acts on the pair $(W,U)$.  Given any output distribution consistent with the source and channel, Theorem~\ref{lemma:cloud general source and channel} bounds the expected total variation between the desired output distribution and the distribution induced by a randomly constructed encoder.}
\label{figure cloud general source and channel}
\end{center}
\end{figure}
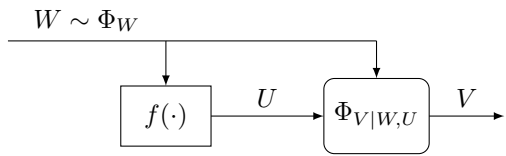

The setting of soft covering for a general source and channel is illustrated in Fig.~\ref{figure cloud general source and channel}.  In order to state the theorem, we first define information density and self-information.

\begin{definition}
The {\em information density} $i_{\distnarg[\Phi]{U,V}}$ for a joint distribution \distnarg[\Phi]{U,V} is a function on the space ${\cal U} \times {\cal V}$ specified by the log-likelihood ratio of the joint distribution to the product distribution:
\begin{eqnarray}
i_{\distnarg[\Phi]{U,V}}(u;v) & \triangleq & \log \frac{\dist[\Phi]{U,V}}{\dist[\Phi]{U} \dist[\Phi]{V}}.
\end{eqnarray}
\end{definition}

\begin{definition}
{\em Self-information} is $i(W) = i(W;W)$.
\end{definition}

Notice that the expected value of information density is mutual information and the expected value of self-information is entropy.
\begin{eqnarray}
\mathbf{E}_{\Phi} \; i_{\distnarg[\Phi]{U,V}}(U;V) & = & I_{\Phi}(U;V), \\
\mathbf{E}_{\Phi} \; i_{\distnarg[\Phi]{W}}(W) & = & H_{\Phi}(W).
\end{eqnarray}

\begin{theorem}[Soft covering - General source and channel]
  \label{lemma:cloud general source and channel}
  For any source distribution \distnarg[\Phi]{W}, codebook distribution \distnarg[\Phi]{U|W}, and channel \distnarg[\Phi]{V|W,U}, we bound the expected total variation error of the distribution of $V$ induced by a randomly constructed codebook.  Let ${\cal B}$ be a randomly generated collection of channel inputs $u(w) \in {\cal U}$, $w \in {\cal W}$, each drawn independently from \distnarg[\Phi]{U|W}.  Let \distnarg{V} be the output distribution induced by applying the codebook, and let $\distnarg[q]{V} = \distnarg[\Phi]{V}$ be the desired output distribution $\sum_{w,u} \distnarg[\Phi]{W} \distnarg[\Phi]{U|W} \distnarg[\Phi]{V|W,U}$.  For any $\tau$,
  \begin{eqnarray}
    \label{eq:cloud mixing general source and channel}
    {\bf E} \left\| \distnarg{V} - \distnarg[q]{V} \right\|_{TV} & \leq & \mathbf{P}_{\Phi} \left( {\cal A}_{\tau}^c \right) + \delta_{\Phi}(\tau),
  \end{eqnarray}
  where ${\cal A}_{\tau}^c$ is the complement of ${\cal A}_{\tau}$, expectation is with respect to the random codebook, and
  \begin{eqnarray}
    \label{definition modified typical set no rate}
    \mkern-36mu {\cal A}_{\tau} & \triangleq & \left\{ (w,u,v) \; : \; i_{\Phi}(w,u;v) - i_{\Phi}(w) \leq \tau \right\}, \\
    \mkern-36mu \delta_{\Phi}(\tau) & \triangleq & \frac{1}{2} \mathbf{E}_{\distnarg[\Phi]{V}} \sqrt{ \mathbf{E}_{\distnarg[\Phi]{W,U|V}} \; 2^{i_{\Phi}(W,U;V) - i_{\Phi}(W)} \mathbf{1}_{{\cal A}_{\tau}} } \label{eq:cloud general tighter bound} \\
    & \leq & \frac{1}{2} 2^{\tau/2} \label{eq:cloud general simpler bound}.
  \end{eqnarray}
\end{theorem}

A simple proof of Theorem~\ref{lemma:cloud general source and channel} is given in \S\ref{subsection cloud proofs}.  The significance of \eqref{eq:cloud general tighter bound} over the simpler relaxation \eqref{eq:cloud general simpler bound} is motivated by Hayashi's derivation in \cite{hayashi06-resolvability} of tighter error exponents in the memoryless channel case based on a bound related to \eqref{eq:cloud general tighter bound}.

Notice that the setting of Theorem~\ref{lemma:cloud general source and channel} is equally general even if the channel \distnarg[\Phi]{V|W,U} does not explicitly depend on $W$.  The random variable $U$ can be chosen to contain $W$ if necessary, producing the same effect.  We choose this presentation because it emphasizes the versatility.

\begin{figure}[ht]
\begin{center}
\begin{tikzpicture}[node distance=2cm]
 \node (src)   [coordinate] {};
 \node (enc)   [node,minimum width=12mm,right=18mm of src] {${\cal B}$};
 \node (ch)    [node,minimum width=14mm,minimum height=1cm,rounded corners,right=15mm of enc] {\distnarg[\Phi]{V|U}};
 \node (outV)  [coordinate,right=1cm of ch] {};

 \draw[arw] (src) to node [midway,above] {$J \sim \text{Unif}$} (enc);
 \draw[arw] (enc) to node [midway,above] {$u(J)$} (ch);
 \draw[arw] (ch) to node [midway,above] {$V$} (outV);
\end{tikzpicture}
\caption{{\em Soft covering - General channel:}  Corollary~\ref{cor:cloud general} arises as a special case of Theorem~\ref{lemma:cloud general source and channel}.  Here, an input to a channel is selected uniformly at random from a codebook ${\cal B}$ in order to induce a desired output distribution.}
\label{figure cloud general channel}
\end{center}
\end{figure}
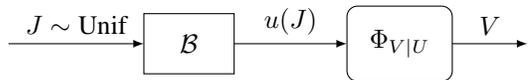

From this theorem we derive a corollary related to known results in the literature.  The setting involves the case where $W$ is independent of $U$ and $V$ and uniformly distributed (represented as $J$ in Fig.~\ref{figure cloud general channel}).

\begin{corollary}[Soft covering - General channel (see Lemma~2 of \cite{hayashi06-resolvability})]
\label{cor:cloud general}
For any channel \distnarg[\Phi]{V|U} and codebook distribution \distnarg[\Phi]{U}, we bound the expected total variation error of the distribution of $V$ induced by a randomly constructed codebook.  Let ${\cal B}$ be a randomly generated collection of channel inputs $u(j) \in {\cal U}$, $j = 1,...,M$, each drawn independently from \distnarg[\Phi]{U}.  Let \distnarg{V} be the output distribution induced by selecting a channel input uniformly at random from the codebook, and let $\distnarg[q]{V} = \distnarg[\Phi]{V}$ be the desired output distribution $\sum_{u \in {\cal U}} \distnarg[\Phi]{U} \distnarg[\Phi]{V|U}$.  For any $\tau$,
\begin{eqnarray}
\label{eq:cloud mixing general}
{\bf E} \left\| \distnarg{V} - \distnarg[q]{V} \right\|_{TV} & \leq & \mathbf{P}_{\Phi} \left( {\cal A}_{\tau}'^c \right) + \delta_{\Phi}'(\tau),
\end{eqnarray}
where ${\cal A}_{\tau}'^c$ is the complement of ${\cal A}_{\tau}'$, expectation is with respect to the random codebook, and
  \begin{eqnarray}
    {\cal A}_{\tau}' & \triangleq & \left\{ (u,v) \; : \; i_{\Phi}(u;v) \leq \tau \right\}, \\
    \mkern-18mu \delta_{\Phi}'(\tau) & \triangleq & \frac{1}{2 \sqrt{M}} \mathbf{E}_{\distnarg[\Phi]{V}} \sqrt{ \mathbf{E}_{\distnarg[\Phi]{U|V}} \; 2^{i_{\Phi}(U;V)} \mathbf{1}_{{\cal A}_{\tau}'} } \label{eq:hayashi tighter bound} \\
    & \leq & \frac{1}{2} \sqrt{ \frac{2^{\tau}}{M} } \label{eq:cloud simpler bound}.
  \end{eqnarray}
\end{corollary}

The above corollary differs from Lemma~2 of \cite{hayashi06-resolvability} only in \eqref{eq:hayashi tighter bound}, which can be relaxed using Jensen's inequality to arrive at the same statement as in \cite{hayashi06-resolvability}, by moving the expectation inside the square root.

A comparison of this bound to the bounds in \cite{han-verdu93-output-statistics} is given in the appendix.

For the next two corollaries we consider an arbitrary sequence of channels and use Theorem~\ref{lemma:cloud general source and channel} to state sufficient conditions for a random codebook to render an output distribution with arbitrarily high fidelity in the limit.  The second of these corollaries specializes to the case of an independent and uniformly distributed source, to recover \cite[Theorem~4]{han-verdu93-output-statistics}.

\begin{definition}
  The {\em limit superior in probability} with respect to $\Phi$ is defined as
  \begin{eqnarray}
    \limsup_{\Phi, \; n \to \infty} W_n & \triangleq & \inf \{ \tau \; : \; \mathbf{P}_{\Phi} (W_n > \tau) \to 0 \}.
  \end{eqnarray}
\end{definition}

\begin{definition}
The {\em sup-information rate} $\bar{I}_{\Phi}(U;V)$ for a sequence of joint distributions \distnarg[\Phi]{U^{(n)},V^{(n)}} of pairs of random variables $(U^{(n)},V^{(n)})$ is defined as
\begin{eqnarray}
\mkern-24mu \bar{I}_{\Phi}(U;V) & \triangleq & \limsup_{\Phi, \; n \to \infty} \frac{1}{n} \; i_{\Phi_{U^{(n)};V^{(n)}}} \left( U^{(n)};V^{(n)} \right).
\end{eqnarray}
\end{definition}

\begin{corollary}[Soft covering - Sequence of sources and channels]
\label{lemma:cloud sequence with source}
Given a sequence of sources, channels, and codebook distributions, specified by \distnarg[\Phi]{W^{(n)}}, \distnarg[\Phi]{V^{(n)}|W^{(n)},U^{(n)}}, and \distnarg[\Phi]{U^{(n)}|W^{(n)}}, respectively, for $n = 1,2,...$, let ${\cal B}^{(n)}$ be a randomly generated collection of channel inputs $u^{(n)}(w^{(n)}) \in {\cal U}^{(n)}$ $\forall w^{(n)} \in {\cal W}^{(n)}$, each drawn independently from \distnarg[\Phi]{U^{(n)}|W^{(n)}}.  Let \distnarg{V^{(n)}} be the output distribution induced by applying the codebook, and let $\distnarg[q]{V^{(n)}} = \distnarg[\Phi]{V^{(n)}}$ be the desired output distribution $\sum_{u^{(n)} \in {\cal U}^{(n)}} \distnarg[\Phi]{U^{(n)}} \distnarg[\Phi]{V^{(n)}|U^{(n)}}$.  The distribution \distnarg{V^{(n)}} is random because the codebook ${\cal B}^{(n)}$ is random.

Then,
\begin{gather*}
  \lim_{\Phi, \; n \to \infty} i_{\Phi}(W^{(n)},U^{(n)};V^{(n)}) - i_{\Phi}(W^{(n)}) \; = \; - \infty \\
  \Downarrow \\
  \lim_{n \to \infty} {\bf E} \left\| \distnarg{V^{(n)}} - \distnarg[q]{V^{(n)}} \right\|_{TV} \; = \; 0.
\end{gather*}
\end{corollary}

\begin{corollary}[Soft covering - Sequence of channels {\cite[Theorem~4]{han-verdu93-output-statistics}}]
\label{lemma:cloud sequence}
Given a sequence of channels and codebook distributions, specified by \distnarg[\Phi]{V^{(n)}|U^{(n)}} and \distnarg[\Phi]{U^{(n)}} for $n = 1,2,...$, let ${\cal B}^{(n)}$ be a randomly generated collection of $2^{nR}$ channel inputs in ${\cal U}^{(n)}$, each drawn independently from \distnarg[\Phi]{U^{(n)}}.  Let \distnarg{V^{(n)}} be the output distribution induced by selecting a channel input uniformly at random from the codebook, and let $\distnarg[q]{V^{(n)}} = \distnarg[\Phi]{V^{(n)}}$ be the desired output distribution $\sum_{u^{(n)} \in {\cal U}^{(n)}} \distnarg[\Phi]{U^{(n)}} \distnarg[\Phi]{V^{(n)}|U^{(n)}}$.  The distribution \distnarg{V^{(n)}} is random because the codebook ${\cal B}^{(n)}$ is random.

Then,
\begin{gather*}
R \; > \;  \bar{I}_{\Phi}(U;V) \\
\Downarrow \\
\lim_{n \to \infty} {\bf E} \left\| \distnarg{V^{(n)}} - \distnarg[q]{V^{(n)}} \right\|_{TV} \; = \; 0.
\end{gather*}
\end{corollary}

\subsection{Implications of Soft Covering}
\label{subsection:cloud variants}

From Theorem~\ref{lemma:cloud general source and channel} we can derive a variety of results about randomly generated but deterministic encoders used to synthesizing a stochastic process.  In this section we highlight some examples involving memoryless channels.  For convenience, we will assume that all random variables are discrete.  However, only corollaries~\ref{cor:cloud local synthesis} and \ref{cor:cloud extension two encoders} require any modification for general distributions.

Through simple entropy arguments, most of the required rates in the corollaries of this section can be shown to be tight, up to a null space in the channel transition matrix, as outlined in the last section of the appendix.

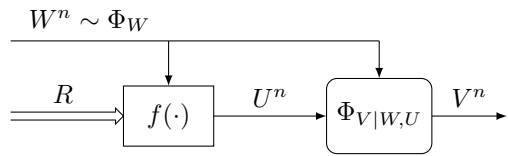
\begin{figure}[ht]
\begin{center}
\begin{tikzpicture}[node distance=2cm]
 \node (src1)   [coordinate] {};
 \node (src2)   [coordinate,below=1cm of src1] {};
 \node (enc)    [node,minimum width=12mm,right=15mm of src2] {$f(\cdot)$};
 \node (ch)     [node,minimum width=14mm,minimum height=1cm,rounded corners,right=15mm of enc] {\distnarg[\Phi]{V|W,U}};
 \node (out)  [coordinate,right=1cm of ch] {};

 \draw (src1) to node [midway,above] {$W^n \sim \distnarg[\Phi]{W}$} (src1-|enc.center);
 \draw (src1-|enc.center) to (src1-|ch.center);
 \draw[arw] (src1-|enc.center) to (enc);
 \draw[arw] (src1-|ch.center) to (ch);
 \rightratearrow{src2.east}{enc.west}{1.5pt}{3pt}{$R$}{3pt};
 \draw[arw] (enc) to node [midway,above] {$U^n$} (ch);
 \draw[arw] (ch) to node [midway,above] {$V^n$} (out);
\end{tikzpicture}
\caption{{\em (Corollary~\ref{cor:cloud sequence to encoder-channel}):}  A deterministic encoder, represented by $f(\cdot)$, is randomly generated according to \distnarg[\Phi]{U|W}.  If $R > I(W,U;V) - H(W)$ then the output $V^n$ is i.i.d. in the limit of large $n$.}
\label{figure cloud sequence to encoder-channel}
\end{center}
\end{figure}

First, consider as a starting point an i.i.d. sequence $W^n$ and a memoryless channel \distnarg[\Phi]{V,U|W} depicted in Fig.~\ref{figure cloud sequence to encoder-channel}.  A deterministic but randomly generated encoder receives both the source $W^n$ and a uniformly distributed variable $J \in [2^{nR}]$.  The following corollary, which serves as a conceptual building block for the remainder of this section, states sufficient rates for the channel output $V^n$ to be i.i.d. in the limit of large $n$.  Notice that the corollary states that the entropy of the source $W$ directly replaced some (or all) of the required random bits $J$ fed into the deterministic encoder.

\begin{corollary}
\label{cor:cloud sequence to encoder-channel}
  Consider any i.i.d. source distribution specified by \distnarg[\Phi]{W}, codebook distribution \distnarg[\Phi]{U|W}, and memoryless channel \distnarg[\Phi]{V|W,U}.  Let ${\cal B}^{(n)}$ be a randomly generated collection of channel inputs $u^n(w^n,j) \in {\cal U}^n$, for all $w^n \in {\cal W}^n$ and $j \in [2^{nR}]$, each drawn independently from $\prod \distnarg[\Phi]{U|W}$.  Let \distnarg{V^n} be the output distribution induced by applying the codebook in the configuration in Fig.~\ref{figure cloud sequence to encoder-channel}, and let $\distnarg[q]{V^n} = \distnarg[\Phi]{V^n}$ be the desired i.i.d. output distribution specified by $\sum_{w,u} \distnarg[\Phi]{W} \distnarg[\Phi]{U|W} \distnarg[\Phi]{V|W,U}$.

  Then,
  \begin{gather*}
  R \; > \; I_{\Phi}(W,U;V) - H_{\Phi}(W) \\
  \Downarrow \\
  \lim_{n \to \infty} {\bf E} \left\| \distnarg{V^n} - \distnarg[q]{V^n} \right\|_{TV} \; = \; 0,
  \end{gather*}
  and convergence occurs exponentially quickly in $n$.
\end{corollary}

The above Corollary~\ref{cor:cloud sequence to encoder-channel} is an immediate consequence of Corollary~\ref{lemma:cloud sequence with source} and the law of large numbers, where $W^n$ and the uniformly distributed index $J$ are together defined as the source in Corollary~\ref{lemma:cloud sequence with source}, and $J$ is independent of the codebook distribution and the channel.  Exponential convergence follows from the technique of \S\ref{subsection exponential convergence}, as with the remaining corollaries of this section.

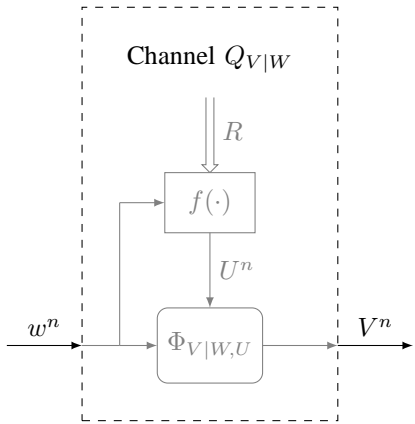
\begin{figure}[ht]
  \begin{center}
    \begin{tikzpicture}[node distance=2cm]
      \node (src) [coordinate] {};

      \begin{scope}[gray]
      \node (bbox1) [coordinate,right=1cm of src] {};
      \node (ch)    [node,minimum width=14mm,minimum height=1cm,rounded corners,right=1cm of bbox1] {\distnarg[\Phi]{V|W,U}};
      \node (enc)   [node,minimum width=12mm,above =1cm of ch] {$f(\cdot)$};
      \node (dummy) [coordinate, above=1cm of enc] {};
      \node (bbox2) [coordinate, right=1cm of ch] {};
      \node (split)  [coordinate, right=5mm of bbox1] {};
      \end{scope}

      \node [rectangle,above=2mm of dummy] {Channel \distnarg[q]{V|W}}; 
      \node (out) [coordinate,right=1cm of bbox2] {};

      \draw[arw] (src) to node [midway,above] {$w^n$} (bbox1);

      \begin{scope}[gray]
      \draw (bbox1) to (split);
      \draw[arw] (split) to (ch);
      \draw (split) to (enc -| split);
      \draw[arw] (enc -| split) to (enc);
      \downratearrow{dummy}{enc.north}{1.5pt}{3pt}{$R$}{3pt};
      \draw[arw] (enc) to node [midway,right] {$U^n$} (ch);
      \draw[arw] (ch) to (bbox2);
      \end{scope}

      \draw[arw] (bbox2) to node [midway,above] {$V^n$} (out);
      \draw[dashed] ([yshift=-1cm] bbox1) rectangle ([yshift=4.5cm] bbox2);
    \end{tikzpicture}
    \caption{{\em Local Channel Synthesis (Corollary~\ref{cor:cloud local synthesis}):}  A memoryless channel \distnarg[q]{V|W} is locally synthesized with $I(U;V|W)$ bits of randomness and a channel \distnarg[\Phi]{V|W,U}.}
    \label{figure local channel}
  \end{center}
\end{figure}

Next, consider locally synthesizing a memoryless channel by making use of a random index and another memoryless channel as the stochastic resources.  The setting is depicted in Fig.~\ref{figure local channel}.  The case where the channel output is equal to the codebook output has been studied in the literature (e.g. \cite{steinberg-verdu94-channel-simulation}).

For the statement of the corollary, we first define empirical distribution.

\begin{definition}
  The {\em empirical distribution} of a sequence $w^n \in {\cal W}^n$ is a probability mass function expressing the frequencies of each $w \in {\cal W}$, denoted
  \begin{eqnarray}
    \mathbb{P}_{w^n}(w) & \triangleq & \frac{1}{n} \sum_{t=1}^n \mathbf{1} (w_t = w).
  \end{eqnarray}
\end{definition}

\begin{corollary}[Local channel synthesis]
\label{cor:cloud local synthesis}
Consider a codebook distribution \distnarg[\Phi]{U|W} and memoryless channel \distnarg[\Phi]{V|W,U}.  Let ${\cal B}^{(n)}$ be a randomly generated collection of channel inputs $u^n(w^n,j) \in {\cal U}^n$, for all $w^n \in {\cal W}^n$ and $j \in [2^{nR}]$, each drawn independently from $\prod \distnarg[\Phi]{U|W}$.  Let \distnarg{V^n|W^n} be the conditional distribution induced by applying the codebook in the configuration in Fig.~\ref{figure local channel}, and let $\distnarg[q]{V^n|W^n} = \distnarg[\Phi]{V^n|W^n}$ be the desired memoryless conditional distribution specified by $\distnarg[\Phi]{V|W} = \sum_{u} \distnarg[\Phi]{U|W} \distnarg[\Phi]{V|W,U}$.

For all $w^n$ having empirical distribution $\mathbb{P}_{w^n}$ such that $R > I_{\mathbb{P}\Phi}(U;V|W) + \gamma_n$, where mutual information is calculated with respect to $\mathbb{P}_{w^n} \distnarg[\Phi]{U,V|W}$, and $\gamma_n \in \omega(1/\sqrt{n})$, the expected value of the total variation between the induced conditional distribution and the desired conditional distribution vanishes uniformly as $n$ grows.  That is, there exists $\epsilon_n$ going to zero, depending only on \distnarg[\Phi]{U,V|W} and $\gamma_n$, such that
  \begin{gather*}
  R \; > \; I_{\mathbb{P}\Phi}(U;V|W) + \gamma_n \\
   \Downarrow \\
   {\bf E} \left\| \distnarg{V^n|W^n=w^n} - \distnarg[q]{V^n|W^n=w^n} \right\|_{TV} \; < \; \epsilon_n \; \to \; 0.
  \end{gather*}
  Furthermore, if $\gamma_n$ is a constant $\gamma > 0$, then $\epsilon_n$ can be chosen to go to zero exponentially fast.
\end{corollary}

Notice that if we had defined $W^n$ to be an i.i.d. source and asked that the induced joint distribution \distnarg{W^n,V^n} approach the desired i.i.d. distribution \distnarg[q]{W^n,V^n}, the result in Corollary~\ref{cor:cloud local synthesis} would be a special case of Corollary~\ref{cor:cloud sequence to encoder-channel}.  This occurs by defining the channel in Corollary~\ref{cor:cloud sequence to encoder-channel} to be \distnarg[\Phi]{W,V|U,W}, which outputs $W$ as well as $V$.  However, Corollary~\ref{cor:cloud local synthesis} is stronger in that it states that the conditional distribution of $V^n$ given $w^n$ will be accurate for all $w^n$ with the appropriate empirical distribution, rather than on average over $w^n$.

Proof of Corollary~\ref{cor:cloud local synthesis} follows from Corollary~\ref{lemma:cloud sequence}.  Define the distribution $\prod_{t=1}^n \distnarg[\Phi]{U|W=w_t}$ as the codebook distribution in Corollary~\ref{lemma:cloud sequence} and the conditional distribution $\prod_{t=1}^n \distnarg[\Phi]{V|U,W=w_t}$ as the channel.  The statement of Corollary~\ref{cor:cloud local synthesis} follows from Chebyshev's inequality, since the input-output information density has mean $n I_{\mathbb{P}\Phi}(U;V|W)$ and standard deviation $O(\sqrt{n})$.

\begin{figure}[ht]
\begin{center}
\begin{tikzpicture}[node distance=2cm]
 \node (src1)   [coordinate] {};
 \node (src2)   [coordinate,below=1cm of src1] {};
 \node (enc)    [node,minimum width=12mm,right=15mm of src2] {$f(\cdot)$};
 \node (ch)     [node,minimum width=14mm,minimum height=1cm,rounded corners,right=15mm of enc] {\distnarg[\Phi]{V|U}};
 \node (out)  [coordinate,right=1cm of ch] {};

 \draw (src1) to node [midway,above] {$W^{r n} \sim \distnarg[\Phi]{W}$} (src1-|enc.center);
 \draw[arw] (src1-|enc.center) to (enc);
 \rightratearrow{src2.east}{enc.west}{1.5pt}{3pt}{$R$}{3pt};
 \draw[arw] (enc) to node [midway,above] {$U^n$} (ch);
 \draw[arw] (ch) to node [midway,above] {$V^n$} (out);
\end{tikzpicture}
\caption{{\em (Corollary~\ref{cor:cloud sequence to encoder}):}  A deterministic encoder, represented by $f(\cdot)$, is randomly generated according to \distnarg[\Phi]{U}.  If $R + r H(W) > I(U;V)$ then the output $V^n$ is i.i.d. in the limit of large $n$.}
\label{figure cloud sequence to encoder}
\end{center}
\end{figure}
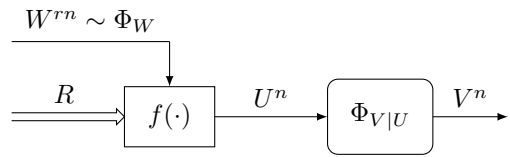

Next, we make a simple extension to Lemma~\ref{lemma:cloud memoryless} that incorporates memoryless sources other than a uniformly distributed random index.  Fig.~\ref{figure cloud sequence to encoder} depicts a deterministic encoder that has access to both a random index at rate $R$ and an i.i.d. source $W^{rn}$ at potentially a different rate than the output, specified by $r$.  The case where the channel \distnarg[\Phi]{V|U} is the identity channel has been studied in depth in the literature (see \cite[Chapter~2]{han02-information-spectrum}).

\begin{corollary}
\label{cor:cloud sequence to encoder}
  Consider any i.i.d. source distribution specified by \distnarg[\Phi]{W}, codebook distribution \distnarg[\Phi]{U}, and memoryless channel \distnarg[\Phi]{V|U}.  Let ${\cal B}^{(n)}$ be a randomly generated collection of channel inputs $u^n(w^{rn},j) \in {\cal U}^n$, for all $w^{rn} \in {\cal W}^{rn}$ and $j \in [2^{nR}]$, each drawn independently from $\prod \distnarg[\Phi]{U}$.  Let \distnarg{V^n} be the output distribution induced by applying the codebook in the configuration in Fig.~\ref{figure cloud sequence to encoder}, and let $\distnarg[q]{V^n} = \distnarg[\Phi]{V^n}$ be the desired i.i.d. output distribution specified by $\sum_{u} \distnarg[\Phi]{U} \distnarg[\Phi]{V|U}$.

  Then
  \begin{gather*}
  R + r H_{\Phi}(W) \; > \; I_{\Phi}(U;V) \\
  \Downarrow \\
  \lim_{n \to \infty} {\bf E} \left\| \distnarg{V^n} - \distnarg[q]{V^n} \right\|_{TV} \; = \; 0,
  \end{gather*}
  and convergence occurs exponentially quickly in $n$.
\end{corollary}

The above Corollary~\ref{cor:cloud sequence to encoder} is an immediate consequence of Corollary~\ref{lemma:cloud sequence with source} and the law of large numbers, where $W^{rn}$ and the uniformly distributed index $J$ are together defined as the source in Corollary~\ref{lemma:cloud sequence with source}, both independent of the codebook distribution and the channel.  Notice that in the synchronous case where $r=1$ there is actually flexibility in designing the codebook.  The result still holds if the codebook is constructed from any conditional distribution \distnarg[\Phi]{U|W} resulting in the same marginal distribution on $U$.

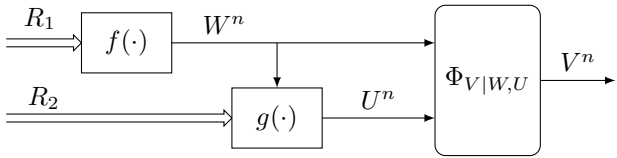
\begin{figure}[ht]
\begin{center}
\begin{tikzpicture}[node distance=2cm]
 \node (src1)   [coordinate] {};
 \node (src2)   [coordinate,below=1cm of src1] {};
 \node (chv)    [coordinate,below=5mm of src1] {};
 \node (enc1)   [node,minimum width=12mm,right=10mm of src1] {$f(\cdot)$};
 \node (enc2)   [node,minimum width=12mm,right=30mm of src2] {$g(\cdot)$};
 \node (split1) [coordinate] at (src1 -| enc2.center) {};
 \node (chh)    [coordinate,right=22mm of enc2] {};
 \node (ch)     [node,minimum width=14mm,minimum height=2cm,rounded corners] at (chv -| chh) {\distnarg[\Phi]{V|W,U}};
 \node (outline)  [coordinate,right=1cm of ch] {};

 \rightratearrow{src1}{enc1.west}{1.5pt}{3pt}{$R_1$}{3pt};
 \draw[draw=none] (src2) to node [midway,above] {$R_2$} (src2 -| enc1.west);
 \rightratearrow{src2}{enc2.west}{1.5pt}{3pt}{}{3pt};
 \draw (enc1) to node [midway,above] {$W^n$} (split1);
 \draw[arw] (split1) to (src1 -| ch.west);
 \draw[arw] (split1) to (enc2);
 \draw[arw] (enc2) to node [midway,above] {$U^n$} (src2 -| ch.west);
 \draw[arw] (ch) to node [midway,above] {$V^n$} (chv -| outline);
\end{tikzpicture}
\caption{{\em Superposition Encoding:}  Two deterministic encoders, represented by $f(\cdot)$ and $g(\cdot)$, are randomly generated according to \distnarg[\Phi]{W} and \distnarg[\Phi]{U|W}.  Sufficient rates for producing an i.i.d. output sequence $V^n$ in the limit of large $n$ are given in Corollary~\ref{cor:cloud extension two encoders}.}
\label{figure cloud extension two encoders}
\end{center}
\end{figure}

In the final derivation of this section, we consider two sources of random bits feeding into two separate deterministic encoders.  The output of the first encoder is fed into the second encoder, as in Fig.~\ref{figure cloud extension two encoders}.  The codebooks together form a superposition codebook.  A similar superposition construction was analyzed in \cite{gohari-anantharam11-multiterminal-strong-coordination-ITW}.

\begin{corollary}[Superposition]
\label{cor:cloud extension two encoders}
  Consider two codebook distributions \distnarg[\Phi]{W} and \distnarg[\Phi]{U|W} and a memoryless channel \distnarg[\Phi]{V|W,U}.  Let ${\cal B}_1^{(n)}$ be a randomly generated collection of sequences $w^n(j) \in {\cal W}^n$, for all $j \in [2^{nR_1}]$, each drawn independently from $\prod \distnarg[\Phi]{W}$, and let ${\cal B}_2^{(n)}$ be a randomly generated collection of sequences $u^n(w^n,k) \in {\cal U}^n$, for all $w^n \in {\cal B}_1^{(n)}$ and $k \in [2^{nR_2}]$, each drawn independently from $\prod \distnarg[\Phi]{U|W}$.  Let \distnarg{V^n} be the output distribution induced by applying the codebooks in the configuration of Fig.~\ref{figure cloud extension two encoders}, and let $\distnarg[q]{V^n} = \distnarg[\Phi]{V^n}$ be the desired i.i.d. output distribution specified by $\sum_{w,u} \distnarg[\Phi]{W} \distnarg[\Phi]{U|W} \distnarg[\Phi]{V|W,U}$.

  Then,
  \begin{gather*}
  \begin{split}
   R_1 & \; > \; I(W;V), \\
   R_2 & \; > \; I(W,U;V) - H(W), \\
   R_1 + R_2 & \; > \; I(W,U;V)
  \end{split} \\
  \Downarrow \\
  \lim_{n \to \infty} {\bf E} \left\| \distnarg{V^n} - \distnarg[q]{V^n} \right\|_{TV} \; = \; 0,
  \end{gather*}
  and convergence occurs exponentially quickly in $n$.
\end{corollary}

To prove Corollary~\ref{cor:cloud extension two encoders}, we need only show how to achieve the two corner points.  First consider rates exceeding $(R_1,R_2) = (I(W;V), I(U;V|W))$.  By Corollary~\ref{cor:cloud local synthesis}, the second encoder $g(\cdot)$, operating at rate $R_2 > I(U;V|W)$, synthesizes a memoryless channel from $W$ to $V$, universally for all $w^n$ with the appropriate empirical distribution, which will occur with exponentially high probability in the random codebook ${\cal B}_1^{(n)}$.  Thus, the first encoder need only operate at a rate $R_1 > I(W;V)$ to induce an i.i.d. output, according to Lemma~\ref{lemma:cloud memoryless}.

Next consider rates exceeding $(R_1,R_2) = (H(W), I(W,U;V) - H(W))$.  According to Corollary~\ref{cor:cloud sequence to encoder} with the identity channel, the first encoder $f(\cdot)$, operating at rate $R_1 > H(W)$, renders the sequence $W^n$ i.i.d. in the limit of large $n$.  With $W^n$ an i.i.d. sequence, Corollary~\ref{cor:cloud sequence to encoder-channel} declares the rate $R_2 > I(W,U;V) - H(W)$ to be sufficient to produce an i.i.d. output.

The proof is completed by drawing on lemmas~\ref{lemma:tv marginal} and \ref{lemma:tv channel} and noting that the expected total variation decays exponentially quickly in each of the supporting lemmas and corollaries.

Notice the similarity between Corollary~\ref{cor:cloud extension two encoders} and the ``generalization of Lemma~6.1 of \cite{cuff08-channel-synthesis-ISIT}'' in \cite{gohari-anantharam11-multiterminal-strong-coordination-ITW}.  The difference is the constraint on $R_2$.  In \cite{gohari-anantharam11-multiterminal-strong-coordination-ITW}, the index $J$ is given to second encoder, instead of only the sequence $w^n(J)$.  Notice that when $R_1 < H(W)$ the index $J$ will be uniquely determined from $w^n(J)$ with high probability.  Thus, the required rate region only differs when $R_1 \geq H(W)$, after which increases to $R_1$ have no further effect in the situation of Corollary~\ref{cor:cloud extension two encoders}.

\subsection{Proof of Theorem~\ref{lemma:cloud general source and channel}}
\label{subsection cloud proofs}

\subsubsection{Explanation}

This proof of soft covering is enabled by two important steps.  First is to define a well-behaved ``typical set'' and separate the induced output distribution into two parts accordingly.  For this we use ${\cal A}_{\tau}$ defined in \eqref{definition modified typical set no rate} of the theorem.  The typical set is intended to contain the bulk of the probability mass.  To analyze the total variation contribution from the typical part, the key step is to use Jensen's inequality in the following way:  $\mathbf{E} | \cdot | \leq \sqrt{ \mathbf{E} ( \cdot )^2 }$.  This allows for the variance analysis in \eqref{eq:variance analysis}.

The steps of this proof are also nearly the same steps used by Hayashi in \cite{hayashi06-resolvability}.  The difference is that Hayashi applies Jensen's inequality to the square-root function twice, which can be avoided by changing the order of summation.

In \cite[Lemma~19]{cuff09-dissertation}, we use these same steps to directly prove the digital rate case in Lemma~\ref{lemma:cloud memoryless}, which is the basic soft covering lemma pertaining to a memoryless channel and uniform and independent source distribution.  Due to those simplifying assumptions, some readers may find the proof in \cite{cuff09-dissertation} easier to internalize because of the reduced notation and more familiar definition of the typical set.

One might also gain intuition by substituting $\dist[\Phi]{W} = \frac{1}{M}$ for $w = 1,...,M$, $\dist[\Phi]{U|W} = \dist[\Phi]{U}$, and $\dist[\Phi]{V|W,U} = \dist[\Phi]{V|U}$ throughout this proof, to directly prove Corollary~\ref{cor:cloud general}, which is the form relevant to channel resolvability.

This proof is stated in terms of discrete random variables, but it can be modified for general random variables with the appropriate use of the Radon-Nikodym derivative.  In the general proof, care must be taken in several places, including expressions such as the definition of ${\cal A}_{\tau}$ in \eqref{definition modified typical set no rate}.  Comparisons of two infinite values are considered to not satisfy the inequality and thus are not included in ${\cal A}_{\tau}$.

\subsubsection{Proof}

Recall that we are given three distributions:  the source distribution \distnarg[\Phi]{W}, the codebook distribution \distnarg[\Phi]{U|W}, and the channel \distnarg[\Phi]{V|W,U}.  The source and channel are stochastic according to their prescribed distributions.  However, the encoder produces a deterministic output $u(w)$.  The induced output distribution is
\begin{eqnarray}
\label{definition induced output no rate}
  \dist{V} & = & \sum_{w \in {\cal W}} \dist[\Phi]{W} \distalt[\Phi]{V|W,U}{v|w,u(w)}.
\end{eqnarray}
The theorem bounds the total variation between the desired output distribution $\distnarg[q]{V} = \distnarg[\Phi]{V}$ that would result from a stochastic encoder that operates according to \distnarg[\Phi]{U|W} and the output distribution induced by the deterministic encoder in \eqref{definition induced output no rate}.  Specifically, we bound the expected total variation when the codebook entries are generated randomly and independently according to the desired conditional distribution \distnarg[\Phi]{U|W}.

For brevity, we omit most subscripts of distributions in this proof, which are clear in context.  Thus, $\Phi(w)$ means \dist[\Phi]{W}.

We start by separating out the contribution to \distnarg{V} coming from typical triples $(w,u,v) \in {\cal A}_{\tau}$.  Let us define two functions on ${\cal V}$ that sum to \distnarg{V}:
\begin{eqnarray}
\mkern-34mu P_1(v) & \triangleq & \sum_{w \in {\cal W}} \Phi(w) \Phi(v|w,U(w)) \; \mathbf{1}_{{\cal A}_{\tau}} \left( w,U(w),v \right), \\
\mkern-34mu P_2(v) & \triangleq & \sum_{w \in {\cal W}} \Phi(w) \Phi(v|w,U(w)) \; \mathbf{1}_{{\cal A}_{\tau}^c} \left( w,U(w),v \right),
\end{eqnarray}
where $\mathbf{1}$ represents the indicator function, ${\cal A}_{\tau}^c$ is the complement of ${\cal A}_{\tau}$, and $U(w)$ is the codebook entry for source $w$ and is a capital letter to represent that the codebook is randomly generated.  Under these definitions,
\begin{eqnarray}
\distnarg{V} & = & P_1 + P_2.
\end{eqnarray}

An important observation about the induced output distribution \distnarg{V}, which is random because the codebook is random, is that it is unbiased with respect to the desired output distribution \distnarg[q]{V}:
\begin{eqnarray}
{\bf E} \dist{V} & \stackrel{(a)}{=} & \sum_{w \in {\cal W}} \Phi(w) {\bf E} \; \Phi(v|w,U(m)) \nonumber \\
& \stackrel{(b)}{=} & \sum_{w \in {\cal W}} \Phi(w) \sum_{u \in {\cal U}} \Phi(u|w)  \Phi(v|w,u) \nonumber \\
& = & \Phi(v) \nonumber \\
& = & \dist[q]{V},
\end{eqnarray}
where (a) is an application of linearity of expectation to the definition of \distnarg{V}, and (b) arises by inserting the distribution of the codebook.

We separate the total variation ${\bf E} \left\| \distnarg{V} - \distnarg[q]{V} \right\|_{TV}$ into two parts:
\begin{eqnarray}
\label{equation separate distribution general no rate}
& & \mkern-54mu {\bf E} \left\| \distnarg{V} - \distnarg[q]{V} \right\|_{TV} \nonumber \\
& = & {\bf E} \left\| \distnarg{V} - {\bf E} \distnarg{V} \right\|_{TV} \nonumber \\
& = & \frac{1}{2} \sum_{v \in {\cal V}} {\bf E} \left| \dist{V} - {\bf E} \dist{V} \right| \nonumber \\
& \stackrel{(a)}{\leq} & \frac{1}{2} \sum_{v \in {\cal V}} {\bf E} \left| P_1(v) - {\bf E} P_1(v) \right| \nonumber \\
& & + \frac{1}{2} \sum_{v \in {\cal V}} {\bf E} \left| P_2(v) - {\bf E} P_2(v) \right|,
\end{eqnarray}
where (a) is due to the triangle inequality.

The first sum in \eqref{equation separate distribution general no rate} is the interesting one to consider, so we save it for last.  The second sum is easy to handle and is small as long as the typical set ${\cal A}_{\tau}$ is likely.  Starting by again making use of the triangle inequality,
\begin{eqnarray}
& & \mkern-54mu \frac{1}{2} \sum_{v \in {\cal V}} {\bf E} \left| P_2(v) - {\bf E} P_2(v) \right| \nonumber \\
& \leq & \sum_{v \in {\cal V}} {\bf E} P_2(v) \nonumber \\
& = & \sum_{v \in {\cal V}} {\bf E} \left( \sum_{w \in {\cal W}} \Phi(w) \Phi(v|w,U(m)) \; \mathbf{1}_{{\cal A}_{\tau}^c} \left( w,U(w),v \right) \right) \nonumber \\
& = & \sum_{(w,v) \in {\cal W}\times{\cal V}} \Phi(w) {\bf E} \left( \Phi(v|w,U(m)) \; \mathbf{1}_{{\cal A}_{\tau}^c} \left( w,U(w),v \right) \right) \nonumber \\
& = & \sum_{(w,v) \in {\cal W}\times{\cal V}} \Phi(w) \sum_{u \in {\cal U}} \Phi(u|w) \Phi(v|w,u) \; \mathbf{1}_{{\cal A}_{\tau}^c} (w,u,v) \nonumber \\
& = & \mathbf{P}_{\Phi} ({\cal A}_{\tau}^c).
\end{eqnarray}

The remaining term in (\ref{equation separate distribution general no rate}) deals with only typical triples.  To bound this term we appeal to a variance bound with the help of Jensen's inequality:
\begin{eqnarray}
{\bf E} \left| P_1(v) - {\bf E} P_1(v) \right| & \leq & \sqrt{{\bf E} \left( P_1(v) - {\bf E} P_1(v) \right)^2} \nonumber \\
& = & \sqrt{{\bf Var} \; P_1(v)}.
\end{eqnarray}
Since the codebook is randomly constructed independently for each $w \in {\cal W}$, the variance of $P_1(v)$ separates as
\begin{eqnarray}
& & \mkern-54mu {\bf Var} \; P_1(v) \nonumber \\
& = & {\bf Var} \left( \sum_{w \in {\cal W}} \Phi(w) \Phi(v|w,U(m)) \; \mathbf{1}_{{\cal A}_{\tau}} \left( w,U(w),v \right) \right) \nonumber \\
& \stackrel{(a)}{=} & \sum_{w \in {\cal W}} {\bf Var} \left( \Phi(w) \Phi(v|w,U(w)) \; \mathbf{1}_{{\cal A}_{\tau}} \left( w,U(w),v \right) \right) \nonumber \\
& \leq & \sum_{w \in {\cal W}} {\bf E} \left( \Phi(w) \Phi(v|w,U(w)) \; \mathbf{1}_{{\cal A}_{\tau}} \left( w,U(w),v \right) \right)^2 \nonumber \\
& = & \sum_{(w,u) \in {\cal W}\times{\cal U}} \Phi(u|w) \left( \Phi(w) \Phi(v|w,u) \; \mathbf{1}_{{\cal A}_{\tau}} (w,u,v) \right)^2 \nonumber \\
& = & \sum_{(w,u) \in {\cal W}\times{\cal U}} \Phi(u|w) \Phi^2(w) \Phi^2(v|w,u) \mathbf{1}_{{\cal A}_{\tau}} \nonumber \\
& = & \sum_{(w,u) \in {\cal W}\times{\cal U}} \Phi(w,u,v) \Phi(w) \Phi(v|w,u) \mathbf{1}_{{\cal A}_{\tau}} \nonumber \\
& = & \Phi^2(v) \sum_{(w,u) \in {\cal W}\times{\cal U}} \Phi(w,u|v) \Phi(w) \frac{\Phi(v|w,u)}{\Phi(v)} \mathbf{1}_{{\cal A}_{\tau}} \nonumber \\
& = & \Phi^2(v) \; \mathbf{E}_{\distnarg[\Phi]{W,U|V}} \; 2^{i_{\Phi}(W,U;V) - i_{\Phi}(W)} \mathbf{1}_{{\cal A}_{\tau}}. \label{eq:variance analysis}
\end{eqnarray}
Equality (a) is due to the independence of the items in the codebook.

The conclusion with respect to the first term in (\ref{equation separate distribution general no rate}) is
\begin{eqnarray}
& & \mkern-54mu \frac{1}{2} \sum_{v \in {\cal V}} {\bf E} \left| P_1(v) - {\bf E} \; P_1(v^n) \right| \nonumber \\
& \leq & \frac{1}{2} \sum_{v \in {\cal V}} \sqrt{ \Phi^2(v) \; \mathbf{E}_{\distnarg[\Phi]{W,U|V}} \; 2^{i_{\Phi}(W,U;V) - i_{\Phi}(W)} \mathbf{1}_{{\cal A}_{\tau}} } \nonumber \\
& = & \frac{1}{2} \mathbf{E}_{\distnarg[\Phi]{V}} \sqrt{ \mathbf{E}_{\distnarg[\Phi]{W,U|V}} \; 2^{i_{\Phi}(W,U;V) - i_{\Phi}(W)} \mathbf{1}_{{\cal A}_{\tau}} } \nonumber \\
& \stackrel{(a)}{\leq} & \frac{1}{2} \mathbf{E}_{\distnarg[\Phi]{V}} \sqrt{ \mathbf{E}_{\distnarg[\Phi]{W,U|V}} 2^{\tau} } \nonumber \\
& = & \frac{1}{2} 2^{\tau/2},
\end{eqnarray}
where (a) results from the definition of ${\cal A}_{\tau}$. \hfill $\blacksquare$

\subsection{Exponents of Total Variation}
\label{subsection exponential convergence}

In this section we derive bounds on the exponential rate of decay of total variation error for soft covering in the memoryless case.  The bounds are stated in terms of discrete distributions but apply generally.

\begin{lemma}[Exponent for Theorem~\ref{lemma:cloud general source and channel}]
\label{lemma:exponent general}
Consider the $n$-fold memoryless extension of Theorem~\ref{lemma:cloud general source and channel}.  That is, the source distribution of Fig.~\ref{figure cloud general source and channel} is the i.i.d. distribution according to \distnarg[\Phi]{W}, the codebook distribution is the memoryless distribution according to \distnarg[\Phi]{U|W}, and the channel is memoryless according to \distnarg[\Phi]{V|W,U}.

If $H_{\Phi}(W) > I_{\Phi}(W,U;V)$, then the expected total variation error vanishes exponentially fast:
\begin{eqnarray}
{\bf E} \left\| \distnarg{V^n} - \distnarg[q]{V^n} \right\|_{TV} & \leq & \frac{3}{2} \exp (-\gamma n),
\end{eqnarray}
where
\begin{eqnarray}
  \label{eq:exponent general}
  \gamma & \triangleq & \max_{\beta, \beta' \geq 0} \; \frac{-\beta'}{2 \beta + \beta'} \log {\bf E}_{\Phi} Z^{\beta} \nonumber \\
  & & + \; \frac{-\beta}{2 \beta + \beta'} \log \left( {\bf E}_{\distnarg[\Phi]{V}} \sqrt{ {\bf E}_{\distnarg[\Phi]{W,U|V}} Z^{1 - \beta'} } \right)^2,
\end{eqnarray}
and
\begin{eqnarray}
  Z & \triangleq & \distalt[\Phi]{W}{W} \frac{ \distalt[\Phi]{V|W,U}{V|W,U} }{ \distalt[\Phi]{V}{V} }. \label{definition Z}
\end{eqnarray}
\end{lemma}

The digital rate case in Lemma~\ref{lemma:cloud memoryless}, which is the basic soft covering lemma, is derived by substituting $\distnarg[\Phi]{W} = 2^{-R}$, $\distnarg[\Phi]{U|W} = \distnarg[\Phi]{U}$, and $\distnarg[\Phi]{V|W,U} = \distnarg[\Phi]{V|U}$.  Technically, $2^R$ is restricted to be an integer for this substitution, but this is merely a technicality that is avoided by a direct proof Lemma~\ref{lemma:cloud memoryless} from Corollary~\ref{cor:cloud general}.  The result of this substitution provides a new achievable exponent for channel resolvability.  We obtain,
\begin{eqnarray}
  \label{eq:exponent digital rate}
  \gamma & = & \max_{\alpha \geq 1, \alpha' \leq 2} \; \frac{\alpha - 1}{2 \alpha - \alpha'} \bigg( R - \breve{I}_{\Phi,\alpha}(U;V) \nonumber \\
  & & + \; (\alpha' - 1) \left( \breve{I}_{\Phi,\alpha}(U;V) - \overline{I}_{\Phi,\alpha'}(U;V) \right) \bigg),
\end{eqnarray}
where $\breve{I}_{\Phi,\alpha}(U;V)$ is the R\'{e}nyi divergence of order $\alpha$ between the joint distribution and the product distribution, defined as
\begin{eqnarray}
  \label{eq:renyi information}
  \mkern-18mu \breve{I}_{\Phi,\alpha}(U;V) & \triangleq & \frac{1}{\alpha - 1} \log {\bf E}_{\Phi} \left( \frac{ \distalt[\Phi]{U,V}{U,V} }{ \distalt[\Phi]{U}{U} \distalt[\Phi]{V}{V} } \right)^{\alpha-1},
\end{eqnarray}
and the formula for $\overline{I}_{\Phi,\alpha'}(U;V)$ is
\begin{equation}
  \label{eq:renyi information modified}
  \frac{1}{\alpha' - 1} \log \left( {\bf E}_{\distnarg[\Phi]{V}} \sqrt{ {\bf E}_{\distnarg[\Phi]{U|V}} \left( \frac{ \distalt[\Phi]{U,V}{U,V} }{ \distalt[\Phi]{U}{U} \distalt[\Phi]{V}{V} } \right)^{\alpha'-1} } \right)^2.
\end{equation}

Notice that the quantity $\overline{I}_{\Phi,\alpha'}(U;V)$ defined above is smaller than $\breve{I}_{\Phi,\alpha'}(U;V)$ and reminiscent of quantities used by Gallager, Arikan, and Arimoto, compared by Csisz\'{a}r in \cite{csiszar95-renyi}.

Some weaker exponents are also of interest.  Define
\begin{eqnarray}
  \hat{\gamma} & \triangleq & \max_{\alpha \in [1,2]} \; \frac{-1}{\alpha} \log {\bf E}_{\Phi} Z^{\alpha-1}, \\
  \hat{\hat{\gamma}} & \triangleq & \max_{\alpha \geq 1} \; \frac{-1}{\alpha + (\alpha - 1)} \log {\bf E}_{\Phi} Z^{\alpha-1}.
\end{eqnarray}
One can verify that these exponents are positive (implying exponential decay) if the condition $H_{\Phi}(W) > I_{\Phi}(W,U;V)$ is satisfied by evaluating the derivative of ${\bf E}_{\Phi} Z^{\alpha-1}$ at $\alpha = 1$.

We arrive at $\hat{\gamma}$ by first relaxing the second term of \eqref{eq:exponent general}, using Jensen's inequality to move the expected value inside the square-root.  We then make the assignment $\alpha - 1 = \beta = 1 - \beta'$.  This assignment can be viewed as an additional relaxation of the bound, although it appears numerically to be optimal and analytically to be at least locally optimal.

When specialized to the digital rate case in Lemma~\ref{lemma:cloud memoryless}, the exponent $\hat{\gamma}$ recovers Hayashi's result in \cite{hayashi06-resolvability}.  That is,
\begin{eqnarray}
  \label{definition gamma hat digital}
  \hat{\gamma} & = & \max_{\alpha \in [1,2]} \; \frac{\alpha - 1}{\alpha} \left( R - \breve{I}_{\Phi,\alpha}(U;V) \right).
\end{eqnarray}
Recall that $\breve{I}_{\Phi,\alpha}(U;V)$ is defined in \eqref{eq:renyi information}.

On the other hand, choosing $\beta' = 1$ eliminates the second term of \eqref{eq:exponent general}, yielding $\hat{\hat{\gamma}}$ as a suboptimal choice.  This exponent corresponds to the relaxed bound in \eqref{eq:cloud general simpler bound}, as is mentioned in the proof below.  Thus, under that simple relaxation, the best exponential bound that can be attained, when specialized to the digital rate case in Lemma~\ref{lemma:cloud memoryless}, is
\begin{eqnarray}
  \label{definition gamma hat hat digital}
  \hat{\hat{\gamma}} & = & \max_{\alpha \geq 1} \; \frac{\alpha - 1}{\alpha + ( \alpha - 1 )} \left( R - \breve{I}_{\Phi,\alpha}(U;V) \right).
\end{eqnarray}

Using the Taylor expansion, it can be shown that $\gamma$, $\hat{\gamma}$, and $\hat{\hat{\gamma}}$ of \eqref{eq:exponent digital rate}, \eqref{definition gamma hat digital}, and \eqref{definition gamma hat hat digital} are approximately equal when $R - I_{\Phi}(U;V)$ is small.  For example, define $\Delta_I$ to be the first derivative of $\breve{I}_{\Phi,\alpha}(U;V)$ with respect to $\alpha$ at $\alpha = 1$.  Assuming $\Delta_I \neq 0$,
\begin{equation}
  \gamma \; \approx \; \hat{\gamma} \; \approx \; \hat{\hat{\gamma}} \; \approx \; \frac{1}{2} \Delta_I \left( R - I_{\Phi}(U;V) \right)^2.
\end{equation}
However, when $R - I_{\Phi}(U;V)$ is large, the optimizing choices of $\alpha$ and $\alpha'$ go to extreme values, and
\begin{eqnarray}
  \gamma & \approx & \frac{1}{2} \left( R - \overline{I}_{\Phi,2}(U;V) \right), \\
  \hat{\gamma} & \approx & \frac{1}{2} \left( R - \breve{I}_{\Phi,2}(U;V) \right), \\
  \hat{\hat{\gamma}} & \approx & \frac{1}{2} \left( R - \breve{I}_{\Phi,\infty}(U;V) \right),
\end{eqnarray}
where $\overline{I}_{\Phi,\alpha}(U;V)$ is defined in \eqref{eq:renyi information modified}.

\begin{IEEEproof}[Proof of Lemma~\ref{lemma:exponent general}]
  Let $\tau$ in the bound of Theorem~\ref{lemma:cloud general source and channel} grow linearly with $n$.  That is,
  \begin{eqnarray}
    \tau & = & n \tau' \label{eq:tau prime}
  \end{eqnarray}

  This proof follows the technique of Hayashi in \cite{hayashi06-resolvability}.  Beginning with the first term of the bound of Theorem~\ref{lemma:cloud general source and channel}, for all $\beta \geq 0$,
  \begin{eqnarray}
    & & \mkern-54mu \mathbf{P}_{\Phi} ( {\cal A}_{\tau}^c ) \nonumber \\
    & = & \mathbf{P}_{\Phi} \left( i_{\Phi} (W^n,U^n;V^n) - i_{\Phi}(W^n) > n \tau' \right) \nonumber \\
    & \stackrel{(a)}{=} & \mathbf{P}_{\Phi} \left( \frac{1}{n} \sum_{t=1}^n \left( i_{\Phi}(W_t,U_t;V_t) - i_{\Phi}(W_t) \right) > \tau' \right) \nonumber \\
    & \stackrel{(b)}{\leq} & \exp \left( n \left( \log \mathbf{E}_{\Phi} 2^{ \beta ( i_{\Phi}(W,U;V) - i_{\Phi}(W) ) - \beta \tau' } \right) \right) \nonumber \\
    & = & \exp \left( n \left( \log {\bf E}_{\Phi} Z^{\beta} - \beta \tau' \log 2 \right) \right), \label{eq:exponent bound part 1}
  \end{eqnarray}
  where (a) is due to the i.i.d. property of the distribution and (b) is the Chernoff bound, which is tight to first order in the exponent.

  Next consider the following upper bound on the indicator function:
  \begin{eqnarray}
    \mathbf{1}_{ \{a \leq b\} } (a,b) & \leq & \left( \frac{b}{a} \right)^{\beta'} \quad \forall a,b,\beta' \geq 0.
  \end{eqnarray}
  By applying
  \begin{eqnarray}
    a & = & 2^{ i_{\Phi}(w^n,u^n;v^n) - i_{\Phi}(w^n) }, \\
    b & = & 2^{n \tau'},
  \end{eqnarray}
  the definition of ${\cal A}_{\tau}$, with the substitution of \eqref{eq:tau prime}, yields
  \begin{eqnarray}
  \label{eq:bound indicator}
    \mkern-28mu \mathbf{1}_{ {\cal A}_{\tau} } (w^n,u^n,v^n) & \leq & \left( \frac{ 2^{n \tau'} } { 2^{ i_{\Phi}(w^n,u^n;v^n) - i_{\Phi}(w^n) } } \right)^{\beta'}, \nonumber \\
    & = & \left( \frac{ 2^{n \tau'} } { \Phi(W^n) \frac{\Phi(V^n|W^n,U^n)}{\Phi(V^n)} } \right)^{\beta'}.
  \end{eqnarray}

  We apply \eqref{eq:bound indicator} to the second term in the bound of Theorem~\ref{lemma:cloud general source and channel}:
  \begin{align}
  \label{eq:exponent bound part 2}
    & \mkern-4mu \delta_{\Phi}(\tau) \nonumber \\
    & = \; \frac{1}{2} \mathbf{E}_{\distnarg[\Phi]{V^n}} \sqrt{ \mathbf{E}_{\distnarg[\Phi]{W^n,U^n|V^n}} \Phi(W^n) \frac{\Phi(V^n|W^n,U^n)}{\Phi(V^n)} \mathbf{1}_{{\cal A}_{\tau}} } \nonumber \\
    & \leq \; \frac{1}{2} \mathbf{E}_{\Phi} \sqrt{ \mathbf{E}_{\distnarg[\Phi]{|V^n}} 2^{n \beta' \tau'} \left( \Phi(W^n) \frac{\Phi(V^n|W^n,U^n)}{\Phi(V^n)} \right)^{1 - \beta'} } \nonumber \\
    & = \; \frac{1}{2} \mathbf{E}_{\Phi} \sqrt{ \mathbf{E}_{\distnarg[\Phi]{|V^n}} 2^{n \beta' \tau'} \left( \prod_{t=1}^n \Phi(W_t) \frac{\Phi(V_t|W_t,U_t)}{\Phi(V_t)} \right)^{1 - \beta'} } \nonumber \\
    & \stackrel{(a)}{=} \; \frac{1}{2} 2^{\frac{n}{2} \beta' \tau'} \prod_{t=1}^n \mathbf{E}_{\Phi} \sqrt{ \mathbf{E}_{\distnarg[\Phi]{|V_t}} \left( \Phi(W_t) \frac{\Phi(V_t|W_t,U_t)}{\Phi(V_t)} \right)^{1 - \beta'} } \nonumber \\
    & \stackrel{(b)}{=} \; \frac{1}{2} 2^{\frac{n}{2} \beta' \tau'} \left( \mathbf{E}_{\distnarg[\Phi]{V}} \sqrt{ \mathbf{E}_{\distnarg[\Phi]{W,U|V}} Z^{1 - \beta'} } \right)^n \\
    & = \; \frac{1}{2} \exp \left( n \left( \frac{1}{2} \beta' \tau'\log 2 + \log \mathbf{E}_{\distnarg[\Phi]{V}} \sqrt{ \mathbf{E}_{\distnarg[\Phi]{W,U|V}} Z^{1 - \beta'} } \right) \right), \nonumber
  \end{align}
  where (a) is from the independence of the distribution at each point in the sequence and (b) uses stationarity.  Notice that $\beta'=1$ gives the relaxation found in \eqref{eq:cloud general simpler bound}.

  Select $\tau'$ so that the exponents of \eqref{eq:exponent bound part 1} and \eqref{eq:exponent bound part 2} are equal:
  \begin{eqnarray}
    \mkern-34mu \tau' & = & \frac{ 2 \log \mathbf{E}_{\Phi} Z^{\beta} - 2 \log \mathbf{E}_{\distnarg[\Phi]{V}} \sqrt{ \mathbf{E}_{\distnarg[\Phi]{W,U|V}} Z^{1 - \beta'} } }{(2\beta + \beta') \log 2}.
  \end{eqnarray}
  The common exponent is then
  \begin{equation}
     \frac{ \beta' \log \mathbf{E}_{\Phi} Z^{\beta} + 2 \beta \log \mathbf{E}_{\distnarg[\Phi]{V}} \sqrt{ \mathbf{E}_{\distnarg[\Phi]{W,U|V}} Z^{1 - \beta'} } }{2\beta + \beta'}.
  \end{equation}
\end{IEEEproof}

\section{Summary}
\label{section summary}

The distributed channel synthesis problem demands unconventional codec constructions, including a stochastic decoder, yet lends itself to a complete information theoretic description of the achievable rate region, found in Theorem~\ref{theorem main result}.  This region reveals that common randomness, independent of the channel input, can replace some of the required communication rate for channel synthesis, reducing the communication rate from Wyner's common information $C(X;Y)$ to Shannon's mutual information $I(X;Y)$.  Also, \S\ref{subsection local randomness} highlights that distributed channel synthesis is as efficient as local channel synthesis in terms of random bits needed by the system.

The main result of Theorem~\ref{theorem main result} can be extended to arbitrary channel input sequences, not necessarily i.i.d., and unknown at the time of the codec design.  This is shown in \cite{bennett-devetak-harrow-shor-winter09-channel-synthesis-Preprint} using a proof based on the method of types and is also obtained in \cite{berta-renes-wilde13-quantum-measurement-Preprint}.  The modification needed in our proof is simple but important.  Rather than make a statement about the expected total variation vanishing with $n$, a stronger soft covering lemma must instead show that the probability of the total variation exceeding a vanishing threshold goes doubly exponentially to zero, which can be accomplished using the Chernoff bound.  This allows the union bound to bridle the exponentially large space of channel input sequences.

Distributed channel synthesis has application to secrecy, game theory, quantum measurements, etc.  Additionally, the proof and coding techniques of this work may be of independent interest.  In particular, the achievability proof embarks on a construction of a feasible joint distribution over all parts of the system, without first specifying the encoder and decoder behavior.  From this, the likelihood encoder is derived.  This approach, and the likelihood encoder, can be utilized for problems in information theory in general.  Furthermore, this work generalizes and extends the concept of soft covering discussed in \S\ref{section:cloud} and \S\ref{section:soft covering continued}, providing a variety of tools for using codebooks and limit randomness to match an output distribution.  In doing so, we derive improved exponents for channel resolvability and, in the appendix, a converse for mean-resolvability of discrete memoryless channels.

\appendix

\subsection{Derivation for Erasure Channel Example of \S\ref{section example erasure channel}}

For any $\distnarg{X,Y,U} \in {\cal D}$, the Markov property constrains that for each value $u$ in the support of $U$ the conditional distribution \distnarg{X,Y|U=u} is a product distribution $\distnarg{X|U=u} \distnarg{Y|U=u}$.  These distributions must fall into three categories, shown in Fig.~\ref{figure erasure three categories}, because the events $(X,Y) = (0,1)$ and $(X,Y) = (1,0)$ have zero probability.

\begin{figure}[ht]
  \begin{center}
    \begin{tikzpicture}
      \node (Actr)   [coordinate] at (0,0) {};
      \node (Ain)    [rectangle,left=1cm of Actr] {$0$};
      \node (Aout1)  [rectangle,right=1cm of Actr] {$0$};
      \node (Aout2)  [rectangle,below=3mm of Aout1] {$\mathsf{e}$};
      \node (Alabel) [rectangle,left=1.5cm of Actr] {Category A};

      \draw (Actr) -- (Ain);
      \draw (Actr) -- (Aout1);
      \draw (Actr) -- (Aout2);

      \node (Bctr)   [coordinate] at (0,-2) {};
      \node (Bdummy) [coordinate,left=1cm of Bctr] {};
      \node (Bin1)    [rectangle,above =3mm of Bdummy] {$0$};
      \node (Bin2)    [rectangle,below =3mm of Bdummy] {$1$};
      \node (Bout)  [rectangle,right=1cm of Bctr] {$\mathsf{e}$};
      \node (Blabel) [rectangle,left=1.5cm of Bctr] {Category C};

      \draw (Bctr) -- (Bin1);
      \draw (Bctr) -- (Bin2);
      \draw (Bctr) -- (Bout);

      \node (Cctr)   [coordinate] at (0,-4) {};
      \node (Cin)    [rectangle,left=1cm of Cctr] {$1$};
      \node (Cout1)  [rectangle,right=1cm of Cctr] {$1$};
      \node (Cout2)  [rectangle,above=3mm of Cout1] {$\mathsf{e}$};
      \node (Clabel) [rectangle,left=1.5cm of Cctr] {Category B};

      \draw (Cctr) -- (Cin);
      \draw (Cctr) -- (Cout1);
      \draw (Cctr) -- (Cout2);
    \end{tikzpicture}
    \caption{{\em Categories of Conditional Distributions:}  Conditional distributions \distnarg{X,Y|U=u} for the erasure channel must fit into one of three categories due to the sparsity of the channel.}
    \label{figure erasure three categories}
  \end{center}
\end{figure}

The three conditional distribution categories are as follows:
\begin{itemize}
\item {\em Category A:}  If $\distnarg{X|U=u} \distnarg{Y|U=u}$ puts positive probability on $Y=0$, then $X=0$ with probability one.
\item {\em Category B:}  The reverse occurs if $\distnarg{X|U=u} \distnarg{Y|U=u}$ puts positive probability on $Y=1$.
\item {\em Category C:}  The only alternative to categories A and B is to put zero probability on $Y \in \{0,1\}$.
\end{itemize}

We now make two observations and prove them out in the subsequent paragraphs.  First, it is sufficient to consider only distributions $\distnarg{X,Y,U} \in {\cal D}$ which have at most one value of $u$ for each of the three categories in Fig.~\ref{figure erasure three categories}.  Notice that this gives us a bound of $|{\cal U}| \leq 3$ for this example, which is less than the nominal bound $|{\cal U}| \leq |{\cal X}||{\cal Y}| + 1 = 7$.  Second, the distribution $\distnarg{X,Y,U}$ is symmetric.  Therefore, the optimal construction of \distnarg{X,Y,U} for synthesizing the symmetric binary erasure channel for symmetric inputs is a concatenation of two symmetric binary erasure channels, as depicted in Fig.~\ref{figure erasure cascade}.

{\em Only one of each category:}  Consider a distribution \distnarg{X,Y,U} where $U$ has two values in its support that are associated with the same category of product distributions (Fig.~\ref{figure erasure three categories}).  Define $W = f(U)$ as the label of the distribution category associated with $U$ (i.e. values of $U$ having the same product distribution category map to the same value of $W$).  The data processing inequality says that $I(X;W) \leq I(X;U)$ and $I(X,Y;W) \leq I(X,Y;U)$.  We simply need to verify that $\distnarg{X,Y,W} \in {\cal D}$---in particular, that the Markov chain property $X-W-Y$ holds.  This follows because in each category either $X$ or $Y$ is deterministic.

{\em Symmetry:}  The desired input-output distribution \distnarg[q]{X,Y} is symmetric.  Consider any candidate distribution $\distnarg{X,Y,U} \in {\cal D}$, where ${\cal U} = \{A,B,C\}$ labels the category of the associated conditional product distribution \distnarg{X,Y|U}.  Define \distnarg{\tilde{X},\tilde{Y},\tilde{U}} to be the flipped distribution where $\tilde{X} = 1-X$, $\tilde{Y} = 1-Y$, and $\tilde{U}$ is equal to $U$ with $A$ and $B$ exchanged.  Clearly \distnarg{\tilde{X},\tilde{Y},\tilde{U}} is also in ${\cal D}$ and produces the same point in ${\cal S}$.  It also has the same property that the value of $\tilde{U}$ correctly labels the category of the product distribution.  Now define the symmetric distribution \distnarg[P']{X,Y,U} to be the average of \distnarg{X,Y,U} and \distnarg{\tilde{X},\tilde{Y},\tilde{U}}.  Noting that the distribution on $(X,Y)$ is constant within ${\cal D}$, the convexity of mutual information with respect to conditional distributions gives $I_{P'}(X;U) \leq I_P(X;U)$ and $I_{P'}(X,Y;U) \leq I_P(X,Y;U)$.  Furthermore, $\distnarg[P']{X,Y,U} \in {\cal D}$ because mixtures of distributions within a category always result in a product distribution, as discussed above.

\subsection{Proofs for \S\ref{section extensions}}

\begin{IEEEproof}[Proof of Lemma~\ref{lemma game theory} in \S\ref{subsection game theory}]
The proof that ${\cal G} \supset \mbox{Convex Hull} ({\cal G}_0)$ follows naturally from Theorem~\ref{theorem main result}.  Notice that any point in ${\cal G}_0$ can be achieved by Player 1 first generating $X^n$ and then using the communication to synthesize a channel with output $Y^n$.  The inequalities in \eqref{definition S} are both satisfied with $(R,R_0) = (C(X;Y),0)$.  The property of total variation stated in \eqref{eq:tv expected value bound} allows us to analyze the payoff as if the actions produced are exactly i.i.d.  Then time sharing gives us the convex hull of ${\cal G}_0$.

For the converse statement, ${\cal G} \subset \mbox{Convex Hull} ({\cal G}_0)$, we need to rule out the possibility that some other use of the communication, not resulting in nearly i.i.d. actions, is more beneficial.  Let $U$ represent the message used for communication.  Notice the following:
\begin{eqnarray}
  n R & \geq & H(U) \nonumber \\
  & \geq & I(U;X^n,Y^n) \nonumber \\
  & = & \sum_{t=1}^n I(U;X_t,Y_t|X^{t-1},Y^{t-1}) \nonumber \\
  & = & n I(U;X_T,Y_T|X^{T-1},Y^{T-1},T),
\end{eqnarray}
and
\begin{eqnarray}
  \Pi & \leq & \frac{1}{n} \sum_{t=1}^n \Pi_t \nonumber \\
  & = & \frac{1}{n} \sum_{t=1}^n \min_{z(\cdot,\cdot)} {\mathbb E} \; \pi(X_t,Y_t,z(X^{t-1},Y^{t-1})) \nonumber \\
  & = & \min_{z(\cdot,\cdot,\cdot)} \frac{1}{n} \sum_{t=1}^n {\mathbb E} \; \pi(X_t,Y_t,z(X^{t-1},Y^{t-1},t)) \nonumber \\
  & = & \min_{z(\cdot,\cdot,\cdot)} {\mathbb E} \; \pi(X_T,Y_T,z(X^{T-1},Y^{T-1},T)),
\end{eqnarray}
where $T$ is an independent random variable uniformly distributed on the set $\{1,...,n\}$.  For simplicity, let us summarize by making the substitution $W = (X^{T-1},Y^{T-1},T)$, $X = X_T$, and $Y = Y_T$.  Notice that we have the Markov chain $X - (U,W) - Y$ by the constraints of the communication.  Then,
\begin{eqnarray}
  R & \geq & I(X,Y;U|W), \\
  \Pi & \leq & \min_{z(\cdot)} {\mathbf E} \; \pi(X,Y,z(W)).
\end{eqnarray}
This is equivalent to the convexification of the points in ${\cal G}_0.$
\end{IEEEproof}

\begin{IEEEproof}[Proof of \S\ref{subsection public channel}]
  Points in ${\cal S}_{PC}$ can be achieved the same way as points in ${\cal S}$ are achieved for the main result of Theorem~\ref{theorem main result} with the additional step of applying a one-time-pad to the communication message.

  To prove that this is optimal (converse), we first use the triangle inequality and Lemma~\ref{lemma:tv marginal} to note that
  \begin{eqnarray}
    & & \mkern-54mu \left\| \distnarg{X^n,Y^n,J} - \distnarg{X^n,Y^n} \distnarg{J} \right\|_{TV} \nonumber \\
    & \leq & \left\| \distnarg{X^n,Y^n,J} - \distnarg{J} \prod \distnarg[q]{X} \distnarg[q]{Y|X}  \right\|_{TV} \nonumber \\
     & & \quad \quad \quad + \; \left\| \distnarg{X^n,Y^n} \distnarg{J} - \distnarg{J} \prod \distnarg[q]{X} \distnarg[q]{Y|X} \right\|_{TV} \nonumber \\
    & = & \left\| \distnarg{X^n,Y^n,J} - \distnarg{J} \prod \distnarg[q]{X} \distnarg[q]{Y|X}  \right\|_{TV} \nonumber \\
     & & \quad \quad \quad + \; \left\| \distnarg{X^n,Y^n} - \prod \distnarg[q]{X} \distnarg[q]{Y|X} \right\|_{TV} \nonumber \\
    & \leq & 2 \left\| \distnarg{X^n,Y^n,J} - \distnarg{J} \prod \distnarg[q]{X} \distnarg[q]{Y|X}  \right\|_{TV}.
  \end{eqnarray}
  By the definition of achievability, the right-hand side can be made arbitrarily small.  We next follow the steps of \S\ref{section converse}.  Notice first that we can use Theorem~17.3.3 of \cite{cover-thomas06-eit} to bound the mutual information,
  \begin{eqnarray}
    & & \mkern-54mu I(X^n,Y^n;J) \nonumber \\
    & = & H(X^n,Y^n) + H(J) - H(X^n,Y^n,J) \nonumber \\
    & \leq & 4 n \epsilon \left( \log |{\cal X}| + \log |{\cal Y}| + R + \log \frac{1}{\epsilon} \right),
  \end{eqnarray}
  where $\epsilon$ is the arbitrarily small total variation tolerance of the synthesis objective.

  Now we replace the steps of \eqref{eq:converse sum rate bound} with
  \begin{eqnarray}
    n R_0 & \geq & H_P(K) \nonumber \\
    & \geq & H_P(K|J) \nonumber \\
    & \geq & I_P(X^n,Y^n;K|J) \nonumber \\
    & \geq & I_P(X^n,Y^n;J,K) \nonumber \\
    & & \quad \quad - \; 4 n \epsilon \left( \log |{\cal X}| + \log |{\cal Y}| + R + \log \frac{1}{\epsilon} \right) \nonumber \\
    & & \vdots
  \end{eqnarray}
  The proof is completed by following the remaining steps of \S\ref{section converse} and altering the definitions of ${\cal S}_{\epsilon}$ and ${\cal D}_{\epsilon}$ in \eqref{definition S epsilon} and \eqref{definition D epsilon} appropriately.
\end{IEEEproof}

\begin{IEEEproof}[Proof of \S\ref{subsection limited memory}]
  To pass statistical tests with limited memory, we use the same construction as in \S\ref{section achievability}.  Notice that \eqref{eq:ach constraint bound} still holds with the rates provided, and we adjust \eqref{eq:ach objective bound} to claim that, uniformly for all $t$,
  \begin{eqnarray}
    \mkern-20mu \lim_{n \to \infty} \mathbf{E} \left\| \distnarg[\Upsilon]{X_{t-B}^t,Y_{t-B}^t} - \prod \distnarg[q]{X} \distnarg[q]{Y|X} \right\|_{TV} & = & 0.
  \end{eqnarray}
  Furthermore, we call on \eqref{eq:cloud mixing memoryless part 2} from Lemma~\ref{lemma:cloud memoryless} to claim that the limits converge exponentially quickly in $n$.

  Taking steps analogous to \eqref{eq:ach summary bound},
  \begin{eqnarray}
    \mkern-20mu & & \mkern-54mu \left\| \distnarg{X_{t-B}^t,Y_{t-B}^t} - \prod \distnarg[q]{X} \distnarg[q]{Y|X} \right\|_{TV} \nonumber \\
    \mkern-20mu & \leq & \left\| \distnarg{X_{t-B}^t,Y_{t-B}^t} - \distnarg[\Upsilon]{X_{t-B}^t,Y_{t-B}^t} \right\|_{TV} \nonumber \\
    \mkern-20mu & & \quad \quad \quad + \; \left\| \distnarg[\Upsilon]{X_{t-B}^t,Y_{t-B}^t} - \prod \distnarg[q]{X} \distnarg[q]{Y|X} \right\|_{TV} \nonumber \\
    \mkern-20mu & \stackrel{(a)}{\leq} & \left\| \distnarg{X^n,Y^n,J,K} - \distnarg[\Upsilon]{X^n,Y^n,J,K} \right\|_{TV} \nonumber \\
    \mkern-20mu & & \quad \quad \quad + \; \left\| \distnarg[\Upsilon]{X_{t-B}^t,Y_{t-B}^t} - \prod \distnarg[q]{X} \distnarg[q]{Y|X} \right\|_{TV} \nonumber \\
    \mkern-20mu & \stackrel{(b)}{=} & \left\| \distnarg{X^n,K} - \distnarg[\Upsilon]{X^n,K} \right\|_{TV} \nonumber \\
    \mkern-20mu & & \quad \quad \quad + \; \left\| \distnarg[\Upsilon]{X_{t-B}^t,Y_{t-B}^t} - \prod \distnarg[q]{X} \distnarg[q]{Y|X} \right\|_{TV} \nonumber \\
    \mkern-20mu & = & \left\| \frac{1}{2^{nR_0}} \left( \prod \distnarg[q]{X} \right) - \distnarg[\Upsilon]{X^n,K} \right\|_{TV} \nonumber \\
    \mkern-20mu & & \quad \quad \quad + \; \left\| \distnarg[\Upsilon]{X_{t-B}^t,Y_{t-B}^t} - \prod \distnarg[q]{X} \distnarg[q]{Y|X} \right\|_{TV},
  \end{eqnarray}
  where (a) is a consequence of Lemma~\ref{lemma:tv marginal}, and (b) uses Lemma~\ref{lemma:tv channel}.

  The expected value of the right-hand side above goes to zero exponentially fast.  Therefore,
  \begin{eqnarray}
    & & \mkern-54mu {\mathbf E} \; \sum_{t=B+1}^n \left\| \distnarg{X_{t-B}^t,Y_{t-B}^t} - \prod \distnarg[q]{X} \distnarg[q]{Y|X} \right\|_{TV} \nonumber \\
    & = & \sum_{t=B+1}^n \; {\mathbf E} \; \left\| \distnarg{X_{t-B}^t,Y_{t-B}^t} - \prod \distnarg[q]{X} \distnarg[q]{Y|X} \right\|_{TV} \nonumber \\
    & \to & 0.
  \end{eqnarray}
  This confirms the existence of a channel synthesis code that passes limited memory statistical tests for all $t$ simultaneously.
\end{IEEEproof}

\begin{IEEEproof}[Proof of \S\ref{subsection local randomness}]
  For limited local randomness, achievability is straightforward.  The construction in \S\ref{section achievability} is valid.  We only need to locally synthesize the channel \distnarg{Y|U}.  This can be done with a rate $R_L > H(Y|U)$ (see Corollary~\ref{cor:cloud local synthesis} of \S\ref{subsection:cloud variants}).

  For the converse, we modify the proof of the epsilon rate region in Lemma~\ref{lemma epsilon rate region} with the following argument, where $K_L$ represents the local randomness available to the decoder:
  \begin{eqnarray}
    n R_L & \geq & H_P(K_L) \nonumber \\
    & \geq & I_P (Y^n;K_L|J,K) \nonumber \\
    & = & H_P (Y^n|J,K) \nonumber \\
    & = & \sum_{t=1}^n H_P (Y_t|J,K,Y^{t-1}) \nonumber \\
    & = & n H_P (Y_T|J,K,Y^{T-1},T).
  \end{eqnarray}
  Notice that $Y^{T-1}$ can be added into \eqref{eq:converse rate bound} alongside $J$, $K$, and $T$, and also it can be left in the derivation of \eqref{eq:converse sum rate bound}.  Also notice that $X_T - (J,K,Y^{T-1},T) - Y_T$ forms a Markov chain.  Therefore, the converse can be complete by replacing $(J,K,Y^{T-1},T)$ with $U$ using Lemma~\ref{lemma cardinality bound}, as in \S\ref{subsection epsilon rate region}.
\end{IEEEproof}

\subsection{Comparison of Soft Covering Lemma to bound in \cite{han-verdu93-output-statistics}}

The proof of Lemma~\ref{lemma:cloud general source and channel} provided in this work, which yields Corollary~\ref{lemma:cloud sequence}, differs from the proof in \cite{han-verdu93-output-statistics}.  Their proof is built around a relationship between the log-likelihood ratio and total variation, encapsulated in Lemma 5 of \cite{han-verdu93-output-statistics}.  Despite some similarity between the proofs, they are fundamentally different and produce different bounds.  We can follow the steps of \cite{han-verdu93-output-statistics} to arrive at an equivalent of Corollary~\ref{cor:cloud general} and make a straightforward comparison to \eqref{eq:cloud mixing general}.  After making appropriate substitutions in \cite{han-verdu93-output-statistics},
\begin{eqnarray}
  \label{eq:han verdu bound}
  & & \mkern-54mu {\bf E} \left\| \distnarg{V} - \distnarg[q]{V} \right\|_{TV} \nonumber \\
  & \leq & \mathbf{P}_{\Phi} \left( i_{\Phi}(U;V) > \tau \right) + 2 \frac{2^\tau}{M} \nonumber \\
  & & + \; \mathbf{P}_{\Phi} \left( i_{\Phi}(U;V) > \log M \right) \\
  & & + \; \left( \frac{M}{2^{\tau}} \right)^2 \mathbf{E} \; 2^{i_{\Phi}(U;V) - \log M} \; \mathbf{1} \left( i_{\Phi}(U;V) \leq \log M \right). \nonumber
\end{eqnarray}

Consider the four terms in \eqref{eq:han verdu bound}.  The first two terms are of similar form to \eqref{eq:cloud mixing general} and in fact smaller for large $M$.  But the second and fourth terms are actually the dominant terms.  Using techniques from the proof of Lemma~\ref{lemma:exponent general} it can be shown that for memoryless sources and channels the above bound proves an exponential decay in total variation with respect to the block-length.  However, the exponent, given below, is smaller than that of Lemma~\ref{lemma:cloud memoryless}:
\begin{equation}
  \min_{\alpha \in [1,2]} \frac{\alpha - 1}{3} \left( R - \breve{I}_{\Phi,\alpha}(U;V) \right),
\end{equation}
where $\breve{I}_{\Phi,\alpha}(U;V)$ is defined in \eqref{eq:renyi information}.

If we follow the method of \cite{han-verdu93-output-statistics} further, we split the fourth term in \eqref{eq:han verdu bound} into two pieces using an indicator function with a carefully chosen threshold.  The result is a simpler bound.  After substituting $\overline{\tau} = \frac{1}{\eta} \tau + \frac{\eta - 1}{\eta} \log M$ for any $\eta \geq 3$ and combining terms, we obtain the following:
\begin{eqnarray}
{\bf E} \left\| \distnarg{V} - \distnarg[q]{V} \right\|_{TV} & \leq & 3 \left( \frac{M}{2^{\overline{\tau}}} \right)^{\frac{\eta-1}{\eta}} \mathbf{P}_{\Phi} \left( i_{\distnarg[\Phi]{U,V}}(U;V) > \overline{\tau} \right) \nonumber \\
& & + \; 3 \left( \frac{M}{2^{\overline{\tau}}} \right)^{-\frac{1}{\eta}} \quad \forall \eta \geq 3.
\end{eqnarray}
In this form, it is easy to verify that this inequality is dominated by \eqref{eq:cloud mixing general} in Corollary~\ref{cor:cloud general}.

\subsection{Mean-resolvability Converse for DMCs}

Many of the soft covering lemmas of \S\ref{section:cloud} and \S\ref{section:soft covering continued} give a tight rate requirement for producing an accurate channel output distribution in the limit of large block-lengths.  This input rate requirement is the definition of channel resolvability \cite{han-verdu93-output-statistics}.  Here we demonstrate a simple converse for memoryless channels based on entropy.  The following method serves also as a converse for mean-resolvability (measured by entropy of the index to the codebook rather than the logarithm of the cardinality), which settles Remark 4 of \cite{han-verdu93-output-statistics}.

Consider the setting of Lemma~\ref{lemma:cloud memoryless}, where $J$ is the stochastic input to a deterministic codebook which produce $u^n(J)$ as the input to a memoryless channel specified by \distnarg[\Phi]{V|U}.  Relax the requirement that $J$ is uniformly distributed, but require that $H(J) \leq n R$, where $n$ is the block-length.  Consider a desired output distribution \distnarg[\Phi]{V} which is uniquely induced through the channel by an input distribution \distnarg[\Phi]{U} (the proof is easily modified if the input distribution is not unique).  We will show that if for all $\epsilon > 0$ there exists a block-length $n$, a stochastic input $J$, and a codebook such at that the induced output distribution \distnarg{V^n} is $\epsilon$-close to the desired output distribution $\distnarg[q]{V^n} = \prod \distnarg[\Phi]{V}$ as measured by total variation, then $R \geq I_{\Phi}(U;V)$.

First consider the following entropy manipulation:
\begin{eqnarray}
  H_P (V^n) & \leq & H_P (J,U^n,V^n) \nonumber \\
  & = & H_P (J,U^n) + H_P (V^n|U^n) \nonumber \\
  & = & H_P (J) + H_P (V^n|U^n) \nonumber \\
  & \leq & nR + H_P (V^n|U^n).
\end{eqnarray}

By Theorem~17.3.3 of \cite{cover-thomas06-eit}, for any $\epsilon < 1/4$,
\begin{eqnarray}
  \mkern-30mu H_P (V^n) & \geq & H_{\Phi} (V^n) - 2 \epsilon \log \left( \frac{|{\cal V}|^n}{\epsilon} \right) \nonumber \\
  & = & n H_{\Phi} (V) - 2 n \epsilon \log |{\cal V}| - 2 \epsilon \log \frac{1}{\epsilon} \nonumber \\
  & \geq & n \left( H_{\Phi} (V) - 2 \epsilon \left( \log |{\cal V}| + \log \frac{1}{\epsilon} \right) \right).
\end{eqnarray}

Finally, interpret $H_P (V^n|U^n)$ as an expected value of the channel entropy over the input distribution.
\begin{eqnarray}
  \mkern-20mu H_P (V^n|U^n) & \stackrel{(a)}{=} & \sum_{t=1}^n H_P (V_t|U_t) \nonumber \\
  & = & \mathbf{E}_{\distnarg{U^n}} \sum_{t=1}^n H_{\Phi} (V|U_t) \nonumber \\
  & = & \sum_{t=1}^n \mathbf{E}_{\distnarg{U_t}} H_{\Phi} (V|U_t),
\end{eqnarray}
where (a) is the memoryless property of the channel.

Now we assert that \distnarg{U_t} is close to \distnarg[\Phi]{U} in total variation.  First notice that $\| \distnarg{V_t} - \distnarg[\Phi]{V} \|_{TV} \leq \| \distnarg{V^n} - \distnarg[q]{V^n} \|_{TV} < \epsilon$ for all $t$ by Lemma~\ref{lemma:tv marginal}.  Also, the channel \distnarg[\Phi]{V|U} acts as a linear function on the input distributions \distnarg{U_t} to produce an output distribution \distnarg{V_t}.  Because the function is continuous on a compact domain, there exists a $\beta(\epsilon)$ which goes to zero as $\epsilon$ goes to zero such that $\| \distnarg{U_t} - \distnarg[\Phi]{U} \|_{TV} < \beta(\epsilon)$.

By the bound in \eqref{eq:tv expected value bound}, for all $t$,
\begin{eqnarray}
  \mkern-36mu \mathbf{E}_{\distnarg{U_t}} H_{\Phi} (V|U_t) & \leq & \mathbf{E}_{\distnarg[\Phi]{U}} H_{\Phi} (V|U) + 2 \beta(\epsilon) \log |{\cal V}|.
\end{eqnarray}

Finally, combining inequalities gives
\begin{eqnarray}
  R & \geq & I_{\Phi}(U;V) - 2 \epsilon \left( \log |{\cal V}| + \log \frac{1}{\epsilon} \right) \nonumber \\
  & & \quad \quad \quad - \; 2 \beta(\epsilon) \log |{\cal V}|.
\end{eqnarray}
Since this statement is true for all $\epsilon \in (0,1/4)$, we conclude that $R \geq I_{\Phi}(U;V)$. \hfill $\blacksquare$

\bibliographystyle{IEEEtran}
\bibliography{it}

\begin{thebibliography}{10}
\providecommand{\url}[1]{#1}
\csname url@samestyle\endcsname
\providecommand{\newblock}{\relax}
\providecommand{\bibinfo}[2]{#2}
\providecommand{\BIBentrySTDinterwordspacing}{\spaceskip=0pt\relax}
\providecommand{\BIBentryALTinterwordstretchfactor}{4}
\providecommand{\BIBentryALTinterwordspacing}{\spaceskip=\fontdimen2\font plus
\BIBentryALTinterwordstretchfactor\fontdimen3\font minus
  \fontdimen4\font\relax}
\providecommand{\BIBforeignlanguage}[2]{{%
\expandafter\ifx\csname l@#1\endcsname\relax
\typeout{** WARNING: IEEEtran.bst: No hyphenation pattern has been}%
\typeout{** loaded for the language `#1'. Using the pattern for}%
\typeout{** the default language instead.}%
\else
\language=\csname l@#1\endcsname
\fi
#2}}
\providecommand{\BIBdecl}{\relax}
\BIBdecl

\bibitem{cuff08-channel-synthesis-ISIT}
P.~Cuff, ``Communication requirements for generating correlated random
  variables,'' in \emph{IEEE Int'l. Symp. on Inf. Theory (ISIT)}, July 2008.

\bibitem{cuff-permuter-cover10-coordination-capacity}
P.~Cuff, H.~Permuter, and T.~Cover, ``Coordination capacity,'' \emph{IEEE
  Trans. Inf. Theory}, vol.~56, no.~9, pp. 4181--4206, Sept. 2010.

\bibitem{anantharam-borkar07-counterexample}
\BIBentryALTinterwordspacing
V.~Anantharam and V.~Borkar, ``Common randomness and distributed control: A
  counterexample,'' \emph{Systems \& Control Letters}, vol.~56, no. 7-8, pp.
  568--572, 2007. [Online]. Available:
  \url{http://www.sciencedirect.com/science/article/pii/S0167691107000540}
\BIBentrySTDinterwordspacing

\bibitem{gilpin-sandholm08-game-incomplete-information-AAMAS}
A.~Gilpin and T.~Sandholm, ``Solving two-person zero-sum repeated games of
  incomplete information,'' in \emph{7th international joint conference on
  autonomous agents and multiagent systems (AAMAS)}, 2008.

\bibitem{gacs-korner73-common-information}
P.~G\'{a}cs and J.~K\"{o}rner, ``Common information is far less than mutual
  information,'' \emph{Problems of Control and Inf. Theory}, vol.~2, pp.
  149--162, 1973.

\bibitem{wyner75-common-information}
A.~Wyner, ``The common information of two dependent random variables,''
  \emph{IEEE Trans. Inf. Theory}, vol.~21, no.~2, pp. 163--179, March 1975.

\bibitem{bennett-shor-smolin-thapliyal02-reverse-shannon-theorem}
C.~Bennett, P.~Shor, J.~Smolin, and A.~Thapliyal, ``Entanglement-assisted
  capacity of a quantum channel and the reverse shannon theorem,'' \emph{IEEE
  Trans. Inf. Theory}, vol.~48, no.~10, pp. 2637--2655, Oct. 2002.

\bibitem{soljanin02}
E.~Soljanin, ``Compressing quantum mixed-state sources by sending classical
  information,'' \emph{IEEE Trans. Inf. Theory}, vol.~48, no.~8, pp.
  2263--2275, 2002.

\bibitem{bennett-shor-smolin-thapliyal99-pre-reverse-shannon}
\BIBentryALTinterwordspacing
C.~Bennett, P.~Shor, J.~Smolin, and A.~Thapliyal, ``Entanglement-assisted
  classical capacity of noisy quantum channels,'' \emph{Phys. Rev. Lett.},
  vol.~83, pp. 3081--3084, Oct. 1999. [Online]. Available:
  \url{http://link.aps.org/doi/10.1103/PhysRevLett.83.3081}
\BIBentrySTDinterwordspacing

\bibitem{bennett-devetak-harrow-shor-winter09-channel-synthesis-Preprint}
C.~Bennett, I.~Devetak, A.~Harrow, P.~Shor, and W.~A., ``Quantum reverse
  shannon theorem,'' April 2012, submitted to IEEE Trans. Inf. Theory,
  arXiv:0912.5537.

\bibitem{berta-christandl-renner11-quantum-reverse-shannon-theorem}
\BIBentryALTinterwordspacing
M.~Berta, M.~Christandl, and R.~Renner, ``\BIBforeignlanguage{English}{The
  quantum reverse shannon theorem based on one-shot information theory},''
  \emph{\BIBforeignlanguage{English}{Communications in Mathematical Physics}},
  vol. 306, no.~3, pp. 579--615, 2011. [Online]. Available:
  \url{http://dx.doi.org/10.1007/s00220-011-1309-7}
\BIBentrySTDinterwordspacing

\bibitem{winter02-reverse-shannon-finite-randomness-Arxiv}
A.~Winter, ``Compression of sources of probability distributions and density
  operators,'' Aug. 2002, arXiv:quant-ph/0208131.

\bibitem{winter04-quantum-measurements}
\BIBentryALTinterwordspacing
------, ``{\em Extrinsic} and {\em intrinsic} data in quantum measurements:
  Asymptotic convex decomposition of positive operator valued measures,''
  \emph{Communications in Mathematical Physics}, vol. 244, pp. 157--185, 2004,
  10.1007/s00220-003-0989-z. [Online]. Available:
  \url{http://dx.doi.org/10.1007/s00220-003-0989-z}
\BIBentrySTDinterwordspacing

\bibitem{wilde-hayden-buscemi-hsieh12-simulating-quantum-measurements}
\BIBentryALTinterwordspacing
M.~Wilde, P.~Hayden, F.~Buscemi, and M.-H. Hsieh, ``The information-theoretic
  costs of simulating quantum measurements,'' \emph{Journal of Physics A:
  Mathematical and Theoretical}, vol.~45, no.~45, p. 453001, 2012. [Online].
  Available: \url{http://stacks.iop.org/1751-8121/45/i=45/a=453001}
\BIBentrySTDinterwordspacing

\bibitem{harsha-jain-mcallester-radhakrishnan07-exact-synthesis-rejection-samp%
ling-ICC}
P.~Harsha, R.~Jain, D.~McAllester, and J.~Radhakrishnan, ``The communication
  complexity of correlation,'' in \emph{Twenty-Second Annual IEEE Conference on
  Computational Complexity (CCC)}, June 2007.

\bibitem{cubitt-leung-matthews-winter11-exact-synthesis}
T.~Cubitt, D.~Leung, W.~Matthews, and A.~Winter, ``Zero-error channel capacity
  and simulation assisted by non-local correlations,'' \emph{IEEE Trans. Inf.
  Theory}, vol.~57, no.~8, pp. 5509--5523, Aug. 2011.

\bibitem{han-verdu93-output-statistics}
T.~Han and S.~Verd\'{u}, ``Approximation theory of output statistics,''
  \emph{IEEE Trans. Inf. Theory}, vol.~39, no.~3, pp. 752--772, May 1993.

\bibitem{gohari-anantharam11-multiterminal-strong-coordination-ITW}
A.~Gohari and V.~Anantharam, ``Generating dependent random variables over
  networks,'' in \emph{IEEE Information Theory Workshop (ITW)}, Oct. 2011.

\bibitem{yassaee-aref-gohari12-random-binning-ISIT}
M.~Yassaee, M.~Aref, and A.~Gohari, ``Achievability proof via output statistics
  of random binning,'' in \emph{IEEE Int'l. Symp. on Inf. Theory (ISIT)}, July
  2012.

\bibitem{yassaee-gohari-aref12-channel-simulation-interactive-ISIT}
M.~Yassaee, A.~Gohari, and M.~Aref, ``Channel simulation via interactive
  communications,'' in \emph{IEEE Int'l. Symp. on Inf. Theory (ISIT)}, July
  2012.

\bibitem{haddadpour-yassaee-gohari-aref12-coordination-via-relay-ISIT}
F.~Haddadpour, M.~Yassaee, A.~Gohari, and M.~Aref, ``Coordination via a
  relay,'' in \emph{IEEE Int'l. Symp. on Inf. Theory (ISIT)}, July 2012.

\bibitem{satpathy-cuff13-cascade-ISIT}
S.~Satpathy and P.~Cuff, ``Secure cascade channel synthesis,'' in \emph{IEEE
  Int'l. Symp. on Inf. Theory (ISIT)}, July 2013.

\bibitem{cuff09-dissertation}
P.~Cuff, ``Communication in networks for coordinating behavior,'' Ph.D.
  dissertation, Stanford University, Aug. 2009.

\bibitem{bennett-devetak-harrow-shor-winter07-reverse-shannon-Presentation}
C.~Bennett, I.~Devetak, A.~Harrow, P.~Shor, and A.~Winter, ``Quantum reverse
  shannon theorem,'' 2007, presentation:
  http://www.research.ibm.com/people/b/bennetc/QRSTonlineVersion.pdf.

\bibitem{witsenhausen76-common-information}
\BIBentryALTinterwordspacing
H.~Witsenhausen, ``Values and bounds for the common information of two discrete
  random variables,'' \emph{SIAM Journal on Applied Mathematics}, vol.~31,
  no.~2, pp. 313--333, 1976. [Online]. Available:
  \url{http://epubs.siam.org/doi/abs/10.1137/0131026}
\BIBentrySTDinterwordspacing

\bibitem{winter05-triples}
A.~Winter, ``Secret, public and quantum correlation cost of triples of random
  variables,'' in \emph{IEEE Int'l. Symp. on Inf. Theory (ISIT)}, Sept. 2005.

\bibitem{cuff10-lossless-secrecy-Globecom}
P.~Cuff, ``A framework for partial secrecy,'' in \emph{IEEE Global
  Telecommunications Conference (GLOBECOM)}, Dec. 2010.

\bibitem{cuff10-partial-secrecy-Allerton}
------, ``Using a secret key to foil an eavesdropper,'' in \emph{48th Annual
  Allerton Conference on Communication, Control, and Computing (Allerton)},
  Oct. 2010.

\bibitem{schieler-cuff12-zero-rate-secrecy-ISIT}
C.~Schieler and P.~Cuff, ``Secrecy is cheap if the adversary must
  reconstruct,'' in \emph{IEEE Int'l. Symp. on Inf. Theory (ISIT)}, July 2012.

\bibitem{steinberg-verdu96-simulation-random-processes}
Y.~Steinberg and S.~Verd\'{u}, ``Simulation of random processes and
  rate-distortion theory,'' \emph{IEEE Trans. Inf. Theory}, vol.~42, no.~1, pp.
  63--86, Jan. 1996.

\bibitem{bloch-laneman11-secrecy-from-resolvability-Arxiv}
M.~Bloch and J.~N. Laneman, ``Secrecy from resolvability,'' 2011, submitted to
  IEEE Trans. Inf. Theory, arXiv:1105.5419.

\bibitem{cover-thomas06-eit}
T.~Cover and J.~Thomas, \emph{Elements of Information Theory (Wiley Series in
  Telecommunications and Signal Processing)}.\hskip 1em plus 0.5em minus
  0.4em\relax Wiley-Interscience, 2006.

\bibitem{cuff09-bayesian-games-Allerton}
P.~Cuff, ``State information in bayesian games,'' Nov. 2009, presented at
  Allerton, arXiv:0911.0874.

\bibitem{bloch-kliewer12-constrained-randomness-Arxiv}
M.~Bloch and J.~Kliewer, ``On secure communication with constrained
  randomization,'' in \emph{IEEE Int'l. Symp. on Inf. Theory (ISIT)}, July
  2012.

\bibitem{steinberg-verdu94-channel-simulation}
Y.~Steinberg and S.~Verd\'{u}, ``Channel simulation and coding with side
  information,'' \emph{IEEE Trans. Inf. Theory}, vol.~40, no.~3, pp. 634--646,
  May 1994.

\bibitem{gray-wyner74-source-coding}
R.~Gray and A.~Wyner, ``Source coding for a simple network,'' \emph{Bell
  Systems Technical Journal}, vol.~53, no.~9, pp. 1681--1721, Nov. 1974.

\bibitem{maurer93-secret-key-agreement}
U.~Maurer, ``Secret key agreement by public discussion from common
  information,'' \emph{IEEE Trans. Inf. Theory}, vol.~39, no.~3, pp. 733--742,
  May 1993.

\bibitem{ahlswede-csiszar93-secret-key-agreement-part-1}
R.~Ahlswede and I.~Csisz\'{a}r, ``Common randomness in information theory and
  cryptography. i. secret sharing,'' \emph{IEEE Trans. Inf. Theory}, vol.~39,
  no.~4, pp. 1121--1132, July 1993.

\bibitem{ahlswede-csiszar98-secret-key-agreement-part-2}
------, ``Common randomness in information theory and cryptography. ii. cr
  capacity,'' \emph{IEEE Trans. Inf. Theory}, vol.~44, no.~1, pp. 225--240,
  Jan. 1998.

\bibitem{csiszar-narayan00-secret-key-agreement}
I.~Csisz\'{a}r and P.~Narayan, ``Common randomness and secret key generation
  with a helper,'' \emph{IEEE Trans. Inf. Theory}, vol.~46, no.~2, pp.
  344--366, March 2000.

\bibitem{maurer-wolf00-strong-secret-key-agreement}
U.~Maurer and S.~Wolf, ``Information-theoretic key agreement: From weak to
  strong secrecy for free,'' in \emph{Advances in Cryptology — EUROCRYPT 2000},
  ser. Lecture Notes in Computer Science, B.~Preneel, Ed.\hskip 1em plus 0.5em
  minus 0.4em\relax Springer Berlin / Heidelberg, 2000, vol. 1807, pp.
  351--368.

\bibitem{hayashi06-resolvability}
M.~Hayashi, ``General nonasymptotic and asymptotic formulas in channel
  resolvability and identification capacity and their application to the
  wiretap channel,'' \emph{IEEE Trans. Inf. Theory}, vol.~52, no.~4, pp.
  1562--1575, 2006.

\bibitem{ahlswede-winter02-quantum-id-converse}
R.~Ahlswede and A.~Winter, ``Strong converse for identification via quantum
  channels,'' \emph{IEEE Trans. Inf. Theory}, vol.~48, no.~3, pp. 569--579,
  2002.

\bibitem{wilde11-quantum-textbook}
M.~Wilde, ``From classical to quantum shannon theory,'' 2011, arXiv:1106.1445.

\bibitem{schieler13-coodination-codes}
C.~Schieler and P.~Cuff, ``A connection between good rate-distortion codes and
  backward dmcs,'' in \emph{IEEE Information Theory Workshop (ITW)}, Sept.
  2013.

\bibitem{caratheodory11-convex-set}
\BIBentryALTinterwordspacing
C.~Carathéodory, ``\"{U}ber den variabilit\"{a}tsbereich der fourier'schen
  konstanten von positiven harmonischen funktionen,'' \emph{Rendiconti del
  Circolo Matematico di Palermo (1884 - 1940)}, vol.~32, pp. 193--217, 1911,
  10.1007/BF03014795. [Online]. Available:
  \url{http://dx.doi.org/10.1007/BF03014795}
\BIBentrySTDinterwordspacing

\bibitem{steinitz13-caratheodory-bound}
E.~Steinitz, ``Bedingt konvergente reihen und konvexe systeme,'' \emph{J. Reine
  Angew. Math.}, vol. 143, pp. 128--175, 1913.

\bibitem{eggleston63-caratheodory-bound}
H.~Eggleston, \emph{Convexity}.\hskip 1em plus 0.5em minus 0.4em\relax
  Cambridge University Press, 1963.

\bibitem{salehi78-cardinality-bound}
M.~Salehi, ``Cardinality bounds on auxiliary variables in multiple-user theory
  via the method of ahlswede and k\"{o}rner,'' \emph{Technical Report, Stanford
  University}, no.~33, Aug. 1978.

\bibitem{csiszar-korner11-information-theory-book}
I.~Csisz\'{a}r and J.~K\"{o}rner, \emph{Information Theory: Coding Theorems for
  Discrete Memoryless Systems}, 2nd~ed.\hskip 1em plus 0.5em minus 0.4em\relax
  Cambridge University Press, 2011.

\bibitem{elgamal-kim11-nit}
A.~El~Gamal and Y.-H. Kim, \emph{Network Information Theory}.\hskip 1em plus
  0.5em minus 0.4em\relax Cambridge University Press, 2011.

\bibitem{han02-information-spectrum}
T.~Han, \emph{Information-spectrum methods in information theory}, ser.
  Applications of Mathematics.\hskip 1em plus 0.5em minus 0.4em\relax Springer,
  2003, vol.~50.

\bibitem{csiszar95-renyi}
I.~Csisz\'{a}r, ``Generalized cutoff rates and {R}\'{e}nyi's information
  measures,'' \emph{IEEE Trans. Inf. Theory}, vol.~41, no.~1, pp. 26--34, 1995.

\bibitem{berta-renes-wilde13-quantum-measurement-Preprint}
M.~Berta, J.~Renes, and M.~Wilde, ``Identifying the information gain of a
  quantum measurement,'' January 2013, arXiv:1301.1594.

\end{thebibliography}

\end{document}